\documentclass[12pt]{article}
\usepackage[centertags]{amsmath}
\usepackage{amsfonts}
\usepackage{amssymb}
\usepackage{bbm}
\usepackage{amsthm}
\usepackage{newlfont}
\usepackage[a4paper]{geometry}
\usepackage{graphicx}
\usepackage{dsfont}
\usepackage[pdftex,bookmarks=true]{hyperref}

\numberwithin{equation}{subsection}
\newcommand{\sEig}[1]{(j_{#1}+1)}
\newcommand{\aEig}[1]{j_{#1}(j_{#1}+2)}
\newcommand{\fEig}[1]{\epsilon_{#1}(j_{#1}+\frac{1}{2})}
\newcommand{\thj}[6]{\begin{pmatrix} #1 & #2 & #3 \\ #4 & #5 & #6 \end{pmatrix}}

\begin{document}
\begin{titlepage}
\begin{center}
\hfill{\tt WIS/11/09-AUG-DPP}\\
\vskip 20mm

{\Large{\bf The 2-loop partition function of large $N$ gauge theories with adjoint matter on $S^3$}}

\vskip 10mm

{\bf Matan Mussel, Ran Yacoby\footnote{Emails: matan.mussel@weizmann.ac.il, ran.yacoby@weizmann.ac.il}
}

\vskip 4mm
{\em Department of Particle Physics,}\\
{\em Weizmann Institute of Science,}\\
{\em Rehovot 76100, Israel}\\
[2mm]

\end{center}
\vskip 1in

\begin{center} {\bf ABSTRACT }\end{center}
\begin{quotation}
\noindent
We compute the 2-loop thermal partition function of Yang-Mills theory on a small 3-sphere, in the large $N$ limit with weak 't~Hooft coupling $\lambda=g_{YM}^2N$. We include $N_s$ scalars and $N_f$ chiral fermions in the adjoint representation of the gauge group $(S)U(N)$, with arbitrary Yukawa and quartic scalar couplings, assuming only commutator interactions. From this computation one can extract information on the perturbative corrections to the spectrum of the theory, and the correction to its Hagedorn temperature. Furthermore, the computation of the 2-loop partition function is a necessary step towards determining the order of the deconfinement phase transition at weak coupling, for which a 3-loop computation is needed.
\end{quotation}

%%%%%%%%%%%%%%
\vfill
%%%%%%%%%%%%%%%%%%
%\flushleft{August 2009}
%%%%%%%%%%%%%%%%%%

\end{titlepage}

\eject

%-----------------------------------------------------------------------------
%-----------------------------------------------------------------------------

\pagebreak\tableofcontents

%-----------------------------------------------------------------------------
%-----------------------   start of introduction   ---------------------------
%-----------------------------------------------------------------------------
\pagebreak\section{Introduction}\label{sec:introduction}

%------------------   start of "Overview and outline"-------------------------
\subsection{Overview and outline}\label{ssec:Overview}
The interest in the thermodynamics of large $N$ Yang-Mills (\textbf{YM}) theories on a small 3-sphere $S^3$, originates in \cite{Witten:1998zw}. In that paper, the AdS/CFT correspondence \cite{Maldacena:1997re, Witten:1998qj, Gubser:1998bc} was used to study the thermal behavior of $\cal{N}=4$ supersymmetric Yang-Mills theory (\textbf{SYM}) on $S^3$, with gauge group $(S)U(N)$. As the correspondence dictates, in the large $N$ limit with strong 't~Hooft coupling $\lambda=g_{YM}^2N$, this behavior can be found by studying classical type IIB supergravity. This study showed that the theory has a low temperature confining phase, in which the free energy scales as $N^0$, and a high temperature deconfining  phase, in which the free energy scales as $N^2$. The two phases were found to be separated by a first order phase transition.

It is then interesting to ask whether this behavior persists even at small 't~Hooft coupling. In that limit the conformal $\cal{N}=4$ theory can be studied by using perturbation theory. However the study needs not be restricted to weakly coupled conformal theories, and can be extended to asymptotically free \textbf{YM} theories. As opposed to conformal field theories, asymptotically free gauge theories generate an infrared scale $\Lambda_{YM}$ which leads to the breakdown of perturbation theory in flat space at low temperatures/energy scales. But on $S^3$ with radius $R_{S^3}$ that bad infrared behavior is cut off by the finite size of space, i.e. if $R_{S^3}\Lambda_{YM}\ll 1$ one can use perturbation theory to compute the partition function at all temperatures.

This has been done in \cite{Sundborg:1999ue},\cite{Aharony:2003sx}, where it was shown that for a wide class of models the first order deconfinement transition persists even in the limit of zero 't~Hooft coupling. The phase transition is signaled by a Hagedorn behavior \cite{Hagedorn:1965st} near the critical temperature, where the density of states grows exponentially with energy. This remarkable behavior can be ascribed to a zero mode of the gauge field (i.e. a mode with vanishing quadratic action) which cannot be integrated out in perturbation theory. Gauge invariance highly constrains the low energy effective action of this mode, resulting in non-trivial self interactions for it even at zero coupling.

To order $\lambda$ the phase transition is of first order and the critical temperature is identical with the Hagedorn temperature. However, in the next order in perturbation theory \cite{Aharony:2003sx}, the transition might be of first order and slightly below the Hagedorn temperature, or second order and at the Hagedorn temperature. In order to determine the order of the phase transition one needs to perform a 3-loop computation in perturbation theory. For pure \textbf{YM} this has been done in \cite{Aharony:2005bq}, where it was found that the transition is of first order. For \textbf{YM} theories with adjoint matter fields (including $\cal{N}=4$ \textbf{SYM}) the order of the phase transition at weak coupling is still an open question and examples which exhibit either behavior can be easily constructed.

In this work we take a step further towards answering this question, by computing in perturbation theory the 2-loop partition function of $(S)U(N)$ \textbf{YM} theory in the large $N$ limit with fixed, small 't Hooft coupling on $S^3$. We include scalars and chiral fermions in the adjoint representation of the gauge group with arbitrary Yukawa and quartic scalar couplings, only assuming commutator interactions (though our computation can be easily generalized to include non-commutator interactions as well). Our computation follows closely the one of \cite{Aharony:2006rf} where the same computation was done for pure \textbf{YM}.

As an immediate application we can compute the order $\lambda$ corrections to the Hagedorn temperature and to the spectrum of the theory. We check our computation by comparing to the known results for $\cal{N}=4$ \textbf{SYM}, which were computed by spin-chain methods in \cite{Spradlin:2004pp}.

The structure of this paper is as follows. In the next subsection we review some of the theoretical background necessary for this work. Section \ref{sec:Setup} reviews our conventions for the action, and various mode expansions on $S^3$. In section \ref{sec:RegCount} we discuss our renormalization scheme, and present the computation of the various counterterms. In section \ref{sec:2LoopComputation} we present the computation of the 2-loop diagrams which contribute to the partition function. Finally, we present our conclusions in section \ref{sec:Conclusions}. Additional information regarding the computation is collected in the appendices.
%-------------------   end of "Overview and outline"-------------------------

%-----------------  start of "Theoretical background"------------------------

\subsection{Theoretical background}\label{ssec:Overview}
In this subsection we summarize briefly some results related to gauge theories on a small $S^3$ that will be used in this work. A more thorough account of these matters can be found in \cite{Sundborg:1999ue},\cite{Aharony:2003sx}. As stated in the introduction we are interested in calculating pertubatively the partition function of large $N$ $(S)U(N)$ \textbf{YM} theory on $S^3$ with additional matter fields in the adjoint representation. This can be done by solving the Euclidean path integral on $S^3\times S^1$ with the inverse temperature $\beta$ as the period of $S^1$, in the weak coupling regime\footnote{For asymptotically free theories this regime is $R_{S^3}\Lambda_{YM}\ll 1$. Conformal field theories have a dimensionless coupling so we just set $\lambda\ll 1.$}.

In order to use perturbation theory we must fix the gauge. It is convenient to use the Coulomb gauge,
\begin{align}\label{eq:CoulombGauge}
\partial^iA_i=0,
\end{align}
where $i$ runs over the directions of $S^3$ and the derivative is assumed to be covariant.
This gauge condition fixes spatial gauge transformations but leaves temporal transformations unfixed. In order to fix those we impose the following additional condition,
\begin{align}
\partial_t\alpha \equiv \partial_t\left(\frac{1}{V_{S^3}}\int_{S^3}A_0\right)=0, \label{eq:ZeroModeFix}
\end{align}
where we have defined the zero mode of the gauge field $\alpha$. The additional condition implies that $\alpha$ is constant on $S^3\times S^1$. The mode $\alpha$ plays a special role in our computation since it is the only zero mode of the theory\footnote{We assume here that all scalar fields are conformally coupled so their zero mode is lifted.}.

As shown in \cite{Aharony:2003sx} the Fadeev-Popov determinant conjugate to (\ref{eq:ZeroModeFix}) can be absorbed in the integration measure of $\alpha$, changing it to integration over $U=e^{i\beta\alpha}$. Furthermore, it was also shown that the effective action arising from integrating out the massive modes is only a function of $U$ and $\beta$,

\begin{align}\label{eq:Z}
\cal{Z}(\beta) = \int[dU]e^{-\mathbf{S}_{eff}(U;\beta)} = \int[dU]\left(\int[dA_0][dA_i][d\phi^i]e^{-\mathbf{S}_E[\alpha,A_0,A_i,\phi^i;\beta]}\right),
\end{align}
where by $A_0$ we mean the part of the zeroth component of the gauge field \underline{without} its zero mode $\alpha$,  $\mathbf{S}_E$ is the gauge fixed Euclidean action on the sphere, and all the other massive fields are collectively denoted as $\phi^i$.

The new integration variable $U$ is related to the Wilson line over the time cycle, also known as the Polyakov loop $\cal{P}=\frac{1}{N}\mathrm{tr}P\mathrm{exp}(\int_0^{\beta}A_0dt)$ \cite{Polyakov:1978vu}. The Polyakov loop measures the free energy $F_q(T)$ of \textbf{YM} theory in the presence of an external quark through the relation $\langle\cal{P}\rangle=\mathrm{exp}(-F_q(T))$, and is a standard order parameter for deconfinement. More precisely, $\frac{1}{N}\mathrm{tr}(U)$ is the Wilson line of the zero mode of the gauge field around $S^1$.

The form of the effective action (\ref{eq:Z}) is highly restricted by gauge invariance. To 2 loops order, writing $S_{eff}(U,\beta)=S_{eff}^{1-loop}+S_{eff}^{2-loop}$, it turns out that

\begin{align}
S_{eff}^{1-loop} &= -\sum_{n=1}^{\infty}\frac{1}{n}z_n(x)\mathrm{tr}(U^n)\mathrm{tr}(U^{\dag n}),\label{eq:1LoopEffAction}\\
S_{eff}^{2-loop} &= \beta g^2\left[N\sum_{n=1}^\infty f_n(x)\left(\mathrm{tr}(U^n)\mathrm{tr}(U^{\dag n})-1\right) +\right. \nonumber\\
&\left.+\sum_{n, m =1}^{\infty}f_{n m}(x)\left(\mathrm{tr}(U^n)\mathrm{tr}(U^m)\mathrm{tr}(U^{-n-m}) + c.c.-2N\right)\right]\label{eq:2LoopEffAction},
\end{align}
where $x\equiv e^{-\beta/R_{S^3}}$ and the function $z_n(x)$ is defined in terms of the  ``1-particle partition sums'' for $N_s$ scalars, 1 vector and $N_f$ chiral fermions on $S^3$,

\begin{eqnarray}
\begin{array}{c}\label{eq:zBzF}
z_n(x)=z_B(x^n)+(-1)^{n+1}z_F(x^n),~~z_B(x)=N_sz_{s}(x)+z_V(x),~~  z_F(x)=N_fz_f(x)\\
z_s(x)=\frac{x+x^2}{(1-x)^3},~~  z_V(x)=\frac{6x^2-2x^3}{(1-x)^3},~~  z_F(x)=\frac{4x^{3/2}}{(1-x)^3}.
\end{array}
\end{eqnarray}

The two and three trace terms in (\ref{eq:2LoopEffAction}) arise from planar diagrams with 3 index loops, while the $U$ independent terms come from non-planar diagrams with a single index loop\footnote{The form of the non-planar terms is as in (\ref{eq:2LoopEffAction}) since we assume having only commutator interactions, so the corresponding $U(1)$ theory is free. Therefore when setting $\mathrm{tr}(U^n)=1$ and $N=1$ the 2-loop effective action should vanish. This together with the usual $N$ counting determines the non-planar terms.}. In the low temperature phase, the three trace terms in the matrix integral (\ref{eq:Z}) contribute only through perturbations around the gaussian model, and can be neglected to the order in the coupling that we are interested in.

The matrix model (\ref{eq:1LoopEffAction})-(\ref{eq:2LoopEffAction}) can be solved in the large $N$ limit by using the saddle point approximation. Below the critical temperature the minimizing eigenvalue distribution of $U$ is uniform. The matrix integral (\ref{eq:Z}) can be evaluated at this saddle point to give\footnote{The planar gaussian action vanishes at this saddle point and only contributes to the partition function through the integral over the fluctuations around it. That contribution is independent of $N$ (up to some irrelevant overall normalization of the partition function). This is why we have to include also the non-planar terms in the partition function around this saddle point, even though in the action they are suppressed by a factor of $N^2$ compared to the planar terms.},

\begin{align}\label{eq:UnZ}
\cal{Z}_{U(N)}=e^{-\lambda\beta \tilde{F}_2^{np}}\prod_{n=1}^{\infty}\frac{e^{\lambda\beta f_n(x)}}{1-z_n(x)+\lambda n\beta f_n(x)},
\end{align}
where $\widetilde{F}_2^{np}$ is the non-planar piece given by,

\begin{align}\label{eq:Fnp}
\widetilde{F}_2^{np}(x) = -2\sum_{n,m=1}^{\infty}f_{nm}(x).
\end{align}

For the $SU(N)$ partition function we divide by the $U(1)$ part (assuming only commutator interactions so the $U(1)$ part is decoupled from the $SU(N)$ part of the theory),

\begin{align}\label{eq:SUnZ}
\cal{Z}_{SU(N)}=\frac{\cal{Z}^{U(N)}}{\cal{Z}^{U(1)}}=e^{-\lambda\beta \tilde{F}_2^{np}}\prod_{n=1}^{\infty}\frac{e^{-z_n(x)/n+\lambda\beta f_n(x)}}{1-z_n(x)+\lambda n\beta f_n(x)}.
\end{align}

The goal of our computation is therefore to compute $f_n(x)$ and $f_{nm}(x)$.

The Hagedorn temperature is defined as the temperature for which the large $N$ partition function diverges. This happens when one of the denominators in (\ref{eq:UnZ}) or (\ref{eq:SUnZ}) vanishes. Since the 1-particle partition sums are monotonic functions with $z(0)=0$ and $z(1)=\infty$ this always occurs first for the $n=1$ denominator in the multiplication of (\ref{eq:UnZ}) or (\ref{eq:SUnZ}), so the Hagedorn temperature is given by $z_1(x_H)=1$ . The order $\lambda$ correction to the Hagedorn temperature is therefore,

\begin{align}\label{eq:HagedornCorrection}
\delta x_H = \beta_H\lambda\frac{f_1(x_H)}{z'_1(x_H)},
\end{align}
where $x_H$ is the Hagedorn temperature of the free theory.

Other quantities that can be extracted from the partition function are the sums of the order $\lambda$ corrections to the spectrum of \textbf{YM} on $S^3\times \mathds{R}$. These can be obtained by expanding the partition function in small $x$ (small temperatures)\footnote{For the $U(N)$ theory there are also a states with energy $1/R_{S^3}$ and $3/2R_{S^3}$ corresponding to the scalar operator $\mathrm{tr}(\phi)$ and fermionic operator $\mathrm{tr}(\psi)$, respectively.},

\begin{align}\label{eq:SmallxExp}
\cal{Z}_{SU(N)} = \sum_{\textrm{all states}}x^{E_i+\lambda\delta E_{i}}=\sum_{n=4}^{\infty}x^{\frac{n}{2}}\left(d\left(\frac{n}{2}\right)+\lambda\ln(x)\sum_i\delta E_{n,i}\right),
\end{align}
where in the second equality the sum is over states with energy $\frac{n}{2R_{S^3}}$ with degeneracy $d\left(\frac{n}{2}\right)$, and $\delta E_{n,i}$ is the order $\lambda$ correction to the state $i$ with energy $\frac{n}{2R_{S^3}}$.

By using the state-operator mapping of conformal field theories, the perturbative energy corrections of the theory on $S^3\times\mathds{R}$ are equivalent to the set of 1-loop anomalous dimensions of the theory on $\mathds{R}^4$. All the theories that we are considering are classically conformally invariant and remain so to leading order in $\lambda$. Therefore we can use the partition function on $S^3$ to compute the sums of anomalous dimensions in flat space.
%------------------ end of "Theoretical background" -------------------------

%-----------------------------------------------------------------------------
%-----------------------   end of introduction   -----------------------------
%-----------------------------------------------------------------------------

%-----------------------------------------------------------------------------
%-------------------   start of Setup of calculation   -----------------------
%-----------------------------------------------------------------------------
\pagebreak\section{Setup of the calculation}\label{sec:Setup}

In this section we present some notational conventions and the basic ingredients used throughout the computation.

%----------------- start of "The action on $S^3$" subsection -----------------

\subsection{The action on $S^3$.}\label{ssec:3SphereAction}
We write the Euclidean action of \textbf{YM} theory on $S^3\times S^1$ with the inverse temperature $\beta$ as the period of $S^1$. We include scalars $\Phi_a$ ($a,b,\ldots = 1,\ldots,N_s$ ) and $SO(4)$ Weyl spinors $\Psi^I$ ($I,J,\ldots=1,\ldots,N_f$), all in the adjoint representation of the gauge group $U(N)$,

\begin{align}\label{eq:FlatAction}
\mathbf{S}_E &= \int_0^{\beta}dt\int_{S^3}d\Omega \textrm{tr}\Big\{\frac{1}{4}F_{\mu\nu}F^{\mu\nu}+\frac{1}{2}\Phi^a\left(-D^2+1\right)\Phi^a+i\Psi^{\dag I}\sigma^{\mu}D_{\mu}\Psi_I-\nonumber\\
&-\frac{1}{4}g^2\mathbf{Q}^{abcd}\Phi_a\Phi_b\Phi_c\Phi_d+ \frac{1}{2}g(\rho^{a\dag})^{IJ}\Psi_{I}\varepsilon[\Phi^a,\Psi_J]+\frac{1}{2}g\rho^a_{IJ}\Psi^{\dag I}\varepsilon[\Phi^a,\Psi^{\dag J}]\Big\},\\
&F_{\mu\nu}\equiv \partial_{\mu}A_{\nu}-\partial_{\nu}A_{\mu}-ig[A_{\mu},A_{\nu}]~,~  D_{\mu}\equiv \partial_{\mu}-ig[A_{\mu},*].
\end{align}

The appropriate veilbein and connection terms for the vector and fermion covariant derivatives should be understood.  We also take the scalar fields to be conformally coupled by adding $\frac{1}{2}\phi^2_a$ to their kinetic terms\footnote{The scalar kinetic term is conformal in 4 dimensions if we add $\frac{1}{12}\cal{R}\phi^2$ to the action. We work in units for which $R_{S^3}=1$ so the Ricci scalar turns out to be $\cal{R}=6$.}.

%The trace is taken in the fundamental representation with normalization such that the kinetic terms are canonical.
Without loss of generallity we take the quartic scalar coupling $\mathbf{Q}^{abcd}$ to be cyclic in its indices, and the Yukawa coupling $\rho^a_{IJ}$ to be antisymmetric in $I,J$. This is the most general action that is classically conformally invariant, up to the assumption of having only commutator interactions. The reader is referred to appendix \ref{app:Notations} for further conventions regarding spinors and the restriction of the above action to the $\cal N = 4$ \textbf{SYM} case.

Next we impose the gauge fixing conditions (\ref{eq:CoulombGauge}) and (\ref{eq:ZeroModeFix}). The Fadeev-Popov determinant conjugate to (\ref{eq:CoulombGauge}) leads to the addition of the ghost action,

\begin{align}\label{eq:GhostAction}
\mathbf{S}_{FP} = -\int_0^{\beta}dt\int d^3x \mathrm{tr}(c^{\dag}\partial_iD^ic).
\end{align}
\newline
The resulting gauge fixed action can be found in appendix \ref{sapp:action}.
Next we expand all the fields in Kaluza-Klein modes on the sphere\footnote{Note that in principle we could have expended the vector field also in $\vec{\partial}\cal{S}^{\alpha}(\Omega)$, but the Coulomb gauge excludes such terms. },

\begin{gather}\label{eq:HarmonicsExpansions}
A_0(t,\Omega)  = \sum_{\alpha}a^{\alpha}(t)\cal{S}^{\alpha}(\Omega), ~~ c(t,\Omega)  = \sum_{\alpha}c^{\alpha}(t)\cal{S}^{\alpha}(\Omega), ~~\vec{A}(t,\Omega)  = \sum_{\alpha}A^\alpha(t)\vec{\cal{V}}^{\alpha}(\Omega),\nonumber\\
\Psi_I(t,\Omega)  = \sum_{\alpha}\psi_I^{\alpha}(t)\cal{Y}^{\alpha}(\Omega),~~\Phi_a(t,\Omega)  = \sum_{\alpha}\phi_a^{\alpha}(t)\cal{S}^\alpha(\Omega),
\end{gather}
where $\cal{S}^{\alpha}(\Omega)$, $\vec{\cal{V}}^{\alpha}(\Omega)$ and $\cal{Y}^{\alpha}(\Omega)$ are scalar, vector and Weyl spinor spherical harmonics. The Greek letter indices collect within them all the relevant angular momenta quantum numbers associated with those functions and summation over all of them is implied. To write the interaction terms we also need to compute various integrals involving three such spherical functions
\footnote{The integrals $H^{\alpha\beta\gamma}$, $J^{\alpha\beta\gamma}$ and $K^{\alpha\beta\gamma\delta}$ needed for the Yukawa and 4-scalar interactions are redundant and can be expressed by other integrals: $H^{\alpha\beta\gamma}=F^{\bar{\alpha}\beta\gamma}$,
$J^{\alpha\beta\gamma}=-F^{\bar{\alpha}\beta\gamma}$, and
$K^{\alpha\beta\gamma\delta}=B^{\bar{\lambda}\alpha\beta}B^{\lambda\gamma\delta}$.},

\begin{align}\label{eq:3SphericalIntegrals}
B^{\alpha\beta\gamma}  &= \int_{S^3}\cal{S}^{\alpha}\cal{S}^{\beta}\cal{S}^{\gamma} &,
C^{\alpha\beta\gamma}  &= \int_{S^3}\cal{S}^{\alpha}\vec{\cal{V}}^{\beta}\cdot\vec{\nabla}\cal{S}^{\gamma} &,
D^{\alpha\beta\gamma}  &= \int_{S^3}\vec{\cal{V}}^{\alpha}\cdot\vec{\cal{V}}^{\beta}\cal{S}^{\gamma},\nonumber\\
E^{\alpha\beta\gamma}  &= \int_{S^3}\vec{\cal{V}}^{\alpha}\cdot(\vec{\cal{V}}^{\beta}\times\vec{\cal{V}}^{\gamma}) &,
F^{\alpha\beta\gamma}  &= \int_{S^3}\cal{Y}^{{\alpha}\dag}\cal{Y}^{\beta}\cal{S}^{\gamma} &,
G^{\alpha\beta\gamma}  &= \int_{S^3}\cal{Y}^{{\alpha}\dag}\sigma^{\mu}\cal{Y}^{\beta}\vec{\cal{V}}^{\gamma}_{\mu},\nonumber\\
H^{\alpha\beta\gamma}  &= \int_{S^3}\cal{Y}^{\alpha}\varepsilon\cal{Y}^{\beta}\cal{S}^{\gamma} &,
J^{\alpha\beta\gamma}  &= \int_{S^3}\cal{Y}^{{\alpha}\dag}\varepsilon\cal{Y}^{{\beta}\dag}\cal{S}^{\gamma} &,
K^{\alpha\beta\gamma\delta} &= \int_{S^3}\cal{S}^{\alpha}\cal{S}^{\beta}\cal{S}^{\gamma}\cal{S}^{\delta}.
\end{align}

Properties of $S^3$ spherical harmonics, and the solutions of the above integrals can be found in appendix \ref{app:harmonics}.

The action in terms of the above expansions and integrals is written in (\ref{eq:QuadraticAction})-(\ref{eq:QuarticAction}). Since the fields $a^{\alpha}$ and $c^{\alpha}$ appear in the action only quadratically, it is convenient to integrate them out first. This was done in the appendix \ref{sapp:IntegrateOut} and the resulting action after integrating out is,

\begin{align}
\mathbf{S}_2 &= \int_0^{\beta} dt \mathrm{tr}\left\{
\frac{1}{2}A^{\bar{\alpha}}(-D_t^2+\sEig{\alpha}^2)A^{\alpha}+\frac{1}{2}\phi_a^{\bar{\alpha}}(-D_t^2+\sEig{\alpha}^2)\phi_a^{\alpha}+\right.\nonumber\\
&~\left.+i\psi^{\dag \alpha I}(D_t+\fEig{\alpha})\psi^{\alpha}_I\right\},\label{eq:S2}\\
\mathbf{S}_3 &= ig\int_0^{\beta} dt \mathrm{tr}\bigg\{\epsilon_{\alpha}(j_{\alpha}+1)E^{\alpha\beta\gamma}A^{\alpha}A^{\beta}A^{\gamma}-C^{\alpha\beta\gamma}[A^{\beta},\phi_a^{\alpha}]\phi_a^{\gamma}-iG^{\alpha\beta\gamma}\psi^{\dag \alpha I}[A^{\gamma},\psi^{\beta}_I]-\nonumber\\
&~\left.-\frac{i}{2}F^{\bar{\alpha}\beta\gamma}\psi_I^{\alpha}(\rho^{a\dag})^{IJ}[\phi_a^{\gamma},\psi_J^{\beta}]+\frac{i}{2}F^{\alpha\bar{\beta}\gamma}\psi^{\dag I\alpha}\rho^a_{IJ}[\phi_a^{\gamma},\psi^{\dag J\beta}]\right\},\label{eq:S3}\\
\mathbf{S}_4 &= -\frac{1}{2}g^2\int_0^{\beta} dt \mathrm{tr}\left\{
D^{\alpha\gamma\lambda}B^{\bar{\lambda}\beta\delta}[A^{\alpha},\phi_a^{\beta}][A^{\gamma},\phi_a^{\delta}]+\frac{1}{2}B^{\alpha\beta\bar{\lambda}}B^{\lambda\gamma\delta}\mathbf{Q}^{abcd}\phi_a^{\alpha}\phi_b^{\beta}\phi_c^{\gamma}\phi_d^{\delta}-\right.\nonumber\\
&-\frac{D^{\alpha \beta \lambda}D^{\gamma \delta \bar{\lambda}}}{\aEig{\lambda}}[A^\alpha,D_t A^\beta][A^{\gamma},D_t A^{\delta}]-\frac{B^{\alpha \beta \lambda}B^{\gamma \delta \bar{\lambda} }}{\aEig{\lambda}}[\phi^\alpha_a,D_t \phi^\beta_a][\phi^{\gamma}_{b},D_t \phi_{b}^{\delta}]+\nonumber\\
&\left.+\frac{F^{\alpha \beta \lambda}F^{\gamma \delta \bar{\lambda}}}{\aEig{\lambda}}\{\psi^\beta_I,\psi^{\dagger \alpha I}\}\{\psi_J^{\delta}, \psi^{\dagger \gamma J}\}\right\},\label{eq:S4}
\end{align}
where $D_t\equiv \partial_t-i[\alpha,*]$.
The other effective vertices which arise from integrating out $a^{\alpha}$ and $c^{\alpha}$ do not contribute to our computation as explained in appendix \ref{sapp:IntegrateOut}.

%------------------ end of "The action on $S^3$" subsection ------------------

%--------------------- Start of "Propagators" subsection ---------------------
\subsection{Propagators}\label{ssec:Propagators}

The propagators of the theory can be computed from (\ref{eq:S2})\footnote{We use $i,j,k,\ldots$, as fundamental representation indices.},

\begin{align}
\langle A^{\alpha}_{ij}(t^{\prime})A^{\beta}_{kl}(t)\rangle &= \Delta^{il,kj}_{\alpha}(t^{\prime}-t)\delta^{\alpha\bar{\beta}},\\
\langle (\phi^{\alpha}_a(t^{\prime}))_{ij}(\phi^{\beta}_b(t))_{kl}\rangle &= \Delta^{il,kj}_{\alpha}(t^{\prime}-t)\delta^{\alpha\bar{\beta}}\delta_{ab},\\
\langle (\psi^{\alpha}_I(t^{\prime}))_{ij}(\psi^{\beta}_J(t))^{\dag}_{kl}\rangle &=
\Theta^{il,kj}_{\alpha}(t^{\prime}-t)\delta^{\alpha\beta}\delta_{I\!J}.
\end{align}

We also used the following correlators,

\begin{align}\label{eq:PropDerivative}
\begin{array}{ll}
\langle D_tA^{\alpha}_{ij}(t^{\prime})A^{\beta}_{kl}(t)\rangle &= -\langle A^{\alpha}_{ij}(t^{\prime})D_tA^{\beta}_{kl}(t)\rangle = D_t\Delta^{il,kj}_{\alpha}(t^{\prime}-t)\delta^{\alpha\bar{\beta}},\\
\langle D_tA^{\alpha}_{ij}(t^{\prime})D_tA^{\beta}_{kl}(t)\rangle &= \delta(t^{\prime}-t)\delta^{\alpha\bar{\beta}}\delta_{il}\delta_{kj}-\sEig{\alpha}^2\Delta^{il,kj}_{\alpha}(t^{\prime}-t)\delta^{\alpha\bar{\beta}},\\
\langle D_t\phi^{a\alpha}_{ij}(t^{\prime})\phi^{b\beta}_{kl}(t))\rangle &=
-\langle \phi^{a\alpha}_{ij}(t^{\prime})D_t\phi^{b\beta}_{kl}(t))\rangle
=D_t\Delta^{il,kj}_{\alpha}(t^{\prime}-t)\delta^{\alpha\bar{\beta}}\delta^{ab},\\
\langle D_t\phi^{a\alpha}_{ij}(t^{\prime})D_t\phi^{b\beta}_{kl}(t)\rangle &=
\delta(t^{\prime}-t)\delta^{\alpha\bar{\beta}}\delta_{il}\delta_{kj}\delta^{ab}-\\
&~~-\sEig{\alpha}^2\Delta^{il,kj}_{\alpha}(t^{\prime}-t)\delta^{\alpha\bar{\beta}}\delta^{ab}.
\end{array}
\end{align}

The bosonic propagator $\Delta(t)$ is a periodic function with period $\beta$. For $0\leq t\leq\beta$ it is defined as,
\begin{align}\label{eq:BosPropOferConvention}
\begin{array}{c}
\Delta^{il,kj}_{\lambda}(t)\equiv\left(\frac{e^{i\bar{\alpha}t}}{2\sEig{\lambda}}\left(\frac{e^{-\sEig{\lambda} t}}{1-e^{i\beta\bar{\alpha}}e^{-\beta\sEig{\lambda}}}\right.\right.
\left.\left.-\frac{e^{\sEig{\lambda} t}}{1-e^{i\beta\bar{\alpha}}e^{\beta\sEig{\lambda}}}\right)\right)^{il,kj},
\end{array}
\end{align}
where $\bar{\alpha}\equiv\alpha\otimes\mathbbm{1}-\mathbbm{1}\otimes\alpha$ and a term in the expansion in powers of $e^{i\beta\bar{\alpha}}=e^{i\beta\alpha}\otimes e^{-i\beta\alpha}$ is understood to carry indices by putting them directly on the tensor product ($(A\otimes B)_{il,kj}\equiv A_{il}B_{kj}$).

The fermionic propagator $\Theta(t)$ is anti-periodic with (anti)period $\beta$. We can define it over $t\in(-\beta,\beta]$ (and continue periodically) ,
\begin{align}
\Theta_{\lambda}(t)=e^{i\bar{\alpha} t}\times\begin{cases} \frac{-ie^{-\fEig{\lambda} t}}{1+e^{i\beta\bar{\alpha}}e^{-\beta\fEig{\lambda}}} & t\in(0,\beta]\\
\frac{ie^{-\fEig{\lambda} t}}{1+e^{-i\beta\bar{\alpha}}e^{\beta\fEig{\lambda}}} & t\in(-\beta,0]\end{cases}
\end{align}
where indices should be put on it by the same prescription as in the bosonic case.

Note that the fermionic propagator, and the derivative of the bosonic propagator, are ambiguous at $t=0$. We found that this ambiguity only leads to an overall normalization of the partition function, which we fix by hand anyway\footnote{Say to $\cal{Z}(T=0)=1$.}.
%---------------------- end of "Propagators" subsection ----------------------

%----------------- start of "Counterterms action" subsection -----------------

\subsection{Counterterms action}\label{ssec:CountertermsAction}

As usual in perturbation theory, our 2-loop computation contains divergences. We use the regularization scheme of \cite{Aharony:2005bq}, which involves cutoff regulators which break gauge invariance. This scheme involves multiplying each vector, scalar and fermion line which carries momentum $\vec{p}$, with smooth regulating functions $R_{g,s,f}\left(\frac{|\vec{p}|}{M}\right)$\footnote{On $S^3$ we used the prescription which attaches $R_{g,s}\left(\frac{j+1}{M}\right)$ for vectors and scalars, or $R_f\left(\frac{j+\frac{1}{2}}{M}\right)$ for fermions with total angular momentum $j$.}. In the argument of the regulators, $M$ denotes the cutoff regularization scale. We demand that all the regulators have Heaviside-like properties: $R(0)=1$, $R'(0)=0$, \linebreak $R(k\rightarrow\infty)=0$, but other than that they are arbitrary.

These regulators preserve rotational symmetry on $S^3$, but break Lorentz and gauge invariance.
It was shown by 't~Hooft \cite{'tHooft:1971fh}, that gauge invariance can be restored by adding the appropriate set of renormalizable local counterterms which may break Lorentz and gauge symmetries, while preserving rotational symmetry. To first order in $\lambda = g^2N$, adding the following counterterms is necessary\footnote{Note that we did not write a $Z_0$ term to the fermions even though it is allowed. It turned out that this term does not contribute in our computation.},

\begin{align}\label{eq:CountertermsAction}
S_{ct} &= \lambda\int_0^{\beta}dt\int_{S^3}\mathrm{tr}^{SU(N)}\bigg\{ A_i(Z_{0g}-Z_{1g}\partial^2-Z_{2g}D_t^2)A_i+ \nonumber\\
&~~~ +\Phi^a(Z^{ab}_{0s}-Z^{ab}_{1s}\partial^2-Z^{ab}_{2s}D_t^2)\Phi^b+\nonumber\\
&~~~ +i\Psi^{\dag I}(Z^{IJ}_{1f}\sigma^i\partial_i+Z^{IJ}_{2f}D_t)\Psi_J\bigg\}=\\
&=\lambda\int_0^{\beta}dt\mathrm{ tr}^{SU(N)}\Big\{A^{\bar{\alpha}}(Z_{0g}+Z_{1g}\sEig{\alpha}-Z_{2g}D_t^2)A^{\alpha}+\nonumber\\
&~~~+\phi^{\bar{\alpha}}_a(Z^{ab}_{0s}+Z^{ab}_{1s}\aEig{\alpha}-Z^{ab}_{2s}D_t^2)\phi^{\alpha}_b+\nonumber\\
&~~~+i\psi^{\dag\alpha I}(Z^{IJ}_{1f}\fEig{\alpha}+Z^{IJ}_{2f}D_t)\psi^{\alpha}_J\Big\}.\nonumber
\end{align}

The notation in (\ref{eq:CountertermsAction}) suggests that we take the trace only over the $SU(N)$ part of the fields, (e.g. $A_i^{SU(N)}=A_i-\frac{1}{N}\mathrm{tr}(A_i)$). Since we assume only commutator interactions, the $U(1)$ part of the theory is free and only the $SU(N)$ part of the fields  contributes to the counterterms.

For the gauge boson and fermions, imposing gauge invariance is sufficient in order for the theory to remain conformally invariant at order $\lambda$. However, for the scalar fields this is not so since their mass terms are gauge invariant and contribute to the partition function at order $\lambda$. As noted in the introduction we want our theories to remain conformally invariant to leading order in $\lambda$. Therefore, we need some renormalization condition that keeps the scalars ``massless'' on $S^3$. In practice, we set the counterterms $Z^{ab}_{0s}$ by requiring that the state-operator mapping works (which is why we wanted our theories to be conformally invariant in the first place). It turns out that the knowledge of the sum of the anomalous dimensions of dimension 2 operators in flat space, is sufficient to determine $Z^{ab}_{0s}$ completely.

%------------------ end of "Counterterms action" subsection ------------------

%-----------------------------------------------------------------------------
%----------------   end of "Setup of calculation" section---------------------
%-----------------------------------------------------------------------------

\pagebreak

%-----------------------------------------------------------------------------
%----------  start of "Regularization and counterterms" section --------------
%-----------------------------------------------------------------------------

\section{Regularization and counterterms}\label{sec:RegCount}
In this section we present the computation of the various counterterms in (\ref{eq:CountertermsAction}). The counterterms $Z_1$ and $Z_2$ of all fields can be computed in flat space, since from dimensional analysis they cannot depend on $R_{S^3}$. The $Z_0$ counterterms on the other hand, can depend on the global properties of $S^3$. For instance the scalar terms may include a dependence on the Ricci scalar, $Z_0^{ab}\phi_a\phi_b\supset Z'_0\cal{R}\phi_a\phi_b$, and vector terms may also depend on the Ricci tensor.

Following \cite{Aharony:2006rf} we determine the counterterms by computing 1-loop diagrams in the cutoff scheme and demanding that the result match dimensional regularization with some convenient subtraction scheme. This guarantees once and for all that the results of the 2-loop calculation are gauge invariant.

%------------ start of "Flat space counterterms" subsection ------------

\subsection{Flat space counterterms}\label{ssec:FlatSpaceCount}

For the dimensionally regularized theory we work in $3+d$ spatial dimensions, and extend the coulomb gauge condition to include the $d$ extra components of the gauge field. The flat Euclidean action is then,

\begin{align}
S_2 &= \int d^{d+4}x\mathrm{tr}\Big\{\frac{1}{2}\dot{A_i}\dot{A_i}+\frac{1}{2}\partial_jA_i\partial_jA_i+\frac{1}{2}\partial_iA_0\partial_iA_0+\partial_ic^{\dag}\partial_ic+\nonumber\\
&\qquad+\frac{1}{2}\partial_{\mu}\phi_a\partial_{\mu}\phi_a+i\psi^{\dag I}\sigma^{\mu}\partial_{\mu}\psi_I\Big\},\label{eq:FlatSpaceAction1}\\
S_3 &= g\int d^{d+4}x\mathrm{tr}\Big\{i\dot{A}_i[A_i,A_0]-i\partial_iA_0[A_i,A_0]-i\partial_iA_j[A_i,A_j]-\nonumber\\
&\qquad-i\partial_ic^{\dag}[A_i,c]+\psi^{\dag I}\sigma^0[A_0,\psi_I]+\psi^{\dag I}\sigma^i[A_i,\psi_I]-i\dot{\phi}_a[A_0,\phi_a]-\nonumber\\
&\qquad-i\partial_i\phi_a[A_i,\phi_a]+\frac{1}{2}\psi_I\varepsilon(\rho^{a\dag})^{IJ}[\phi_a,\psi_J]+\frac{1}{2}\psi^{\dag I}\varepsilon\rho^a_{IJ}[\phi_a,\psi^{\dag J}]\Big\},\label{eq:FlatSpaceAction2}\\
S_4 &= -\frac{1}{2}g^2\int d^{d+4}x\mathrm{tr}\Big\{[A_0,A_i]^2+\frac{1}{2}[A_i,A_j]^2+[A_0,\phi_a]^2+\nonumber\\
&\qquad+[A_i,\phi_a]^2+\frac{1}{2}\mathbf{Q}^{abcd}\phi_a\phi_b\phi_c\phi_d\Big\}\label{eq:FlatSpaceAction3}.
\end{align}

We define the spinor matrices to satisfy $\sigma^{\mu}\bar{\sigma}^{\nu}+\sigma^{\nu}\bar{\sigma}^{\mu}=2\delta^{\mu\nu}\mathds{1}$, with \newline $\mu,\nu= 0,\ldots, d\!+\!3$. We also use the conventions $\mathrm{tr}(\mathds{1})=2$, and $\sigma^0=\bar{\sigma}^0=\mathds{1}$, which are consistent with the $d\rightarrow 0$ limit.
The propagators of the fields are,

\begin{align}
\langle A_i(\nu,k)A_j(-\nu,-k)\rangle \equiv \Delta_{ij}(\nu,k) &= \frac{k^2\delta_{ij}-k_ik_j}{k^2(\nu^2+k^2)},\label{eq:AiProp}\\
\langle A_0(\nu,k)A_0(-\nu,-k)\rangle &= \frac{1}{k^2}\label{eq:A0Prop},\\
\langle c(\nu,k)c^{\dag}(\nu,k)\rangle &= \frac{1}{k^2},\label{eq:cProp}\\
\langle \phi^a(\nu,k)\phi^b(-\nu,-k)\rangle &= \frac{1}{\nu^2+k^2}\delta^{ab},\label{eq:sProp}\\
\langle\psi_I(\nu,k)\psi^{\dag J}(\nu,k)\rangle &= \frac{\nu+k_i\bar{\sigma}^i}{\nu^2+k^2},\label{eq:fProp}.
\end{align}

In order to compute $Z_1,Z_2$, we'll compute the 1PI self-energies of the various fields. The self energy diagrams that contribute are depicted in figure 1. No quartic vertex contributes to these counterterms since such diagrams don't depend on external momenta. Also for the gluon self energy the $A_0$ loop cancels with the ghost loop, so we don't have to compute diagrams \textbf{SE1e} and \textbf{SE1f}.

\begin{figure}[!ht]\label{fig:SEFlat}
\centering
  \includegraphics[width=0.5\textwidth]{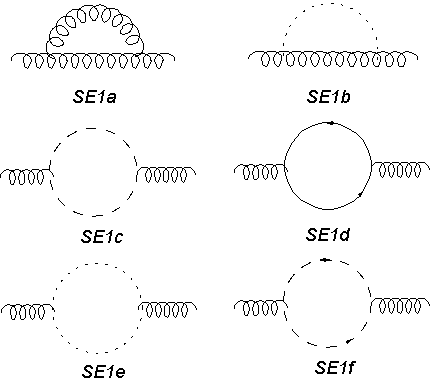}\\
  \includegraphics[width=0.5\textwidth]{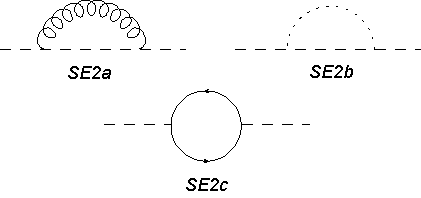}\\
  \includegraphics[width=0.5\textwidth]{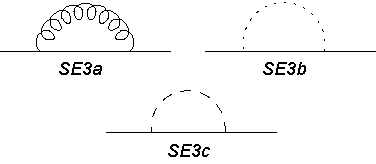}\\
  \caption{Self energy diagrams contributing to Z1 and Z2. Dashed (dotted) lines stand for the scalar ($A_0$) field, wiggly line for gluon and solid for the fermion. The dashed arrowed line stands for the ghost.}
\end{figure}

Diagrams $\mathbf{SE1a}$ and $\mathbf{SE1b}$ were evaluated in \cite{Aharony:2006rf}. The expressions for the other diagrams are (omitting $\lambda$ and group indices throughout.),
\paragraph{\underline{Vector self-energy}: $-\frac{1}{2}\langle A_i(\omega,p)A_j(-\omega,-p)\rangle^{1PI}$.}

\begin{align}
\mathbf{SE1c} &= N_s\int\frac{d\nu d^{3+d}k}{(2\pi)^{4+d}}\frac{k^ip^j-2k^ik^j}{(\nu^2+k^2)((\omega-\nu)^2+(p-k)^2)}~~,\label{eq:SE1cFlat}\\
\mathbf{SE1d} &= 2N_f\int\frac{d\nu d^{3+d}k}{(2\pi)^{4+d}}\frac{[\nu(\omega-\nu)+k\cdot (p-k)]\delta^{ij}+2k^i(k-p)^j}{(\nu^2+k^2)((\omega-\nu)^2+(p-k)^2)}~~.\label{eq:SE1dFlat}
\end{align}

\paragraph{\underline{Scalar self-energy}: $-\frac{1}{2}\langle \phi_a(\omega,p)\phi_b(-\omega,-p)\rangle^{1PI}$.}

\begin{align}
\mathbf{SE2a} &= -\delta^{ab}\int\frac{d\nu d^{3+d}k}{(2\pi)^{4+d}}\frac{(2p-k)\cdot\Delta\cdot (2p-k)}{(\omega-\nu)^2+(p-k)^2}~~,\label{eq:SE2aFlat}\\
\mathbf{SE2b} &= -\delta^{ab}\int\frac{d\nu d^{3+d}k}{(2\pi)^{4+d}}\frac{(2\omega-\nu)^2}{k^2((\omega-\nu)^2+(p-k)^2)}~~,\label{eq:SE2bFlat}\\
\mathbf{SE2c} &= \mathrm{tr}(\rho^{a\dag}\rho^b+\rho^{b\dag}\rho^a)\int\frac{d\nu d^{3+d}k}{(2\pi)^{4+d}}\frac{\nu(\omega-\nu)+k\cdot (p-k)}{(\nu^2+k^2)((\omega-\nu)^2+(p-k)^2)}~~.\label{eq:SE2cFlat}
\end{align}

\paragraph{\underline{Fermion self-energy}: $\langle \psi_I(\omega,p)\psi^{\dag J}(\omega,p)\rangle^{1PI}$.}

\begin{align}
\mathbf{SE3a} &= 2\delta_I^J\int\frac{d\nu d^{3+d}k}{(2\pi)^{4+d}}\frac{2[(p-k)\cdot\Delta]_i\sigma^i-\Delta^{jj}[\omega-\nu+(p-k)_i\sigma^i]}{(\omega-\nu)^2+(p-k)^2}~~,\label{eq:SE3aFlat}\\
\mathbf{SE3b} &= 2\delta_I^J\int\frac{d\nu d^{3+d}k}{(2\pi)^{4+d}}\frac{\omega-\nu-(p-k)_i\sigma^i}{k^2((\omega-\nu)^2+(p-k)^2)}~~,\label{eq:SE3bFlat}\\
\mathbf{SE3c} &= -2(\rho^a\rho^{a\dag})_I^J\int\frac{d\nu d^{3+d}k}{(2\pi)^{4+d}}\frac{\omega-\nu+(p-k)_i\sigma^i}{(\nu^2+k^2)((\omega-\nu)^2+(p-k)^2)}~~.\label{eq:SE3cFlat}
\end{align}

For the $Z_1,Z_2$ counterterms we need to extract the coefficients of $p^2,\omega^2$ in the vector and scalar fields self-energies, while for the fermions those would be the coefficients of $\omega,p_i\sigma^i$. In order to obtain the cutoff scheme integrals we can just set $d\rightarrow 0$ and multiply the propagators by the relevant regulator functions. We will start with the dimensional regularization expressions.

\subsubsection{Dimensional regularization in flat space}\label{sssec:FlatSpaceDimReg}

The coefficients of external momenta can all be expressed by the following integral,

\begin{align}\label{eq:Imn}
I_{m,n}(a) &\equiv \int\frac{d\nu d^{3+d}k}{(2\pi)^{4+d}}\frac{\nu^{2m}k^{2n-2m-2}}{(k^2+a^2)(\nu^2+k^2)^n}=\nonumber\\
&=\frac{1}{4\pi^3}\frac{\Gamma(n-m-\frac{1}{2})\Gamma(m+\frac{1}{2})}{\Gamma(n)}\left(\ln\left(\frac{\mu}{a}\right)+\left[\frac{1}{\epsilon}+\frac{1}{2}\ln(\pi)-\frac{\gamma}{2}+1\right]\right).
\end{align}

Here, $\mu$ is the regularization scale and $\epsilon=-d$. The subtraction scheme we use is to set the square brackets term in $I(a)$ to zero. The relevant coefficients of $p,\omega$ in terms of $I_{m,n}$'s , for vector, scalar and fermion self energy diagrams, can be found in appendix \ref{ssapp:1LoopFlatDR}.

Note that $I_{m,n}(0)$ contains IR divergences, which of course do not exist on $S^3$. To deal with this we can write $I(0)=I(a)+[I(0)-I(a)]$. Since the term in square brackets contains no UV divergences it must match between the two schemes. So we are allowed to use only the IR regulated term $I(a)$ in order to compare between the two regularization schemes. The full expressions for the IR regulated terms after applying the subtraction are,

\begin{align}
-\frac{1}{2}\langle A_i(0,p)A_j(0,-p)\rangle^{1PI}_{p^2\delta_{ij}} &=
p^2\delta_{ij}\left\{\left[-\frac{1}{8\pi^2}+\frac{1}{48\pi^2}\left(4N_f+N_s\right)\right]\ln\left(\frac{\mu}{a}\right)+\right.\nonumber\\
&\left.+\frac{29}{240\pi^2}-\frac{1}{720\pi^2}\left(N_s+28N_f\right)\right\},\label{eq:SE1pFlatDR}
\end{align}
\begin{align}
-\frac{1}{2}\langle A_i(\omega,0)A_j(-\omega,0)\rangle^{1PI}_{\omega^2\delta_{ij}} &= \omega^2\delta_{ij}\left\{\left[-\frac{1}{8\pi^2}+\frac{1}{48\pi^2}\left(N_s+4N_f\right)\right]\ln\left(\frac{\mu}{a}\right)\right.+\nonumber\\
&\left.+\frac{1}{48\pi^2}+\frac{1}{144\pi^2}\left(N_s-2N_f\right)\right\},\label{eq:SE1wFlatDR}
\end{align}
\begin{align}
-\frac{1}{2}\langle \phi^a(0,p)\phi^b(0,-p)\rangle^{1PI}_{p^2} &= p^2\left\{\left[-\frac{1}{4\pi^2}\delta^{ab}+\frac{1}{16\pi^2}\mathrm{tr}\left(\rho^{a\dag}\rho^b+\rho^{b\dag}\rho^a\right)\right]\ln\left(\frac{\mu}{a}\right)+\right.\nonumber\\
&\left.+\frac{1}{6\pi^2}\delta^{ab}-\frac{1}{48\pi^2}\mathrm{tr}\left(\rho^{a\dag}\rho^b+\rho^{b\dag}\rho^a\right) \right\},\label{eq:SE2pFlatDR}
\end{align}
\begin{align}
-\frac{1}{2}\langle \phi^a(\omega,0)\phi^b(-\omega,0)\rangle^{1PI}_{\omega^2} &= -\omega^2\left[\frac{1}{4\pi^2}\delta^{ab}-\frac{1}{16\pi^2}\mathrm{tr}\left(\rho^{a\dag}\rho^b+\rho^{b\dag}\rho^a\right)\right]\ln\left(\frac{\mu}{a}\right),\label{eq:SE2wFlatDR}
\end{align}
\begin{align}
\langle \psi_I(0,p)\psi^{\dag J}(0,p)\rangle^{1PI}_{p_i\sigma^i} &= p_i\sigma^i\left\{\left[-\frac{1}{4\pi^2}\delta_I^{~J}-\frac{1}{8\pi^2}\left(\rho^a\rho^{a\dag}\right)_I^{~J}\right]\ln\left(\frac{\mu}{a}\right)+\right.\nonumber\\
&\left.+\frac{5}{24\pi^2}\delta_I^{~J}+\frac{1}{24\pi^2}\left(\rho^a\rho^{a\dag}\right)_I^{~J}\right\},\label{eq:SE3pFlatDR}
\end{align}
\begin{align}
\langle \psi_I(\omega,0)\psi^{\dag J}(\omega,0)\rangle^{1PI}_{\omega} &= \omega\left\{\left[-\frac{1}{4\pi^2}\delta_I^{~J}-\frac{1}{8\pi^2}\left(\rho^a\rho^{a\dag}\right)_I^{~J}\right]\ln\left(\frac{\mu}{a}\right)+\right.\nonumber\\
&\left.+\frac{1}{8\pi^2}\delta_I^{~J}\right\}.\label{eq:SE3wFlatDR}
\end{align}

\subsubsection{Cutoff regularization in flat space}\label{sssec:FlatSpaceCutoff}
We take $d\rightarrow 0$ in the dimensional regularization expressions (\ref{eq:SE1cFlat})-(\ref{eq:SE3cFlat}), and multiply each diagram by the appropriate regulator functions as described in section \ref{ssec:CountertermsAction}. We now repeat the $p$ and $\omega$ expansions, including the regulator function expansion when necessary,

\begin{align}
R\left(\frac{|p-k|}{M}\right) &= R\left(\frac{k}{M}\right)-\frac{R'\left(\frac{k}{M}\right)(p\cdot k)}{Mk}+\nonumber\\
&~~+\frac{R''\left(\frac{k}{M}\right)(p\cdot k)^2}{2M^2k^2}-\frac{R'\left(\frac{k}{M}\right)(p\cdot k)^2}{2Mk^3}+\frac{R'\left(\frac{k}{M}\right)p^2}{2Mk}+\cdots.
\end{align}

After picking up only the relevant coefficients of the external momenta, the integration over $\nu$ is always easy to perform. We then write our results in terms of regulator dependent integrals defined in (\ref{eq:RegDepFunA})-(\ref{eq:RegDepFunF}).

We also include the counterterm diagrams arising from (\ref{eq:CountertermsAction}), which contribute to the self energies, in the total result (figure 2).

\begin{figure}[!ht]\label{fig:SEct}
\centering
  \includegraphics[width=0.2\textwidth]{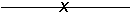}\\
  \caption{Counterterm contribution to 1-loop self energy diagrams.}
\end{figure}

\begin{align}\label{eq:SE1pFlatCO}
-\frac{1}{2}\langle A_i(0,p)A_j(0,-p)\rangle^{1PI}_{p^2\delta_{ij}} &= p^2\delta^{ij}\left[Z_{1g}+\frac{N_s+8N_f-6}{240\pi^2}+\frac{N_s+4N_f-6}{48\pi^2}\ln\left(\frac{M}{a}\right)\right.+\nonumber\\ &~~~+\frac{1}{40\pi^2}\ln\left(\frac{\cal{A}_{100}^4}{\cal{A}_{200}^9}\right)+\frac{N_s}{48\pi^2}\ln\left(\cal{A}_{020}\right)+\frac{N_f}{12\pi^2}\ln\left(\cal{A}_{002}\right)-\\
&~~~\left.-\frac{1}{15}\cal{F}_2^g-\frac{N_s}{30}\cal{F}_2^s-\frac{4N_f}{15}\cal{F}_2^f\right]\nonumber
\end{align}

\begin{align}\label{eq:SE1wFlatCO}
-\frac{1}{2}\langle A_i(\omega,0)A_j(-\omega,0)\rangle^{1PI}_{\omega^2\delta_{ij}} &= \omega^2\delta^{ij}\left[Z_{2g}+\frac{N_s+4N_f-6}{48\pi^2}\ln\left(\frac{M}{a}\right)+\frac{1}{24\pi^2}\ln\left(\frac{\cal{A}_{200}}{\cal{A}_{100}^4}\right)+\right.\nonumber\\
&~~~\left.+\frac{N_s}{48\pi^2}\ln\left(\cal{A}_{020}\right)+\frac{N_f}{12\pi^2}\ln\left(\cal{A}_{002}\right)\right].
\end{align}

\begin{align}\label{eq:SE2pFlatCO}
-\frac{1}{2}\langle \phi^a(0,p)\phi^b(0,-p)\rangle^{1PI}_{p^2} &= p^2\left\{Z_{1s}-\delta^{ab}\left[\frac{1}{4\pi^2}\ln\left(\frac{M}{a}\right)-\frac{1}{12\pi^2}\ln\left(\frac{\cal{A}_{010}}{\cal{A}_{110}^4}\right)\right]+\right.\nonumber\\
&~~~\left.+\mathrm{tr}\left(\rho^{a\dag}\rho^b+\rho^{b\dag}\rho^a\right)\left[\frac{1}{16\pi^2}\ln\left(\frac{\cal{A}_{002}M}{a}\right)+\frac{1}{48\pi^2}-\frac{\cal{F}^f_2}{6}\right]\right\}.
\end{align}

\begin{align}\label{eq:SE2wFlatCO}
-\frac{1}{2}\langle \phi^a(\omega,0)\phi^b(-\omega,0)\rangle^{1PI}_{\omega^2} &= \omega^2\left[Z_{2s}^{ab}-\frac{1}{4\pi^2}\delta^{ab}\ln\left(\frac{\cal{A}_{010}M}{a}\right)+\right.\nonumber
\\&~~~\left.+\frac{1}{16\pi^2}\mathrm{tr}\left(\rho^{a\dag}\rho^b+\rho^{b\dag}\rho^a\right)\ln\left(\frac{\cal{A}_{002}M}{a}\right)\right].
\end{align}

\begin{align}\label{eq:SE3pFlatCO}
\langle \psi_I(0,p)\psi^{\dag J}(0,p)\rangle^{1PI}_{p_i\sigma^i} &= p_i\sigma^i\left\{-Z^{IJ}_{1f}+\delta_I^{~J}\left[-\frac{1}{4\pi^2}\ln\left(\frac{M}{a}\right)+\frac{1}{12\pi^2}\ln\left(\frac{\cal{A}_{101}}{\cal{A}_{001}^4}\right)-\frac{1}{3\pi}\cal{B}_1^g+\frac{1}{6\pi^2}\right]-\right.\nonumber\\
&~~~\left.-(\rho^a\rho^{a\dag})_I^{~J}\left[\frac{1}{8\pi^2}\ln\left(\frac{\cal{A}_{011}M}{a}\right)+\frac{1}{6\pi}\cal{B}_1^s\right]\right\}
\end{align}

\begin{align}\label{eq:SE3wFlatCO}
\langle \psi_I(\omega,0)\psi^{\dag J}(\omega,0)\rangle^{1PI}_{\omega}&= \omega\left[-Z^{IJ}_{2f}-\frac{1}{4\pi^2}\delta_I^{~J}\ln\left(\frac{\cal{A}_{101}M}{a}\right)-\frac{1}{8\pi^2}(\rho^a\rho^{a\dag})_I^{~J}\ln\left(\frac{\cal{A}_{011}M}{a}\right)\right].
\end{align}

\subsubsection{Flat space counterterms summary}\label{sssec:FlatSpaceCountSummary}
We solve for the $Z$'s by equating the expressions in (\ref{eq:SE1pFlatDR})-(\ref{eq:SE3wFlatDR}) with the ones in (\ref{eq:SE1pFlatCO})-(\ref{eq:SE3wFlatCO}). This determines the flat space counterterms,

\begin{align}
Z_{1g} &=\frac{6-N_s-4N_f}{48\pi^2}\ln\left(\frac{M}{\mu}\right)+\frac{105-4N_s-52N_f}{720\pi^2}+\frac{1}{15}\cal{F}_2^g+\frac{N_s}{30}\cal{F}_2^s+\frac{4N_f}{15}\cal{F}_2^f-\nonumber\\
&~~~-\frac{1}{40\pi^2}\ln\left(\frac{\cal{A}_{100}^4}{\cal{A}_{200}^9}\right)-\frac{N_s}{48\pi^2}\ln\left(\cal{A}_{020}\right)-\frac{N_f}{12\pi^2}\ln\left(\cal{A}_{002}\right),\label{eq:Z1g}
\end{align}
\begin{align}
Z_{2g} &= \frac{6-N_s-4N_f}{48\pi^2}\ln\left(\frac{M}{\mu}\right)+\frac{3+N_s-2N_f}{144\pi^2}-\frac{1}{24\pi^2}\ln\left(\frac{\cal{A}_{200}}{\cal{A}_{100}^4}\right)-\nonumber\\
&~~~-\frac{N_s}{48\pi^2}\ln\left(\cal{A}_{020}\right)-\frac{N_f}{12\pi^2}\ln\left(\cal{A}_{002}\right),\label{eq:Z2g}
\end{align}
\begin{align}
Z_{1s}^{ab} &= \left[\frac{1}{4\pi^2}\delta^{ab}-\frac{1}{16\pi^2}\mathrm{tr}\left(\rho^{a\dag}\rho^b+\rho^{b\dag}\rho^a\right)\right]\ln\left(\frac{M}{\mu}\right) +\frac{1}{6\pi^2}\delta^{ab}-\frac{1}{24\pi^2}\mathrm{tr}\left(\rho^{a\dag}\rho^b+\rho^{b\dag}\rho^a\right)-\nonumber\\
&~~~-\frac{1}{16\pi^2}\mathrm{tr}\left(\rho^{a\dag}\rho^b+\rho^{b\dag}\rho^a\right)\ln\left(\cal{A}_{002}\right)-\frac{1}{12\pi^2}\ln\left(\frac{\cal{A}_{010}}{\cal{A}_{110}^4}\right)+\frac{1}{6}\mathrm{tr}\left(\rho^{a\dag}\rho^b+\rho^{b\dag}\rho^a\right)\cal{F}_2^f,\label{eq:Z1s}
\end{align}
\begin{align}
Z_{2s}^{ab} &= \left[\frac{1}{4\pi^2}\delta^{ab}-\frac{1}{16\pi^2}\mathrm{tr}\left(\rho^{a\dag}\rho^b+\rho^{b\dag}\rho^a\right)\right]\ln\left(\frac{M}{\mu}\right)+\frac{1}{4\pi^2}\delta^{ab}\ln\left(\cal{A}_{010}\right)-\nonumber\\
&~~~-\frac{1}{16\pi^2}\mathrm{tr}\left(\rho^{a\dag}\rho^b+\rho^{b\dag}\rho^a\right)\ln\left(\cal{A}_{002}\right),\label{eq:Z2s}
\end{align}
\begin{align}
-Z_{1f}^{IJ} &= \left[\frac{1}{4\pi^2}\delta_I^{~J}+\frac{1}{8\pi^2}(\rho^a\rho^{a\dag})_I^{~J}\right]\ln\left(\frac{M}{\mu}\right)+\frac{1}{24\pi^2}\delta_I^{~J}+\frac{1}{24\pi^2}(\rho^a\rho^{a\dag})_I^{~J}+\frac{1}{3\pi}\delta_I^{~J}\cal{B}_1^g+\nonumber\\
&~~~+\frac{1}{6\pi}(\rho^a\rho^{a\dag})_I^{~J}\cal{B}_1^s-\frac{1}{12\pi^2}\ln\left(\frac{\cal{A}_{101}}{\cal{A}_{001}^4}\right)+\frac{1}{8\pi^2}(\rho^a\rho^{a\dag})_I^{~J}\ln\left(\cal{A}_{011}\right),\label{eq:Z1f}
\end{align}
\begin{align}
-Z_{2f}^{IJ} &= \left[\frac{1}{4\pi^2}\delta_I^{~J}+\frac{1}{8\pi^2}(\rho^a\rho^{a\dag})_I^{~J}\right]\ln\left(\frac{M}{\mu}\right)+\frac{1}{8\pi^2}\delta_I^{~J}+\frac{1}{4\pi^2}\delta_I^{~J}\ln\left(\cal{A}_{101}\right)+\nonumber\\
&~~~+\frac{1}{8\pi^2}(\rho^a\rho^{a\dag})_I^{~J}\ln\left(\cal{A}_{011}\right).\label{eq:Z2f}
\end{align}
%------------- end of "Flat space counterterms" subsection -------------

%---------- start of "Curved space counterterms" subsection ------------

\subsection{Curved space counterterms}\label{ssec:CurvedSpaceCount}

In this section we'll compute the $Z_{0}$ counterterms in (\ref{eq:CountertermsAction}). As described in the beginning of this section the $Z_0$ counterterms may depend on the global properties of $S^3$, so it is necessary to perform a computation on $S^3$ to determine them. For the gauge field counterterm $Z_{0g}$ we apply our gauge invariant renormalization scheme, i.e., we compute a set of correlation functions on $S^3$ using both dimensional regularization and cutoff scheme, and then match the results. We do that by computing the 1-loop 1PI self-energies of the lowest total angular momentum mode of the gauge field.

For the scalar field, using this scheme to determine the counterterm $Z_{0s}^{ab}$ leads to the breaking of conformal invariance at order $\lambda$. We will determine $Z_{0s}^{ab}$ by demanding that our final result is conformally invariant in section \ref{ssec:NumericsChecks}. In this section we will only compute the structure of divergences subtracted by $Z_{0s}^{ab}$, by computing the 1-loop self-energies of the lowest total angular momentum mode of the scalar fields in the cutoff scheme.

Throughout this subsection we work in units where $\mu=R_{S^3}=1$, so the reader should not be alarmed when seeing functions which depend on dimensionful quantities. These always consistently vanish between the diagrams. As before we start with dimensional regularization before moving on to the cutoff regularization method.

\subsubsection{Dimensional regularization on $S^3$}\label{sssec:CurvedSpaceDimReg}
To apply dimensional regularization on curved space we let our theory live on $S^3\times\mathds{R}^{d+1}$ with $d=-\epsilon$. We extend the Coulomb gauge condition to involve all components of the gauge field other than $A_0$,
\footnote{In this section $A_a$ will denote the $\mathds{R}^d$ components of the gauge field, $A_0$ and $A_i$ mean the same as before. The scalars internal indices will be denoted by $\bar{a},\bar{b}...$}

\begin{align}\label{eq:ExtendedCoulombGauge}
\partial_iA_i+\partial_aA_a = 0.
\end{align}

Now $A_i$ has a different mode expansion than before since its divergence doesn't vanish anymore. The expansions of $A_i$ and $A_a$ are,

\begin{align}
A_a &= \sum_{\alpha,j_{\alpha}\geq 0}\cal{A}^{\alpha}_aS^{\alpha},\\
A_i &= \sum_{\alpha,j_{\alpha}>0}\left(A^{\alpha}\cal{V}_i^{\alpha} +\frac{1}{\aEig{\alpha}}\partial_a\cal{A}_a^{\alpha}\partial_iS^{\alpha}\right).
\end{align}

The quadratic part of the action for the gauge field and scalars are then

\begin{align}
\mathbf{S}^g_2+\mathbf{S}^s_2 &= \int dtd^dx\mathrm{tr}\left(\frac{1}{2}A^{\bar{\alpha}}(-\partial_t^2-\partial_a^2+\sEig{\alpha})A^{\alpha}+\frac{1}{2}a^{\bar{\alpha}}(-\partial_a^2+\aEig{\alpha})a^{\alpha}+\right.\nonumber\\
&~~~+\frac{1}{2}\cal{A}_a^{\bar{\alpha}}(-\partial^a_t-\partial_c^2+\aEig{\alpha})(\delta^{ab}-\frac{\partial_a\partial_b}{\aEig{\alpha}})\cal{A}^{\alpha}_b+\\
&~~~\left.+\frac{1}{2}\phi_{\bar{a}}^{\bar{\alpha}}(-\partial_t^2-\partial_a^2+\sEig{\alpha})\phi^{\alpha}_{\bar{a}}\right).\nonumber
\end{align}
The additional interaction vertices are spelled out in appendix \ref{ssapp:1LoopCurvedDR}.

For the spinor part a convenient basis for the Dirac matrices of $SO(d+4)$ is\footnote{Here $a=0,\ldots,d$ label $\mathds{R}^{d+1}$ components. Hopefully there would be no confusion with the decomposition of the gauge field in which $a$ labeled $d$ directions in $\mathds{R}^{d+1}$, since the other component ($0$) does not participate in the coulomb gauge condition.},

\begin{align}
\bar{\Gamma}^{i}=\tau^i\otimes\Gamma,~&~ \bar{\Gamma}^a=\mathds{1}\otimes{\Gamma}^a,
\end{align}
where $\left\{\Gamma,\Gamma^a\right\}=0$ and $\left\{\Gamma^a,\Gamma^b\right\}=2\delta^{ab}\mathds{1}$.\footnote{We used $\text{tr}(\mathds{1})=1$ for the unit matrix in $\Gamma^a$ space. This is consistent with the $d\rightarrow 0$ limit, since the action should then reduce to the 1d action we had after the mode expansions on $S^3$.}
The  $\Gamma^a$ and $\Gamma$ can be taken to be in the Dirac representation of $SO(d+2)$.

The only terms we will need in the Dirac Lagrangian on $\mathds{R}^{d+1}$ are then (vielbein and spin connection should be understood.),

\begin{align}
\cal{L}_{\mathds{R}^{d+1}} &= \int_{S^3}\text{tr}\left\{i\bar{\lambda}\left(\bar{\Gamma^i}\partial_i+\bar{\Gamma^a}\partial_a\right)\lambda +g\bar{\lambda}\bar{\Gamma}^i\left[A_i,\lambda\right]\right\}= \nonumber\\ &= \int_{S^3}\text{tr}\left\{i\lambda^{\dag}\left(\tau^{i}\otimes\Gamma^0\Gamma\partial_{i}+\mathds{1}\otimes\Gamma^0\Gamma^a\partial_a\right)\lambda+g\lambda^{\dag}\tau^i\otimes\Gamma^0\Gamma\left[A_i,\lambda\right]\right\}=\nonumber\\
&=\text{tr}\left\{\bar{\psi}^{\alpha}\left(i\Gamma^a\partial_a+\epsilon_{\alpha}(j_{\alpha}+1)\Gamma\right)\psi^{\alpha}-igG^{\alpha\beta\gamma}\bar{\psi}^{\alpha}\Gamma\left[A^{\gamma},\psi^{\beta}\right]\right\},
\end{align}
where in the last line we expanded $\lambda$ in $S^3$ spinor spherical harmonics ,
\begin{align}
\lambda = \cal{Y}^{\alpha}(\Omega)\otimes\psi^{\alpha}(\vec{y}).
\end{align}

From the quadratic parts we extract the propagators,

\begin{align}
\langle a^{\alpha}(\nu,k)a^{\beta}(-\nu,-k)\rangle &\equiv\delta^{\alpha\bar{\beta}}\bar{\Delta}^{\alpha}(k) = \frac{\delta^{\alpha\bar{\beta}}}{k^2+\aEig{\alpha}}, \\
\langle A^{\alpha}(\nu,k)A^{\beta}(-\nu,-k)\rangle &\equiv\delta^{\alpha\bar{\beta}}\Delta^{\alpha}(\nu,k)= \frac{\delta^{\alpha\bar{\beta}}}{\nu^2+k^2+\sEig{\alpha}}, \\
\langle\cal{A}^{\alpha}_a(\nu,k)\cal{A}^{\beta}_b(-\nu,-k) \rangle &\equiv\delta^{\alpha\bar{\beta}}\hat{\Delta}_{ab}^{\alpha}(\nu,k)= \nonumber\\
&~~~= \frac{\delta^{\alpha\bar{\beta}}}{\nu^2+k^2+\aEig{\alpha}}(\delta_{ab}-\frac{k_ak_b}{k^2+\aEig{\alpha}}),\\
\langle \phi^{\alpha}_{\bar{a}}(\nu,k)\phi^{\beta}_{\bar{b}}(-\nu,-k)\rangle &\equiv\delta^{\alpha\bar{\beta}}\delta_{\bar{a}\bar{b}}\Delta^{\alpha}(\nu,k)= \frac{\delta^{\alpha\bar{\beta}}\delta_{\bar{a}\bar{b}}}{\nu^2+k^2+\sEig{\alpha}}, \\
\langle \psi^{\alpha}_I(\nu,k)\psi^{\dag\beta J}(\nu,k)\rangle &\equiv\delta^{\alpha\beta}\delta_I^{~J}\Theta^{\alpha}(\nu,k)=\frac{\delta^{\alpha\beta}\delta_I^{~J}}{k_a\Gamma^a+\Gamma\lambda_{\alpha}}.
\end{align}

Now, besides the diagrams in figure 1 we have additional diagrams which are shown in Figure 3.

\begin{figure}[!hc]\label{fig:CurvedSpace}
\centering
  %Requires \usepackage{graphicx}
  \includegraphics[width=1\textwidth]{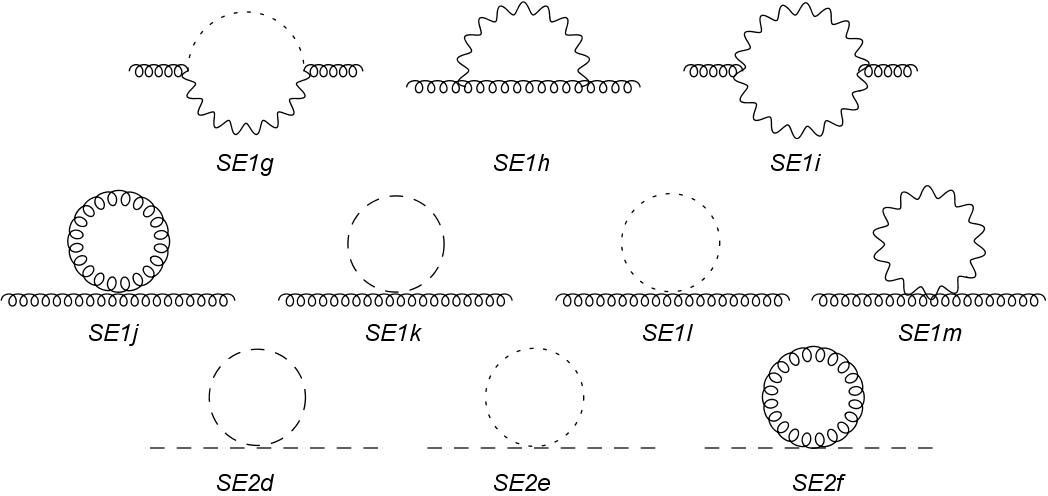}\\
 \caption{Additional self energy diagrams when using dimensional regularization on $S^3$. Dashed (dotted) lines stand for the scalar ($a$) field, wiggly (wavy) line for $A$ ($\cal{A}$) and solid for the fermion.}
\end{figure}

We compute the gauge field 1PI self-energy $\langle A^{\alpha}(0)A^{\beta}(0)\rangle$, with external momenta in the $\mathds{R}^{d+1}$ directions all set to zero.

The expressions for diagrams which were not already evaluated in \cite{Aharony:2006rf} are,

\begin{align}
\mathbf{SE1c} &= 4N_sC^{\gamma\alpha\delta}C^{\bar{\gamma}\beta\bar{\delta}}\int \frac{d\nu d^dk}{(2\pi)^{d+1}}\Delta^{\gamma}(\nu,k)\Delta^{\delta}(\nu,k),\label{eq:SE1cInt}\\
\mathbf{SE1d} &= -2N_fG^{\gamma\delta\alpha}G^{\delta\gamma\beta}\int \frac{d\nu d^dk}{(2\pi)^{d+1}}\mathrm{tr}\left(\Theta^{\gamma}(\nu,k)\Theta^{\delta}(\nu,k)\right),\label{eq:SE1dInt}\\
\mathbf{SE1k} &= -2N_sD^{\alpha\beta\gamma}B^{\bar{\gamma}\bar{\delta}\delta}\int \frac{d\nu d^dk}{(2\pi)^{d+1}}\Delta^{\delta}(\nu,k),\label{eq:SE1kInt}
\end{align}

To evaluate these diagrams we set the quantum numbers of the external lines to $\beta=\bar{\alpha}$ and the total angular momenta on the external lines to $j_{\alpha}=j_{\beta}=1$. We can then sum over all quantum numbers (using appendix \ref{sapp:AMSums}) which do not involve the propagators. This usually leaves us with simpler integrals due to the selection rules imposed by the 3-j symbols. After solving the integrals the summation over the rest of the quantum numbers will be regulated by $d$.

For example, in $\mathbf{SE1c}$ the selection rules constrains the summation to $j_{\gamma}=j_{\delta}$ and $j_{\gamma}\geq 1$. The angular momentum sum over all quantum numbers but $j_{\gamma}$ gives:
$\sum C^{\gamma\alpha\delta}C^{\bar{\gamma}\bar{\alpha}\bar{\delta}}=2R_c^2(j_{\gamma},1,j_{\gamma})=\frac{j_{\gamma}(j_{\gamma}+1)^2(j_{\gamma}+2)}{\pi^2}$
The integral is easily performed and we get,

\begin{align}\label{eq:SE1cCurvedDRExm}
\frac{4N_s}{\pi^2}\frac{\Gamma(\frac{3-d}{2})}{(4\pi)^{\frac{1+d}{2}}}\sum_{j_\gamma=1}^{\infty}j_{\gamma}(j_{\gamma}+2)(j_{\gamma}+1)^{d-1}=\frac{4N_s}{\pi^2}\frac{\Gamma(\frac{3-d}{2})}{(4\pi)^{\frac{1+d}{2}}}(\zeta(-1-d)-\zeta(1-d))=\nonumber\\
=-\frac{N_s}{\pi^2}([\frac{1}{\epsilon}+\frac{1}{2}\ln(\pi)-\frac{\gamma}{2}+1]+\gamma+\frac{1}{12})+\cal O(\epsilon).
\end{align}

Other diagrams can be evaluated in a similar way. The results diagram by diagram are given in appendix \ref{ssapp:1LoopCurvedDR}. After employing our subtraction scheme (see below (\ref{eq:Imn})), the total result for the 1-loop 1PI self-energy of the gauge field is,

\begin{align}
G_{DR} &=\frac{6-N_s-4N_f}{\pi^2}\gamma+\frac{9+N_s+16N_f}{6\pi^2}-\frac{10}{\pi^2}\zeta(3).\label{eq:GluonCurvedDR}
\end{align}

\subsubsection{Cutoff regularization on $S^3$}\label{sssec:CurvedSpaceCutoff}
For the gauge field we obtain the cutoff regularization expressions by taking the $d\rightarrow 0$ limit in the dimensional regularization expressions, and multiplying with the appropriate regulator functions. We also compute the 1PI self-energy diagrams (figures 1 and 3) of the scalar fields, $\langle\phi^{\alpha}_{\bar{a}}(0)\phi^{\beta}_{\bar{b}}(0)\rangle$ with $\beta=\bar{\alpha}$ and $j_{\alpha}=j_{\beta}=0$. The expressions for those (using the notation from the last section for the propagators, and suppressing the regulators) are\footnote{$\mathbf{Q}^{ab(cd)}\equiv\mathbf{Q}^{abcd}+\mathbf{Q}^{abdc}$.},

\begin{align}
\mathbf{SE2a} &= 8\delta^{ab}C^{\alpha\gamma\delta}C^{\beta\bar{\gamma}\bar{\delta}}\int \frac{d\nu }{2\pi}\Delta^{\gamma}(\nu,0)\Delta^{\delta}(\nu,0)\label{eq:SE2aInt},\\
\mathbf{SE2b} &= 2\delta^{ab}B^{\alpha\gamma\delta}B^{\beta\bar{\gamma}\bar{\delta}}\int \frac{d\nu }{2\pi}\nu^2\bar{\Delta}^{\gamma}(0)\Delta^{\delta}(\nu,0),\label{eq:SE2bInt}\\
\mathbf{SE2c} &= -\mathrm{tr}\left(\rho^{a\dag}\rho^b+\rho^{b\dag}\rho^a\right)F^{\bar{\gamma}\delta\alpha}F^{\delta\bar{\gamma}\beta}\int \frac{d\nu }{2\pi}\mathrm{tr}\left(\Theta^{\delta}(\nu,0)\Theta^{\gamma}(-\nu,0)\right),\label{eq:SE2cInt}\\
\mathbf{SE2d} &= \frac{Q^{(ab)cc}}{2}\left(B^{\alpha\beta\bar{\delta}}B^{\delta\gamma\bar{\gamma}}+B^{\alpha\bar{\gamma}\bar{\delta}}B^{\delta\gamma\beta}\right)\int \frac{d\nu}{2\pi}\Delta^{\gamma}(\nu,0),\label{eq:SE2fInt}\\
\mathbf{SE2e} &= -2\delta^{ab}B^{\alpha\gamma\delta}B^{\beta\bar{\gamma}\bar{\delta}}\int \frac{d\nu }{2\pi}\bar{\Delta}^{\gamma}(0),\label{eq:SE2gInt}\\
\mathbf{SE2f} &= -2\delta^{ab}B^{\alpha\beta\delta}D^{\gamma\bar{\gamma}\bar{\delta}}\int \frac{d\nu }{2\pi}\Delta^{\gamma}(\nu,0).\label{eq:SE2eInt}
\end{align}

Note that \textbf{SE2e} has a $\delta(0)$ divergence. This term exactly cancels with a similar term in \textbf{SE2b}.
We express the resulting angular momentum summations with the regulator dependent integrals defined in (\ref{eq:RegDepFunA})-(\ref{eq:RegDepFunF}), by using the Euler-Mclaurin formula\footnote{For $f^{(n)}(\infty)=0$ the formula is $\sum_{a=0}^{\infty}f(a)=\int_0^{\infty}f(x)dx+\frac{1}{2}f(0)-\frac{1}{12}f'(0)+\frac{1}{720}f^{(3)}(0)+\cdots$. We never need more terms in the expansion since each derivative lowers the degree of divergence by one.}. For instance for $\mathbf{SE1c}$ (\ref{eq:SE1cCurvedDRExm}) we have,

\begin{align}
SE1c &= \frac{N_s}{\pi^2}\left(\sum_{a=1}^{\infty} aR_s^2\left(\frac{a}{M}\right) - \sum_{a=1}^{\infty} \frac{1}{a}R_s^2\left(\frac{a}{M}\right) \right) = \nonumber\\
&=\frac{N_s}{\pi^2}\left[4\pi^2\cal{C}_2^{100}-\frac{1}{12}-\ln(\cal{A}_{100}M)-\gamma+\cal{O}(M^{-1})\right].
\end{align}

The rest of the expressions are given in appendix \ref{ssapp:1LoopCurvedCO}. We denote the total result for the self-energies by $G_{CO}$ and $S_{CO}$ for the vector and scalar respectively, including the counterterm diagram contribution in the total answer (figure 2),

\begin{align}
G_{CO} &= -12\left(Z_{0g}+4Z_{1g}\right)+4M^2\left(2\cal{C}_2^{200}-6\cal{C}_2^{100}+N_s\left(\cal{C}_2^{020}-3\cal{C}_2^{010}\right)+4N_f\cal{C}_2^{002}\right)+\nonumber\\
~~~&+8N_f\cal{F}^f_2+\frac{1}{\pi^2}\left[(6-N_s-4N_f)(\ln(M)+\gamma)+\ln\left(\frac{\cal{A}^8_{200}}{\cal{A}^2_{100}\cal{A}^{N_s}_{020}\cal{A}^{4N_f}_{002}}\right))+\right.\nonumber\\
~~~&\left.+\frac{71+N_s-5N_f}{6}-10\zeta(3)\right],\label{eq:1LoopSE1CurvedCO}\\
S_{CO} &= -2Z_{0s}^{ab}+M^2\left[(\mathbf{Q}^{(ab)cc}-2\delta^{ab})\cal{C}_2^{010}-4\delta^{ab}\cal{C}_2^{100}\right]+\frac{\delta^{ab}}{2\pi^2}\left[\ln\left(\frac{\cal{A}^2_{100}M}{\cal{A}_{010}}\right)+\gamma+2\right]-\nonumber\\
&~~~-\frac{\mathbf{Q}^{(ab)cc}}{48\pi^2}+\frac{\mathrm{tr}(\rho^{a\dag}\rho^b+\rho^{b\dag}\rho^a)}{2\pi^2}\left[4\pi^2M^2\cal{C}_2^{002}+\frac{5}{12}-\frac{1}{4}\left(\ln(4\cal{A}_{002}M)+\gamma\right)\right].\label{eq:1LoopSE2CurvedCO}
\end{align}

\subsubsection{Curved space counterterms summary}
We extract the $Z_{0g}$ counterterm by comparing (\ref{eq:1LoopSE1CurvedCO}), with (\ref{eq:GluonCurvedDR}),

\begin{align}
Z_0^g &= M^2\left[\frac{2}{3}\cal{C}_2^{200}-2\cal{C}_2^{100}+\frac{N_s}{3}\cal{C}_2^{020}- N_s \cal{C}_2^{010}\right]+\frac{7}{30\pi^2}\ln\left(\frac{\cal{A}_{100}}{\cal{A}_{200}}\right)-\frac{2}{15}\left(2\cal{F}_2^g+N_s\cal{F}_2^s\right)+\nonumber\\
&+\frac{5}{18\pi^2}+\frac{N_s}{45\pi^2}+N_f\left[\frac{4}{3}M^2\cal{C}_2^{002}-\frac{1}{360\pi^2}-\frac{2}{5}\cal{F}_2^f\right].\label{eq:Z0gCount}
\end{align}

The divergent piece of $Z_{0s}^{ab}$ is determined by (\ref{eq:1LoopSE2CurvedCO}). The finite piece on the other hand is arbitrary if we only demand gauge invariance and we parameterize it as,

\begin{align}
Z_{0s}^{ab} &= \frac{M^2}{2}\left[\left(\mathbf{Q}^{(ab)cc} - 2\delta^{ab} \right)\cal{C}_2^{010} - 4\delta^{ab}\cal{C}_2^{100} \right] + \frac{\delta^{ab}}{4\pi^2}\ln\left(\frac{\cal A_{100}^2M}{\cal A_{010}}\right) +\nonumber \\
&+ \mathrm{tr}(\rho^{a\dag }\rho^b + \rho^{b\dag }\rho^a)\left[M^2\cal{C}_2^{002} - \frac{1}{16\pi^2}\ln(\cal{A}_{002}M)\right]+\nonumber\\
&+\frac{1}{\pi^2}(C_1\delta^{ab}+C_2\mathrm{tr}(\rho^{a\dag}\rho^b+\rho^{b\dag}\rho^a)+C_3\mathbf{Q}^{(ab)cc}). \label{eq:Z0sCount}
\end{align}

The constants $C_1$, $C_2$ and $C_3$ will be determined in section \ref{ssec:NumericsChecks} by demanding conformal invariance to order $\lambda$.
%----------- end of "Curved space counterterms" subsection -------------

%-----------------------------------------------------------------------------
%-----------  end of "Regularization and counterterms" section ---------------
%-----------------------------------------------------------------------------

\pagebreak

%-----------------------------------------------------------------------------
%-------------  start of "Perturbative calculation" section ------------------
%-----------------------------------------------------------------------------

\section{Perturbative calculation of the effective action}\label{sec:2LoopComputation}
In this section we present the computation of the large $N$ effective action $S_{eff}(U)$ defined in (\ref{eq:Z}), to 2-loops in perturbation theory.

%--------------- start of "2-loop diagrams" subsection ----------------

\subsection{2-loop vacuum diagrams}
The 2-loop diagrams which we have to evaluate are depicted in figure 4.

\begin{figure}[!hc]\label{fig:2Loop}
\centering
  %Requires \usepackage{graphicx}
  \includegraphics[width=9cm,height=10cm]{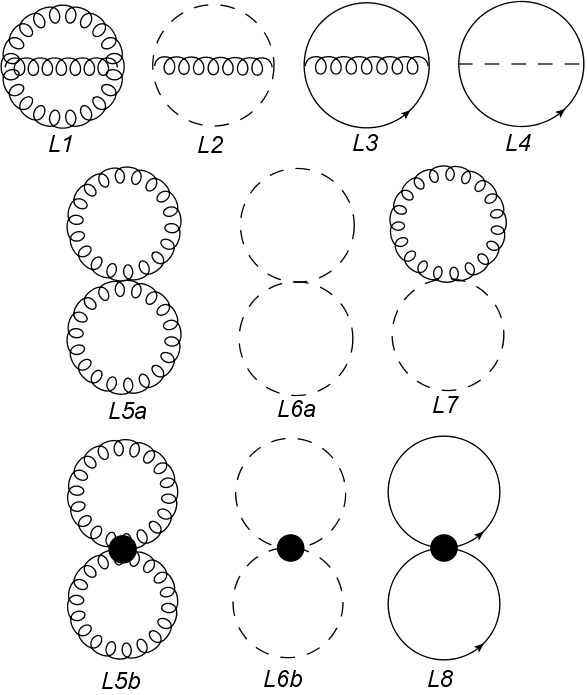}\\
 \caption{The 2-loop diagrams obtained after integrating $c$ and $a$. Vertices with a little circle correspond to terms we got after integrating the linear $a$ vertex. }
\end{figure}

Diagrams $\mathbf{L_1}$, $\mathbf{L_{5a}}$ and $\mathbf{L_{5b}}$ were already evaluated in \cite{Aharony:2006rf}. The expressions for the other diagrams that contribute in the planar limit are,

\begin{align}
\mathbf{L_2} &= -2\beta g^2 N_s C^{\alpha\beta\gamma}C^{\bar{\alpha}\bar{\beta}\bar{\gamma}}\int_0^{\beta} dt \Delta_{\beta}^{ij}(t)\Delta_{\alpha}^{jk}(t)\Delta_{\gamma}^{ki}(t),\label{eq:L2}\\
\mathbf{L_3} &= \frac{1}{2}\beta g^2 N_f G^{\alpha\beta\gamma}G^{\beta\alpha\bar{\gamma}}\int_0^{\beta} dt \left\{\Theta_{\alpha}^{ij}(-t)\Theta_{\beta}^{ik}(t)\Delta_{\gamma}^{kj}(t)+\right.\nonumber\\
       &~~~\left.+\Theta_{\alpha}^{ij}(-t)\Theta_{\beta}^{kj}(t)\Delta_{\gamma}^{ik}(t)\right\},\label{eq:L3}
\end{align}
\begin{align}
\mathbf{L_4} &= -\frac{1}{2}\beta g^2 \mathrm{tr}(\rho^{a\dag}\rho^a) F^{\bar{\alpha}\beta\gamma}F^{\alpha\bar{\beta}\bar{\gamma}}\int_0^{\beta} dt \left\{\Theta_{\alpha}^{ij}(t)\Delta_{\gamma}^{jk}(t)\Theta_{\beta}^{ki}(t)+\Theta_{\alpha}^{ij}(t)\Delta_{\gamma}^{ki}(t)\Theta_{\beta}^{jk}(t)\right\},\label{eq:L4}\\
\mathbf{L_{6a}} &= -\frac{1}{4}\beta g^2 \mathbf{Q}^{aabb}\left(B^{\alpha\bar{\alpha}\gamma}B^{\beta\bar{\beta}\bar{\gamma}}+B^{\alpha\beta\gamma}B^{\bar{\alpha}\bar{\beta}\bar{\gamma}}\right)\Delta^{ij}_{\alpha}(0)\Delta^{jk}_{\beta}(0),\label{eq:L6a}\\
\mathbf{L_{6b}} &= \beta g^2 N_s \frac{B^{\alpha\beta\gamma}B^{\bar{\alpha}\bar{\beta}\bar{\gamma}}}{j_\gamma(j_\gamma+2)}\left(D_{\tau}\Delta^{ij}_{\alpha}(0)D_{\tau}\Delta^{ik}_{\beta}(0)+(j_{\beta}+1)^2\Delta^{ij}_{\alpha}(0)\Delta^{jk}_{\beta}(0)\right),\label{eq:L6b}\\
\mathbf{L_{7}} &= \beta g^2 N_s D^{\alpha\bar{\alpha}\gamma}B^{\beta\bar{\beta}\bar{\gamma}}\Delta^{ij}_{\alpha}(0)\Delta^{jk}_{\beta}(0),\label{eq:L7}\\
\mathbf{L_8} &=\beta g^2 N_f\frac{F^{\alpha\beta\gamma}F^{\beta\alpha\bar{\gamma}}}{2j_{\gamma}(j_{\gamma}+2)} \left( \Theta^{ij}_{\alpha}(0)\Theta^{ik}_{\beta}(0)+\Theta^{ij}_{\alpha}(0)\Theta^{kj}_{\beta}(0)\right).\label{eq:L8}
\end{align}

Each diagram has three index loops, and the short-hand indexing on the propagators indicates how a term in their expansion in powers of $e^{i\beta\alpha}\otimes e^{-i\beta\alpha}$ should be grouped into traces. The notation $\Delta^{ij}$ means that the left side of the tensor product in the expansion belongs to the index loop $i$ and the right side to the index loop $j$. Also, the propagators above should be understood to be multiplied by the appropriate regulators.

We can sum analytically over all angular momentum quantum numbers, except the total angular momentum, by using the identities derived in appendix \ref{sapp:AMSums}. We then have to expand the propagators in powers of $e^{i\beta\alpha}\otimes e^{-i\beta\alpha}$ and extract the coefficients $f_n(x)$ and $f_{nm}(x)$ defined in (\ref{eq:2LoopEffAction}). The results are expressed as sums over the total angular momentum numbers constrained by some selection rules from the integrals over 3 spherical harmonics.

In all diagrams we find that we can write $f_n(x)=f_{1,+}(x^n)+(-1)^{n+1}f_{1,-}(x^n)$. The expressions for those are long and not very illuminating. They are summarized in appendix \ref{app:2LoopDiagrams}.

\subsection{Counterterm vacuum diagrams}
The counterterm diagrams which contribute to our computation are depicted in figure 5. For the gauge field and scalar we find that $f_n^{ct}(x) = f_1^{ct}(x^n)\equiv f^{ct}(x^n)$. After summation over angular momentum we obtain\footnote{In deriving these expressions we have used the equation of motion of the propagator to determine the coefficient of $Z_2$, e.g. in the bosonic case $(-D_t^2+(j+1)^2)\Delta(t)=\delta(t)$. In principle we should have added the $\delta(0)$ term but this term cancels with the corresponding diagram which involves the $U(1)$ part of the fields.}

\begin{align}\label{eq:fctg}
f_{g}^{ct}(x) = 2\left(Z_{0g} f(x) + (Z_{1g} - Z_{2g})g_1(x)\right),
\end{align}
\begin{equation}\label{eq:fcts}
f_{s}^{ct}(x) = \left(k(x)Z^{aa}_{0s} + g_1(x)Z^{aa}_{1s} - \left(g_1(x) + k(x)\right)Z_{2s}^{aa}\right),
\end{equation}
where,

\begin{align}\label{eq:xFunctions}
k(x) = \frac{x}{(1-x)^2}, ~&~ f(x)=k(x)+\ln(1-x), & g_1(x) = 6k(x)^2.
\end{align}

For the fermion we find $f_n^{ct}(x) = (-1)^{n-1}f_1^{ct}(x^n)$. After summation over angular momentum we obtain,

\begin{equation}\label{eq:fctf}
f_{f}^{ct}(x) = 6\frac{x^{\frac{3}{2}}(1+x)}{(1-x)^4}\left(Z_{1f}^{II} - Z_{2f}^{II}\right).
\end{equation}

\begin{figure}[!hc]\label{fig:2LoopCT}
\centering
  %Requires \usepackage{graphicx}
  \includegraphics[width=6cm,height=2.5cm]{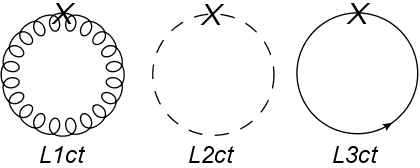}\\
 \caption{Counterterm vacuum diagrams which contribute to the 2-loop computation. }
\end{figure}

Plugging in the expressions for the counterterm coefficients we have computed (\ref{eq:Z1g})-(\ref{eq:Z2f}), (\ref{eq:Z0gCount})-(\ref{eq:Z0sCount}) we obtain,

\begin{align}
f_{g}^{ct}(x) &= f(x)\left\{M^2\left[\frac{4}{3}\cal{C}_2^{200} - 4\cal{C}_2^{100} + \frac{2}{3}N_s\cal{C}_2^{020} - 2N_s \cal{C}_2^{010} + \frac{8}{3}N_f \cal{C}_2^{002}\right] - \right. \nonumber \\
&\left. - \frac{1}{15}\left[8 \cal{F}_2^g + 4\cal F_2^s + 12\cal F_2^f\right] + \frac{7}{15\pi^2}\ln\left(\frac{\cal A_{100}}{\cal A_{200}}\right) + \frac{1}{\pi^2}\left[\frac{5}{9} + \frac{2N_s}{45} - \frac{N_f}{180} \right]\right\} + \nonumber \\
& + g_1(x)\left\{\frac{1}{15}\left[2\cal F_2^g + N_s\cal F_2^s + 8N_f\cal F_2^f\right] + \frac{8}{15\pi^2}\ln\left(\frac{\cal A_{200}}{\cal A_{100}}\right) + \frac{30 - 3N_s - 14N_f}{120\pi^2} \right\},\label{eq:fgct}\\
f_s^{ct}(x) &= k(x)\bigg\{M^2\left[Q^{aabb}\cal{C}_2^{010} - N_s\left(\cal{C}_2^{010} + 2\cal{C}_2^{100}\right) + 2\text{tr}(\rho^\dag\rho)\cal{C}_2^{002}\right] + \frac{N_s}{2\pi^2}\ln\left(\frac{\cal A_{100}}{\cal A_{010}}\right) + \nonumber \\
& + \frac{1}{\pi^2}\left[N_sC_1+2\text{tr}(\rho^{a\dag}\rho^a)C_2+2\mathbf{Q}^{aabb}C_3)\right]\bigg\}  +  \nonumber \\
& +  g_1(x)\left\{\frac{1}{3}\text{tr}(\rho^\dag\rho)\cal F_2^f + \frac{N_s}{3\pi^2}\ln\left(\frac{\cal A_{110}}{\cal A_{010}}\right) + \frac{N_s - \text{tr}(\rho^\dag\rho)}{24\pi^2} \right\},\label{eq:fsct}\\
f_f^{ct}(x) &= - 6\frac{x^{\frac{3}{2}}(1+x)}{(1-x)^4}\left\{\frac{N_f}{3\pi}\cal B_1^g + \frac{\text{tr}(\rho^\dag\rho)}{6\pi}\cal B_1^s + \frac{N_f}{3\pi^2}\ln\left(\frac{\cal A_{001}}{\cal A_{101}}\right) + \frac{-2N_f + \text{tr}(\rho^\dag\rho)}{24\pi^2}  \right\}.\label{eq:ffct}
\end{align}

In principle we are done and all we have to do is to plug the expressions for $f_n$ and $f_{nm}$ we found (\ref{eq:fnL1})-(\ref{eq:fnL8m}), (\ref{eq:fnmL1})-(\ref{eq:fnmL8}) summed with the counterterm expressions, (\ref{eq:fgct})-(\ref{eq:ffct}) inside the partition function formula (\ref{eq:SUnZ}) and perform the angular momentum summations numerically.

However our regularization involved several subtleties, and it is useful to check explicitly that all regulator dependence cancels between the counterterm vacuum diagrams and the 2-loop vacuum diagrams.

In order to see this cancelation we first extract the regulator dependent part of the 2-loop diagrams by expanding asymptotically in large angular momentum as was done in \cite{Aharony:2006rf}. The expressions we found for the regulator dependent part of each 2-loop diagram are listed in appendix \ref{sapp:RegDep2Loop}. One can readily check that indeed $f^{ct}_g+f^{ct}_s+\sum_if^{reg}_{L_i,+}$ and $f^{ct}_f+\sum_if^{reg}_{L_i,-}$ are regulator independent. This is a useful intermediate self consistency check on our computation.
%--------------- end of "2-loop diagrams" subsection ----------------

%-------- start of "Numerical evaluation and checks" subsection ---------
\subsection{Numerical evaluation and checks}\label{ssec:NumericsChecks}
As explained in the introduction there are two immediate applications to our calculation. One involves computing the order $\lambda$ correction to the Hagedorn temperature  (\ref{eq:HagedornCorrection}) of the theory. The second involves computing the order $\lambda$ corrections to the spectrum of the theory on $S^3$, (\ref{eq:SmallxExp}).

Both have already been computed for the $\cal{N}=4$ \textbf{SYM} case in \cite{Spradlin:2004pp} by using spin-chain methods.
Their result for the small temperature expansion is\footnote{The $\cal N=4$ partition function in (\ref{eq:Polya}) is adopted to our conventions. Our coupling constant is defined to be twice the one in \cite{Spradlin:2004pp}.},
\begin{equation}\label{eq:Polya}\begin{split}
\cal{Z}_{\mathbf{SYM}}(x) &= 1+\left(21+\frac{3 \lambda  \ln(x)}{2 \pi ^2}\right) x^2+\left(96+\frac{24 \lambda  \ln(x)}{\pi ^2}\right) x^{5/2}+\left(376+\frac{192 \lambda  \ln(x)}{\pi ^2}\right) x^3+\\
&+\left(1344+\frac{1032 \lambda  \ln(x)}{\pi ^2}\right) x^{7/2}+\left(4836+\frac{4440 \lambda  \ln(x)}{\pi ^2}\right) x^4+\\
&+\left(17472+\frac{17712 \lambda  \ln(x)}{\pi ^2}\right) x^{9/2}+\left(64608+\frac{71508\lambda  \ln(x)}{\pi ^2}\right) x^5+\\
&+432 \left(560+\frac{341 \lambda  \ln(x)}{\pi ^2}\right) x^{11/2}+\left(906741+\frac{2430477 \lambda  \ln(x)}{2 \pi ^2}\right) x^6+\cdots ,
\end{split}
\end{equation}
while the correction to the Hagedorn temperature is,
\begin{align}
\delta x_H=-x_H\frac{\lambda\ln(x_H)}{4\pi^2}\label{eq:N4Hagedorn}.
\end{align}

Using our computation method we can perform this expansion and compute the perturbative correction to the Hagedorn temperature numerically. As a self consistency check we indeed found that the numerics do not depend on the choice of regulating function.

However before we compare our computation to (\ref{eq:Polya}) and (\ref{eq:N4Hagedorn}), we must fix the arbitrary constant piece in (\ref{eq:fsct}). Recall that the source of the constants $C_1$, $C_2$ and $C_3$ in $f_s^{ct}(x)$ is the mass renormalization of the scalar fields (\ref{eq:Z0sCount}). Since a mass term for the scalars is gauge invariant, these constants are arbitrary. In particular to order $x^{2}$ our computation yields the following partition function,

\begin{align}\label{eq:SmallxExpRes}
\cal{Z}(x) &= 1+\bigg[\frac{N_s(N_s+1)}{2}+\frac{\lambda  \ln(x)}{48 \pi ^2 } \big(\left(N_s+1\right)\left(N_s(48C_1-33)+(96C_2+11)\mathrm{tr}(\rho^{a\dag}\rho^a)+\right.\nonumber\\
&+\left.(96C_3+1)\mathbf{Q}^{aabb}\right)+6\mathbf{Q}^{aabb}\big)\bigg] x^2+\cdots.
\end{align}

If our theory is conformally invariant to order $\lambda$, then the coefficient of $x^n\ln(x)$ in (\ref{eq:SmallxExpRes}) is the sum of the anomalous dimensions of operators with classical dimension $n$ in the flat space theory (see below (\ref{eq:SmallxExp})). In particular, the operators of dimension 2 are,

\begin{align}
\hat{\cal{O}}_2^{ab}\equiv\mathrm{tr}(\phi^a\phi^b).%,& ~~~\hat{\cal{O}}^{Ia}_{\frac{5}{2}}\equiv\mathrm{tr}(\psi^I\phi^a).
\end{align}

We computed the anomalous dimension matrix of those operators in the flat space theory corresponding to (\ref{eq:FlatAction})\footnote{The assumption of having only commutator interactions is implicit in (\ref{eq:ADtr2}) since only the combination $\mathbf{Q}^{aabb}$ appears in it. Under this assumption the two possible combinations are related by $\mathbf{Q}^{abab}=-\frac{1}{2}\mathbf{Q}^{aabb}$.},

\begin{align}
\sum_{\mathrm{dim}\left[\hat{\cal{O}}\right]=2}\gamma_{\hat{\cal{O}}} &=-\frac{\lambda}{8\pi^2}\left[3N_s(N_s+1)-(N_s+1)\mathrm{tr}(\rho^{a\dag}\rho^a)-\mathbf{Q}^{aabb}\right],\label{eq:ADtr2} %\\
%\sum_{\mathrm{dim}\left[\hat{\cal{O}}\right]=5/2}\gamma_{\hat{\cal{O}}} &= -\frac{\lambda}{4\pi^2}\left[6N_sN_f+\mathrm{tr}(\rho^a\rho^{a\dag})(4-N_s-2N_f)\right]\label{eq:ADtr52} .
\end{align}

Comparing the coefficient of the $x^2\ln(x)$ in (\ref{eq:SmallxExpRes}) to (\ref{eq:ADtr2}) we can solve for $C_1$, $C_2$ and $C_3$,

\begin{align}
C_1=\frac{5}{16},~~C_2=-\frac{5}{96},~~C_3=-\frac{1}{96}.
\end{align}

We evaluated the angular momentum sums in the 2-loop diagrams numerically up to order $x^6$ (we could easily have reached higher orders with more computer time but it was not needed in this work),

\begin{align}
f_{1,+}(x)&\equiv \sum_i f_{1,+}^{Li}(x)+f^{ct}_g(x)+f^{ct}_s(x)=\nonumber\\
&=\frac{1}{\pi^2}\left[\left(-\frac{3}{8}N_s + \frac{1}{8}\mathrm{tr}(\rho^{a\dag}\rho^a)\right)x\right.+\nonumber\\
&+\left(\frac{1}{4}+\frac{1}{2}N_f+\frac{1}{8}N_s-\frac{1}{8}\mathbf{Q}^{abba}+\frac{1}{2}\mathrm{tr}(\rho^{a\dag}\rho^a)\right)x^2+\nonumber\\
&+\left(4+2N_f+\frac{25}{8}N_s-\frac{1}{2}\mathbf{Q}^{abba}+\frac{11}{8}\mathrm{tr}(\rho^{a\dag }\rho^a)\right)x^3+\nonumber\\
&+\left(\frac{55}{4}+5N_f+\frac{41}{4}N_s-\frac{5}{4}\mathbf{Q}^{abba}+3\mathrm{tr}(\rho^{a\dag}\rho^a)\right)x^4+\nonumber\\
&+\left(32+10N_f+\frac{185}{8}N_s-\frac{5}{2}\mathbf{Q}^{abba}+\frac{45}{8}\mathrm{tr}(\rho^{a\dag}\rho^a)\right)x^5+\nonumber\\
&\left.+\left(\frac{245}{4}+\frac{35}{2}N_f+\frac{347}{8} N_s-\frac{35}{8}\mathbf{Q}^{abba}+\frac{19}{2}\mathrm{tr}(\rho^{a\dag}\rho^a)\right)x^6+\cdots\right],\label{eq:f1p}\\
f_{1-}(x) &\equiv \sum_i f_{1,-}^{Li}(x)+f^{ct}_f(x)= \nonumber\\
&=\frac{1}{\pi^2}\left[\frac{1}{4}\mathrm{tr}(\rho^{a\dag}\rho^a) x^{\frac{3}{2}} + (3 N_f +\frac{5}{4} \mathrm{tr}(\rho^{a\dag}\rho^a)) x^{\frac{
 5}{2}}+\right.\nonumber\\
 & + (12 N_f +\frac{7}{2}\mathrm{tr}(\rho^{a\dag}\rho^a)) x^{\frac{7}{2}} + (30N_f +\frac{15}{2} \mathrm{tr}(\rho^{a\dag}\rho^a)) x^{\frac{9}{2}} + \nonumber\\
 &\left.+(60N_f +\frac{55}{4}\mathrm{tr}(\rho^{a\dag}\rho^a)) x^{\frac{11}{2}}+\cdots\right],\label{f1m}\\
\widetilde{F}_2^{np}(x) &\equiv\sum_i\widetilde{F}_{2,Li}^{np}=\nonumber\\ &=\frac{\mathbf{Q}^{aabb}}{8\pi^2}x^2-\frac{\mathrm{tr}(\rho^{a\dag }\rho^a)}{\pi^2}x^{\frac{5}{2}}+\nonumber\\
&+\frac{\left[4\mathbf{Q}^{aabb}\!-\!2\mathrm{tr}(\rho^{a\dag }\rho^a)\!-\!6N_f\!-\!9N_s\right]}{4\pi^2}x^3\!-\!\frac{4\mathrm{tr}(\rho^{a\dag }\rho^a)}{\pi^2}x^{\frac{7}{2}}+\nonumber\\
&+\frac{18\!+\!23\mathbf{Q}^{aabb}\!-\!48N_s}{8\pi^2}x^4\!+\!\frac{3(N_f\!-3\mathrm{tr}(\rho^{a\dag }\rho^a))}{\pi^2}x^{\frac{9}{2}}+\nonumber\\
&+\!\frac{\!14\mathbf{Q}^{aabb}\!-\!\mathrm{tr}(\rho^{a\dag }\rho^a)\!-9N_f-\!39N_s}{2\pi^2}x^5-\frac{20\mathrm{tr}(\rho^{a\dag }\rho^a)}{\pi^2}x^{\frac{11}{2}}+\nonumber\\
&+\frac{42+50\mathbf{Q}^{aabb}-6\mathrm{tr}(\rho^{a\dag }\rho^a)-18N_f-117N_s}{4\pi^2}x^6+\cdots .\label{eq:F2np}
\end{align}

One can plug those in (\ref{eq:SUnZ}) and see that for $\cal{N}=4$ (see appendix \ref{app:Notations}) we reproduce (\ref{eq:Polya}) and (\ref{eq:N4Hagedorn}). This is a very non-trivial check of our computation.

It is also important to have a check for the generic theory which is not conformally invariant, in order to make sure that conformal invariance is preserved to order $\lambda$ by our renormalization scheme. To do so we computed the anomalous dimensions of dimension 5/2 and 3 operators,

\begin{align}
\hat{\cal{O}}^{Ia}_{\frac{5}{2}}&\equiv\mathrm{tr}(\psi^I\phi^a),& \hat{\cal{O}}_3^{a\mu\nu}&\equiv i\mathrm{tr}(\phi^aF^{\mu\nu}),&
\hat{\cal{O}}_3^{I\alpha J\beta}&\equiv \mathrm{tr}(\psi^I_{\alpha}\psi^J_{\beta})\nonumber\\
\hat{\cal{O}}_3^{abc}&\equiv\mathrm{tr}(\phi^a\phi^b\phi^c),&
\hat{\cal{O}}_3^{1ab}&\equiv i\partial_{\mu}\mathrm{tr}(\phi^a\phi^b),&
\hat{\cal{O}}_3^{2ab}&\equiv i \mathrm{tr}(\phi^{[a}D_{\mu}\phi^{b]}),
\end{align}
for which we found,

\begin{align}
\sum_{\mathrm{dim}\left[\hat{\cal{O}}\right]=5/2}\gamma_{\hat{\cal{O}}} &= -\frac{\lambda}{4\pi^2}\left(6N_sN_f+\mathrm{tr}(\rho^a\rho^{a\dag})(4-N_s-2N_f)\right)\\
\sum_{\mathrm{dim}\left[\hat{\cal{O}}\right]=3}\gamma_{\hat{\cal{O}}}&=\frac{\lambda}{\pi^2}\left(-5N_s-\frac{11}{8}N_s^2-\frac{3}{8}N_s^3-\frac{3}{2}N_f+\frac{1}{2}N_sN_f+\frac{1}{4}\mathrm{tr}(\rho^{a\dag}\rho^a)+N_s\mathrm{tr}(\rho^{a\dag}\rho^a)+\right.\nonumber\\
&\left.+N_f\mathrm{tr}(\rho^{a\dag}\rho^a)+\frac{1}{8}N_s^2\mathrm{tr}(\rho^{a\dag}\rho^a)+\mathbf{Q}^{aabb}-\frac{1}{8}N_s\mathbf{Q}^{aabb}\right),
\end{align}
in agreement with the results derived from (\ref{eq:f1p})-(\ref{eq:F2np}). This gives a check that by applying our renormalization scheme for the scalar field mass terms, the general theory is indeed conformally invariant to order $\lambda$.

In the pure \textbf{YM} and $\cal N =4$ \textbf{SYM} theories $\Delta T_H\equiv\pi^2\delta T_H/T_H$ was found to be positive and rational. In the more general cases that we considered, we found that this quantity can be either positive or negative and that it is not necessarily rational. For example, consider gauge theories with $N_s+N_f=10$, $\mathbf{Q}^{aabb}=N_s(1-N_s)$ corresponding to $\text{tr}([\phi^a,\phi^b]^2)$ self interaction, and $\text{tr}(\rho^a\rho^{a\dag})=N_sN_f$ Yukawa couplings. Strangely, it turns out that all these theories have $x_H(\lambda=0)=7-4\sqrt{3}$. The results for $\Delta T_H$ for these theories are,

\begin{table}[h]\label{tab1}
\begin{center}
\begin{tabular}{|c||c|c|c|c|c|} \hline
\rule{0pt}{1.2em}%
$N_s$        & $10$ & $9$ & $8$ & $7$ & $6$  \cr \hline
$N_f$        & $0$  &  $1$ & $2$ & $3$ & $4$  \cr \hline
$\Delta T_H$ & $-1/8$ & $0.017045$ & $1/8$ & $0.201786$ & $1/4$ \cr \hline
\end{tabular}
\end{center}
\end{table}

For pure scalar theories with the same self interactions as above, $\Delta T_H$ becomes negative for $N_s>2$ but is not necessarily rational. For instance,

 \begin{table}[h]\label{tab1}
 \footnotesize{
\begin{center}
\begin{tabular}{|c||c|c|c|c|c|c|c|c|c|c|c|} \hline
\rule{0pt}{1.2em}%
$N_s$ & $10$ & $11$ & $12$  & $13$  & $14$  & $15$  & $16$  & $17$  & $18$  & $19$  & $20$ \cr\hline
$\Delta T_H$ & $-\frac{1}{8}$ & $-0.132353$ & $-\frac{5}{36}$ & $-0.144737$ & $-\frac{3}{20}$ & $-0.154762$ & $-\frac{7}{44}$ & $-0.163043$ & $-\frac{1}{6}$ & $-\frac{17}{100}$ & $-0.173077$ \cr\hline
\end{tabular}
\end{center}}
\end{table}

For generic couplings the results are of course irrational, but as we see in the examples above, rational results are not so difficult to construct. It would be interesting to understand the reason for $\Delta T_H$ to be rational.

%--------- end of "Numerical evaluation and checks" subsection ----------
%-----------------------------------------------------------------------------
%--------------  end of "Perturbative calculation" section -------------------
%-----------------------------------------------------------------------------
\pagebreak

\section{Conclusions}\label{sec:Conclusions}
In this work we have computed the full large $N$ partition function of $SU(N)$ \textbf{YM} theory on $S^3$ in perturbation theory, to 2-loops order. We included scalar and spinor fields in the adjoint representation of the gauge group, with arbitrary commutator interactions. The computation involved expanding those fields in Kaluza-Klein modes on $S^3$. Then, instead of solving momentum integrals, we had to perform sums over the angular momentum of those modes and derive various identities in order to perform these sums. Furthermore, we used a cutoff regularization scheme which is not gauge invariant. To deal with this we added non-gauge invariant counterterms to the theory and demanded that their value precisely compensates for the non-gauge invariance introduced by the regulator.

Furthermore, we demanded that our theory is conformally invariant to order $\lambda$, so that we can relate the perturbative corrections to the energy spectrum on $S^3\times\mathds{R}$, to the sums of anomalous dimensions on $\mathds{R}^4$, by using the state-operator mapping. Though generically our theories have a non-zero beta function, it contributes to the partition function only at higher orders. The only obstruction to achieving conformal invariance at order $\lambda$ is the generation of mass terms for the scalar fields. By demanding that the state-operator mapping works for dimension 2 operators, we can fix the mass renormalization of the scalar fields such that the theory is still conformal at order $\lambda$.

Our computation agrees with known results for the 1-loop anomalous dimensions of $\cal{N}=4$ \textbf{SYM} which were computed in \cite{Spradlin:2004pp}. This is a highly non-trivial check on our computation which was done by a very different method. Our results can be used to numerically compute the partition function in the confinement phase of the large $N$ gauge theories which we considered. Furthermore, the order $\lambda$ correction to the Hagedorn temperature can be computed very easily for those theories.

This computation also has a direct application towards determining the order of the deconfinement phase transition. As shown in \cite{Aharony:2003sx} this involves a 3-loop computation, and the 2-loop computation of $f_{11}(x)$. The latter can be computed from our results for the $f_{nm}^{Li}(x)$'s (\ref{eq:fnmL1})-(\ref{eq:fnmL8}).

\section*{Acknowledgments}
We would like to thank Ofer Aharony for suggesting this work as an M.Sc project, and for his help and guidance throughout. We would also like to thank Mark
Van Raamsdonk for comments on a draft of this paper. This work was supported in part by the Israel-U.S. Binational Science Foundation, by a center of excellence supported by the Israel Science Foundation (grant number 1468/06), by a grant (DIP H52) of the German Israel Project Cooperation, by the European network MRTN-CT-2004-512194, and by Minerva.
\pagebreak
%-------------------------------------------------
%----------------  Appendix  ---------------------
%-------------------------------------------------
\numberwithin{equation}{subsection}
\begin{appendix}

%-----------------------------------------------------------------------------
%-----   start of "Additional notations" section -------
%-----------------------------------------------------------------------------
\section{Additional notations}\label{app:Notations}
Having enough indices as it is in our computation, we use matrix notation for spinors. Our notation for the Weyl spinor matrices is such that,

\begin{align}
\sigma^{\mu}\equiv (1,i\tau^i)~~,~~&\bar{\sigma}^{\mu}\equiv (1,-i\tau^i),
\end{align}
where $\tau^i$ are the usual Pauli matrices. Also we use the charge conjugation matrix in the Yukawa terms,

\begin{align}
\varepsilon\equiv i\tau^2.
\end{align}

In order to apply our computation to $\cal{N}=4$ \textbf{SYM}, take $N_s=6$ and $N_f=4$. Furthermore, in that case the Yukawa matrices satisfy,

\begin{align}
\rho^{a\dag}\rho^b+\rho^{b\dag}\rho^a = 2\delta^{ab}\mathds{1}_{4\times4},
\end{align}
while the quartic scalar coupling satisfies,

\begin{align}
\mathbf{Q}^{abcd}=2\delta^{ac}\delta^{bd}-\delta^{ad}\delta^{cb}-\delta^{ab}\delta^{dc}.
\end{align}
%-----------------------------------------------------------------------------
%-----   start of "Additional notations" section -------
%-----------------------------------------------------------------------------

%-----------------------------------------------------------------------------
%-----   start of "The gauge fixed Yang-Mills action on $S^3$" section -------
%-----------------------------------------------------------------------------

\section{The gauge fixed Yang-Mills action on $S^3$}
\label{app:Action}

%------------------ start of "The action" subsection ------------------

\subsection{The action}\label{sapp:action}
Imposing the gauge conditions (\ref{eq:CoulombGauge}), (\ref{eq:ZeroModeFix}) and adding the conformal coupling for the scalar fields, the Euclidean action on $S^3$ can be written as $\mathbf{S}_E = \mathbf{S}_g+\mathbf{S}_s+\mathbf{S}_f$ where,

\begin{align}
\mathbf{S}_g &= \int_0^{\beta}dt\int_{S^3}\textrm{tr}\left\{-\frac{1}{2}A_i(D_t^2+\partial_j^2)A^i-\frac{1}{2}A_0\partial^2A_0-c^{\dag}\partial^2c+\right.\nonumber\\
&+igD_tA^i[A_i,A_0]-ig[A^i,A_0]\partial_iA_0-ig\partial_iA_j[A^i,A^j]+\\
&\left.+\frac{1}{4}g^2[A_i,A_j][A^j,A^i]-\frac{1}{2}g^2[A_0,A_i][A^0,A^i]-ig\partial_ic^{\dag}[A_i,c]\right\},\nonumber
\end{align}
where we define $D_t\equiv\partial_t-[\alpha,*]$. For the pure conformally coupled scalar (kinetic plus quartic interaction),

\begin{align}
\mathbf{S}_s=&\int_0^{\beta}dt\int_{S^3}\textrm{tr}\left\{-\frac{1}{2}\Phi_a(D_t^2+\partial_i^2-1)\Phi_a-\right.\nonumber\\
&-ig[A_0,\Phi_a]D_t\Phi_a-ig[A_i,\Phi_a]\partial^i\Phi_a\\
&\left.-\frac{1}{2}g^2[A_i,\Phi_a]^2-\frac{1}{2}g^2[A_0,\Phi_a]^2-\frac{1}{4}g^2\mathbf{Q}^{abcd}\Phi_a\Phi_b\Phi_c\Phi_d\right\},\nonumber
\end{align}
and for the fermions (kinetic plus Yukawa interaction) we write,

\begin{align}
\mathbf{S}_f &= \int_0^{\beta}dt\int_{S^3}\textrm{tr}\left\{i\Psi^{\dag I}(D_t+\sigma^i\partial_i)\Psi_I+\right.\nonumber\\
&+g\Psi^{\dag I}[A_0,\Psi_I]+g\Psi^{I\dag}\sigma^i[A_i,\Psi_I]+\\
&\left.+\frac{1}{2}g\Psi_I^T\varepsilon(\rho^{a\dag})^{IJ}[\Phi_a,\Psi_J]+\frac{1}{2}g\Psi^{\dag I}\varepsilon\rho^a_{IJ}[\Phi_a,(\Psi^{\dag J })^T]\right\}.\nonumber
\end{align}
\end{appendix}

In the above the spatial derivative is assumed to be the covariant one on $S^3$ including the appropriate connection terms for vectors and spinors.

Using the properties listed in appendix \ref{app:harmonics} we can write the action in terms of the expansions of the fields in Kaluza-Klein modes (\ref{eq:HarmonicsExpansions}), and in terms of the spherical harmonics integrals (\ref{eq:3SphericalIntegrals}) as $\mathbf{S}_E=\mathbf{S}_2+\mathbf{S}_3+\mathbf{S}_4$, with

\begin{align}
\mathbf{S}_2 &= \int_0^{\beta} dt \mathrm{tr}\left\{
\frac{1}{2}A^{\bar{\alpha}}(-D_t^2+(j_{\alpha}+1)^2)A^{\alpha}+\frac{1}{2}a^{\bar{\alpha}}j_{\alpha}(j_{\alpha}+2)a^{\alpha}+c^{\dag \alpha}j_{\alpha}(j_{\alpha}+2)c^{\alpha}+\right.\nonumber\\
&~\left.+\frac{1}{2}\phi_a^{\bar{\alpha}}(-D_t^2+(j_{\alpha}+1)^2)\phi_a^{\alpha}+i\psi^{\dag \alpha I}(D_t+\epsilon_{\alpha}(j_{\alpha}+\frac{1}{2}))\psi^{\alpha}_I\right\}.\label{eq:QuadraticAction}\\
\mathbf{S}_3 &= ig\int_0^{\beta} dt \mathrm{tr}\bigg\{C^{\bar{\alpha}\beta\gamma}c^{\dag \alpha}[A^{\beta},c^{\gamma}]+2C^{\alpha\beta\gamma}a^{\alpha}A^{\beta}a^{\gamma}-\nonumber\\
&~-D^{\alpha\beta\gamma}[A^{\alpha},D_tA^{\beta}]a^{\gamma}+\epsilon_{\alpha}(j_{\alpha}+1)E^{\alpha\beta\gamma}A^{\alpha}A^{\beta}A^{\gamma}-B^{\alpha\beta\gamma}[a^{\alpha},\phi^{\beta}_a]D_t\phi_a^{\gamma}-\nonumber\\
&~-C^{\alpha\beta\gamma}[A^{\beta},\phi_a^{\alpha}]\phi_a^{\gamma}-iF^{\alpha\beta\gamma}\psi^{\dag \alpha I}[a^{\gamma},\psi_I^{\beta}]-iG^{\alpha\beta\gamma}\psi^{\dag \alpha I}[A^{\gamma},\psi^{\beta}_I]-\nonumber\\
&~\left.-\frac{i}{2}F^{\bar{\alpha}\beta\gamma}\psi_I^{\alpha}(\rho^{a\dag})^{IJ}[\phi_a^{\gamma},\psi_J^{\beta}]+\frac{i}{2}F^{\alpha\bar{\beta}\gamma}\psi^{\dag I\alpha}\rho^a_{IJ}[\phi_a^{\gamma},\psi^{\dag J\beta}]\right\}.\label{eq:CubicAction}\\
\mathbf{S}_4 &= -\frac{1}{2}g^2\int_0^{\beta} dt \mathrm{tr}\left\{
\left(D^{\beta\bar{\lambda}\alpha}D^{\delta\lambda\gamma}+\frac{1}{j_{\lambda}(j_{\lambda}+2)}C^{\alpha\beta\bar{\lambda}}C^{\gamma\delta\lambda}\right)[a^{\alpha},A^{\beta}][a^{\gamma},A^{\delta}]+\right.\nonumber\\
&~+\left(D^{\alpha\gamma\bar{\lambda}}D^{\beta\delta\lambda}-D^{\alpha\delta\bar{\lambda}}D^{\beta\gamma\lambda}\right)A^{\alpha}A^{\beta}A^{\gamma}A^{\delta}+B^{\alpha\beta\bar{\lambda}}B^{\lambda\gamma\delta}[a^{\alpha},\phi_a^{\beta}][a^{\gamma},\phi_a^{\delta}]+\nonumber\\
&~\left.+D^{\alpha\gamma\lambda}B^{\bar{\lambda}\beta\delta}[A^{\alpha},\phi_a^{\beta}][A^{\gamma},\phi_a^{\delta}]+\frac{1}{2}B^{\alpha\beta\bar{\lambda}}B^{\lambda\gamma\delta}\mathbf{Q}^{abcd}\phi_a^{\alpha}\phi_b^{\beta}\phi_c^{\gamma}\phi_d^{\delta}\right\}.\label{eq:QuarticAction}
\end{align}

%------------------ end of "The action" subsection ------------------

%------- start of "Integrating out $a$ and $c$" subsection ----------

\subsection{Integrating out $a$ and $c$}\label{sapp:IntegrateOut}
The fields $a$ and $c$ appear only quadratically in the gauge fixed action (\ref{eq:QuadraticAction})-(\ref{eq:QuarticAction}) and it is convenient to integrate them out first. Their propagators are,

\begin{align}
\langle c^{\alpha}_{ij}(t^{\prime})c^{\dag \beta}_{kl}(t)\rangle &= \frac{1}{j_{\alpha}(j_{\alpha}+2)}\delta(t^{\prime}-t)\delta^{\alpha\beta}\delta_{il}\delta_{kj},\\
\langle a^{\alpha}_{ij}(t^{\prime})a^{\beta}_{kl}(t) &= \frac{1}{j_{\alpha}(j_{\alpha}+2)}\delta(t^{\prime}-t)\delta^{\alpha\bar{\beta}}\delta_{il}\delta_{kj}.
\end{align}

From loops of $a$'s and $c$'s we get the effective vertices,

\begin{align}
A_2^A &= Ng^2\frac{D^{\beta\bar{\lambda}\alpha}D^{\delta\lambda\bar{\alpha}}}{j_\alpha(j_\alpha+2)}\int_0^{\beta} dt \delta(0)\textrm{tr}(A^{\beta}A^{\delta})\label{eq:A2A}\\
A_2^{\phi} &= Ng^2\frac{B^{\alpha\beta\bar{\lambda}}B^{\bar{\alpha}\delta\lambda}}{j_\alpha(j_\alpha+2)}\int_0^{\beta} dt \delta(0)\textrm{tr}(\phi^\beta_a\phi^\delta_a)\label{eq:A2Phi}.
\end{align}

The other diagrams arise from strings of $a$'s using the linear $a$ vertices in (\ref{eq:CubicAction}),
\begin{align}
B^A_4 &= \frac{g^2}{2}\frac{D^{\alpha \beta \lambda}D^{\gamma \delta \bar{\lambda}}}{j_{\lambda}(j_{\lambda}+2)}\int^{\beta}_{0}dt \mathrm{tr}\left([A^\alpha,D_\tau A^\beta][A^{\gamma},D_\tau A^{\delta}]\right),\label{eq:B4A} \\
B^\phi_4 &= \frac{g^2}{2}\frac{B^{\alpha \beta \lambda}B^{\gamma \delta \bar{\lambda} }}{j_{\lambda}(j_{\lambda}+2)}\int^{\beta}_{0}dt \mathrm{tr}\left([\phi^\alpha_a,D_\tau \phi^\beta_a][\phi^{\gamma}_{b},D_\tau \phi_{b}^{\delta}]\right),\label{eq:B4Phi} \\
B^\psi_4 &= -\frac{g^2}{2}\frac{F^{\alpha \beta \lambda}F^{\gamma \delta \bar{\lambda}}}{j_{\lambda}(j_{\lambda}+2)}\int^{\beta}_{0}dt \textrm{tr}\left(\{\psi^\beta_I,\psi^{\dagger \alpha I}\}\{\psi_J^{\delta}, \psi^{\dagger \gamma J}\}\right),\label{eq:B4Psi} \\
B^{A \phi}_{2,2} &= g^2\frac{D^{\alpha \beta \lambda}B^{\gamma \delta \bar{\lambda}}}{j_{\lambda}(j_{\lambda}+2)}\int^{\beta}_{0}dt \mathrm{tr}\left([A^\alpha,D_\tau A^\beta][\phi^{\gamma}_a,D_\tau \phi^{\delta}_a]\right), \label{eq:B2A2Phi} \\
B^{A \psi}_{2,2} &= ig^2\frac{D^{\alpha \beta \lambda}F^{\gamma \delta \bar{\lambda}}}{j_{\lambda}(j_{\lambda}+2)}\int^{\beta}_{0}dt \mathrm{tr}\left([A^\alpha,D_\tau A^\beta]\{\psi_I^{\delta}, \psi^{\dagger \gamma I }\}\right),\label{eq:B2A2Psi} \\
B^{\phi \psi}_{2,2} &= ig^2\frac{B^{\alpha \beta \lambda}F^{\gamma \delta \bar{\lambda}}}{j_{\lambda}(j_{\lambda}+2)}\int^{\beta}_{0}dt \mathrm{tr}\left([\phi_a^\alpha,D_\tau \phi_a^\beta]\{\psi_I^{\delta}, \psi^{\dagger \gamma I }\}\right).\label{eq:B2Phi2Psi}
\end{align}

The $\delta(0)$ divergence in (\ref{eq:A2A}) and (\ref{eq:A2Phi}) is an artifact of the Coulomb gauge and arises since $a$ and $c$ are not dynamical fields. We see that those vertices cancel by adding to them the expressions obtained by contracting the two covariant derivatives, in (\ref{eq:B4A}) and (\ref{eq:B4Phi}) (see (\ref{eq:PropDerivative})). Therefore we can forget about $\delta(0)$ factors for diagrams which involve those vertices.

Also the vacuum diagrams which arise from one insertion of any of the vertices (\ref{eq:B2A2Phi}), (\ref{eq:B2A2Psi}) or (\ref{eq:B2Phi2Psi}) vanish for kinematical reasons. After the self contractions the summation over spherical harmonics (see (\ref{eq:DB}), (\ref{eq:DF}) ,(\ref{eq:BF})) vanishes unless $j_{\lambda} = 0$, which can not be since this is the angular momenta of $a^{\lambda}$, ($A_0$) whose zero mode has been factored out.

%------- start of "Integrating out $a$ and $c$" subsection ----------

%-----------------------------------------------------------------------------
%------   end of "The gauge fixed Yang-Mills action on $S^3$" section --------
%-----------------------------------------------------------------------------

%-----------------------------------------------------------------------------
%-------  start of "Properties of $S^3$ spherical harmonics" section ---------
%-----------------------------------------------------------------------------

\section{Properties of $S^3$ spherical harmonics}\label{app:harmonics}

%--------- start of "$S^3 spherical harmonics" subsection -----------
\subsection{ $S^3$ spherical harmonics}
In this section we list some properties of spherical harmonics that we used. Due to lack of space this is far from a complete review, and the interested reader is referred to \cite{Aharony:2005bq}, \cite{Cutkosky:1983jd} for scalar and vector spherical harmonics, and to \cite{Sen:1985dc},\cite{Pais:1954} for spinor spherical harmonics.

Spherical harmonics sit in representations of the isometry group of $S^3$, which is $SO(4)\simeq SU(2)\times SU(2)$.
The scalar spherical harmonics transform in the $(j/2,j/2)$ representation, where $j\geq 0$ and $-j/2\leq m,m^{\prime}\leq j/2$.

They form a complete orthonormal set of eigenfunctions of the corresponding laplacian on $S^3$,

\begin{equation}\label{eq:ScalarEigenvalue}
\nabla^2\cal{S}^{m,n}_j =-j(j+2)\cal{S}^{m,n}_j,
\end{equation}
and satisfy the conjugation relation,

\begin{align}
(\cal{S}^{m,n}_j)^* & =(-1)^{m+n}\cal{S}^{-m,-n}_j.
\end{align}

The vector spherical harmonics sit in the $((j+\epsilon)/2,(j-\epsilon)/2)$ representation, where $\epsilon=\pm 1$ and $j\geq 1$. This set of functions satisfies the eigenvalue equations,

\begin{eqnarray}
\begin{array}{lll}\label{eq:VectorEigenvalue}
\nabla^2\vec{\cal{V}}^{m,n}_{j,\epsilon} & = &-(j+1)^2\vec{\cal{V}}^{m,n}_{j,\epsilon},\\
\vec{\nabla}\times\vec{\cal{V}}^{m,n}_{j,\epsilon} & = &-\epsilon(j+1)\vec{\cal{V}}^{m,n}_{j,\epsilon},\\
\vec{\nabla}\cdot\vec{\cal{V}}^{m,n}_{j,\epsilon} & = & 0,
\end{array}
\end{eqnarray}
and the conjugation relation,

\begin{align}
(\vec{\cal{V}}^{m,n}_{j,\epsilon})^* & =  (-1)^{m+n+1} \vec{\cal{V}}^{-m,-n}_{j,\epsilon}.
\end{align}

The spinor spherical harmonics are two-component Weyl spinors, and transform in the representations $(\frac{j-(1-\epsilon)/2}{2},\frac{j-(1+\epsilon)/2}{2})$ for $\epsilon=\pm 1$ and $j\geq 1$. They can be found from the scalar functions by tensor multiplying a $(j/2,j/2)$ representation, with a $(1/2,0)$, (or $(0,1/2)$) basis spinor with the right Clebsch-Gordan coefficients. The properly normalized result is

\begin{eqnarray}
\begin{array}{lll}\label{eq:SpinorHarmonicDef}
\cal{Y}^{m,n}_{j,+} & = &
~~~\frac{1}{\sqrt{j}}\begin{pmatrix}
\sqrt{j/2+m}\cal{S}^{m-1/2,n}_{j-1} \\
\sqrt{j/2-m}\cal{S}^{m+1/2,n}_{j-1}
\end{pmatrix}, \\

\cal{Y}^{m,n}_{j,-} & = &
\frac{1}{\sqrt{j+1}}\begin{pmatrix}
-\sqrt{j/2-m+1/2}\cal{S}^{m-1/2,n}_{j} \\
\sqrt{j/2+m+1/2}\cal{S}^{m+1/2,n}_{j}
\end{pmatrix}.

\end{array}
\end{eqnarray}

From the properties of the scalar functions one can check that those are normalized correctly

\begin{equation}\label{eq:SpinorOrthonormality}
\int_{S^3}\cal{Y}^{m,n\dag}_{j,\epsilon}\cal{Y}^{m^{\prime},n^{\prime}}_{j^{\prime},\epsilon^{\prime}} =  \delta_{j,j^{\prime}}\delta_{m,m^{\prime}}\delta_{n,n^{\prime}}\delta_{\epsilon,\epsilon^{\prime}},
\end{equation}
and that under conjugation those obey

\begin{equation}\label{eq:SpinorConjugate}
\left(\cal{Y}^{m,n}_{j,\epsilon}\right)^{*}= (-1)^{m+n-\epsilon/2}\varepsilon\left(\cal{Y}^{-m,-n}_{j,\epsilon}\right).
\end{equation}

The spinor functions satisfy the eigenvalues equation\footnote{The inclusion of the appropriate veilbein and spin connection should be understood. Also recall that in our convention for spinor matrices the $\sigma^i$ are anti-hermitian, hence the real eigenvalue in the above equation.},

\begin{align}\label{eq:SpinorEigenvalue}
\sigma^i\partial_i\cal{Y}^{m,n}_{j,\epsilon}=\epsilon(j+\frac{1}{2})\cal{Y}^{m,n}_{j,\epsilon}.
\end{align}

In the text we collect all the angular momentum indices of a scalar, vector or spinor spherical harmonic by a single Greek index. A barred Greek index on a scalar or vector spherical harmonic means the corresponding function is conjugated, while by a barred Greek index on a spinor spherical harmonic we mean $\cal{Y}^{\bar{\alpha}}\equiv(-1)^{m_{\alpha}+n_{\alpha}-\epsilon_{\alpha}/2}\left(\cal{Y}^{-m_{\alpha},-n_{\alpha}}_{j_{\alpha},\epsilon_{\alpha}}\right)$.
%---------- end of "$S^3 spherical harmonics" subsection ------------

%------- Start of "Spherical harmonics integrals" subsection --------
\subsection{Spherical harmonics integrals}
While expanding our fields in Kaluza-Klein modes on $S^3$ we defined integrals over three spherical harmonics   (\ref{eq:3SphericalIntegrals}). These integrals can be expressed as a reduced matrix element times the appropriate 3-j symbols,

\begin{align}
B^{\alpha\beta\gamma} &= R_{B}(j_{\alpha},j_{\beta},j_{\gamma})\thj{\frac{j_{\alpha}}{2}}{\frac{j_{\beta}}{2}}{\frac{j_{\gamma}}{2}}{m_{\alpha}}{m_{\beta}}{m_{\gamma}}\thj{\frac{j_{\alpha}}{2}}{\frac{j_{\beta}}{2}}{\frac{j_{\gamma}}{2}}{n_{\alpha}}{n_{\beta}}{n_{\gamma}} ,\nonumber\\
C^{{\alpha}{\beta}{\gamma}} &= R_{C}(j_{\alpha},j_{\beta},j_{\gamma})\thj{\frac{j_{\alpha}}{2}}{\frac{j_{\beta}+\epsilon_{\beta}}{2}}{\frac{j_{\gamma}}{2}}{m_{\alpha}}{m_{\beta}}{m_{\gamma}}\thj{\frac{j_{\alpha}}{2}}{\frac{j_{\beta}-\epsilon_{\beta}}{2}}{\frac{j_{\gamma}}{2}}{n_{\alpha}}{n_{\beta}}{n_{\gamma}} ,\nonumber\\
D^{{\alpha}{\beta}{\gamma}} &= R_{D\epsilon_{\alpha}\epsilon_{\beta}}(j_{\alpha},j_{\beta},j_{\gamma})\thj{\frac{j_{\alpha}+\epsilon_{\alpha}}{2}}{\frac{j_{\beta}+\epsilon_{\beta}}{2}}{\frac{j_{\gamma}}{2}}{m_{\alpha}}{m_{\beta}}{m_{\gamma}}\thj{\frac{j_{\alpha}-\epsilon_{\alpha}}{2}}{\frac{j_{\beta}-\epsilon_{\beta}}{2}}{\frac{j_{\gamma}}{2}}{n_{\alpha}}{n_{\beta}}{n_{\gamma}},\\
E^{{\alpha}{\beta}{\gamma}} &=  R_{E\epsilon_{\alpha}\epsilon_{\beta}\epsilon_{\gamma}}(j_{\alpha},j_{\beta},j_{\gamma})\thj{\frac{j_{\alpha}+\epsilon_{\alpha}}{2}}{\frac{j_{\beta}+\epsilon_{\beta}}{2}}{\frac{j_{\gamma}+\epsilon_{\gamma}}{2}}{m_{\alpha}}{m_{\beta}}{m_{\gamma}}\thj{\frac{j_{\alpha}-\epsilon_{\alpha}}{2}}{\frac{j_{\beta}-\epsilon_{\beta}}{2}}{\frac{j_{\gamma}-\epsilon_{\gamma}}{2}}{n_{\alpha}}{n_{\beta}}{n_{\gamma}} ,\nonumber \\
H^{{\alpha}{\beta}{\gamma}} &= R_{H\epsilon_{\alpha} \epsilon_{\beta}} (j_{\alpha},j_{\beta},j_{\gamma}) \thj{\frac{(j_{\alpha}-(1-\epsilon_{\alpha})/2)}{2}}{\frac{(j_{\beta}-(1-\epsilon_{\beta})/2)}{2}}{\frac{j_{\gamma}}{2}}{m_{\alpha}}{m_{\beta}}{m_{\gamma}}\thj{\frac{(j_{\alpha}-(1+\epsilon_{\alpha})/2)}{2}}{\frac{(j_{\beta}-(1+\epsilon_{\beta})/2)}{2}}{\frac{j_{\gamma}}{2}}{n_{\alpha}}{n_{\beta}}{n_{\gamma}} ,\nonumber \\
\tilde{G}^{{\alpha}{\beta}{\gamma}} &= R_{\tilde{G}\epsilon_{\alpha}\epsilon_{\beta}\epsilon_{\gamma}}(j_{\alpha},j_{\beta},j_{\gamma})\thj{\frac{(j_{\alpha}-(1-\epsilon_{\alpha})/2)}{2}}{\frac{(j_{\beta}-(1-\epsilon_{\beta})/2)}{2}}{\frac{j_{\gamma}+\epsilon_{\gamma}}{2}}{m_{\alpha}}{m_{\beta}}{m_{\gamma}}\thj{\frac{(j_{\alpha}-(1+\epsilon_{\alpha})/2)}{2}}{\frac{(j_{\beta}-(1+\epsilon_{\beta})/2)}{2}}{\frac{j_{\gamma}-\epsilon_{\gamma}}{2}}{n_{\alpha}}{n_{\beta}}{n_{\gamma}} .\nonumber
\end{align}
where,

\begin{equation}
\tilde{G}^{{\alpha}{\beta}{\gamma}}=\int d\Omega (\cal{Y}^{{\alpha}})^T\varepsilon\sigma^{i}\cal{Y}^{{\beta}}\cal{V}_i^{\gamma}.
\end{equation}

This can be related to the fermionic integrals listed in (\ref{eq:3SphericalIntegrals}), by using the conjugation relations of the spinor spherical functions (\ref{eq:SpinorConjugate}).

We computed the reduced matrix elements which were not already evaluated in \cite{Aharony:2005bq} explicitly,

\begin{eqnarray}
R_{H+}(x,y,z) & = & \frac{(-1)^{\sigma+1}}{\pi}\left[ \frac{(z+1)(\sigma-z)(\sigma+1)}{2} \right]^\frac{1}{2},  \nonumber\\
R_{H-}(x,y,z) & = & \frac{(-1)^{\tilde{\sigma}+1}}{\pi}\left[ \frac{(z+1)(\tilde{\sigma}-x)(\tilde{\sigma}-y)}{2} \right]^\frac{1}{2}, \nonumber \\
R_{\tilde{G}0}(x,y,z) & = & \frac{(-1)^{\tilde{\sigma}+1}}{\pi}\left[ \frac{(\tilde{\sigma}-x)(\tilde{\sigma}-y)\tilde{\sigma}(\tilde{\sigma}+1)}{(z+1)} \right]^\frac{1}{2},  \\
R_{\tilde{G}1}(x,y,z) & = & \frac{(-1)^{\tilde{\sigma}+1}}{\pi}\left[ \frac{(\tilde{\sigma}-x)(\tilde{\sigma}-y)(\tilde{\sigma}-z-1)(\tilde{\sigma}-z)}{(z+1)} \right]^\frac{1}{2}, \nonumber \\
R_{\tilde{G}2}(x,y,z) & = & \frac{(-1)^{\sigma}}{\pi}\left[ \frac{(\sigma-z)(\sigma+1)(\sigma-y)(\sigma-y+1)}{(z+1)} \right]^\frac{1}{2}, \nonumber
\end{eqnarray}
where $\sigma \equiv (x+y+z)/2$ and $\tilde{\sigma} \equiv (x+y+z+1)/2$ are integers and,
\begin{align}
R_{H+} \equiv R_{H++} = R_{H--},~&~ R_{H-} \equiv R_{H+-} = -R_{H-+},\nonumber\\
R_{\tilde{G}0} \equiv R_{\tilde{G}+++} = R_{\tilde{G}---}, ~&~ R_{\tilde{G}1} \equiv R_{\tilde{G}--+} = -R_{\tilde{G}++-},\\ R_{\tilde{G}2} \equiv R_{\tilde{G}+-+} = R_{\tilde{G}-+-}, ~&~ R_{\tilde{G}2}(x\leftrightarrow y) = -R_{\tilde{G}-++} = -R_{\tilde{G}+--}\nonumber.
\end{align}

%-------- end of "Spherical harmonics integrals" subsection ---------

%------- start of "Angular momentum summations" subsection ----------

\subsection{Performing angular momentum sums} \label{sapp:AMSums}
We computed various sums over spherical harmonics quantum numbers, which appear in the expressions for our diagrams. We list those identities which were not already derived in \cite{Aharony:2006rf}\footnote{Translating the notation in \cite{Aharony:2006rf} to ours: $R_2\rightarrow R_C$ , $R_{3\epsilon_{\alpha}\epsilon_{\beta}}\rightarrow R_{D\epsilon_{\alpha}\epsilon_{\beta}}$, and $R_{4\epsilon_{\alpha}\epsilon_{\beta}\epsilon_{\gamma}}\rightarrow R_{E\epsilon_{\alpha}\epsilon_{\beta}\epsilon_{\gamma}}$. },

\begin{align}
\sum_{m's}G^{\alpha\beta\gamma}G^{\beta\alpha\bar{\gamma}} &= (-1)^{\frac{\epsilon_{\alpha}+\epsilon_{\beta}}{2}}R_{\tilde{G}\epsilon_{\alpha}\epsilon_{\beta}\epsilon_{\gamma}}(j_{\alpha},j_{\beta},j_{\gamma})R_{\tilde{G}\epsilon_{\beta}\epsilon_{\alpha}\epsilon_{\gamma}}(j_{\beta},j_{\alpha},j_{\gamma})\label{eq:GG},\\
\sum_{m's}F^{\bar{\alpha}\beta\gamma}F^{\alpha\bar{\beta}\bar{\gamma}} &= (-1)^{j_{\alpha}+j_{\beta}+j_{\gamma}+\frac{\epsilon_{\alpha}+\epsilon_{\beta}}{2}}R_{H\epsilon_{\alpha}\epsilon_{\beta}}^2(j_{\alpha},j_{\beta},j_{\gamma}),\label{eq:FF}
\end{align}
\begin{align}
\sum_{m's} B^{\alpha\beta\gamma}B^{\bar{\alpha}\bar{\beta}\bar{\gamma}} &= R_B^2(j_{\alpha},j_{\beta},j_{\gamma}), \label{eq:BB1}\\
\sum_{m's,\epsilon's}D^{\alpha\bar{\alpha}\gamma}B^{\beta\bar{\beta}\bar{\gamma}} &= 2(-1)^{j_{\alpha}+j_{\beta}+1}\delta_{j_{\gamma},0}\times\nonumber\\ &\times\sqrt{j_\alpha(j_\alpha+2)}(j_\beta+1)R_{D+}(j_{\alpha},j_{\alpha},0)R_B(j_{\beta},j_{\beta},0),\label{eq:DB}\\
\sum_{m's}F^{\alpha\beta\gamma}F^{\beta\alpha\bar{\gamma}} &= (-1)^{1+\frac{\epsilon_{\alpha}+\epsilon_{\beta}}{2}}R_{H\epsilon_{\alpha}\epsilon_{\beta}}(j_{\alpha},j_{\beta},j_{\gamma})R_{H\epsilon_{\beta}\epsilon_{\alpha}}(j_{\beta},j_{\alpha},j_{\gamma}),\label{eq:FF1}\\
\sum_{m's}F^{\alpha\alpha\gamma}F^{\beta\beta\bar{\gamma}} &= (-1)^{j_{\alpha}+j_{\beta}+j_{\gamma}+1+\frac{\epsilon_{\alpha}+\epsilon_{\beta}}{2}}(j_{\alpha}(j_{\alpha}+1)j_{\beta}(j_{\beta}+1))^{\frac{1}{2}}\delta_{j_{\gamma},0}\times\nonumber\\
&~~~\times R_{H+}(j_{\alpha},j_{\alpha},j_{\gamma})R_{H+}(j_{\beta},j_{\beta},j_{\gamma}), \label{eq:FF2}\\
\sum_{m's,\epsilon_{\alpha}}D^{\alpha\bar{\alpha}\gamma}F^{\beta\beta\bar{\gamma}} &= 2(-1)^{j_{\alpha}+j_{\beta}+\frac{1+\epsilon_{\beta}}{2}}(j_{\alpha}(j_{\alpha}+2)j_{\beta}(j_{\beta}+1))^{\frac{1}{2}}\delta_{j_{\gamma},0}\times\nonumber\\
&~~~\times R_{D+}(j_{\alpha},j_{\alpha},j_{\gamma})R_{H+}(j_{\beta},j_{\beta},j_{\gamma}),\label{eq:DF}\\
\sum_{m's}B^{\alpha\bar{\alpha}\gamma}F^{\beta\beta\bar{\gamma}} &= (-1)^{j_{\alpha}+j_{\beta}+\frac{1+\epsilon_{\beta}}{2}}(j_{\alpha}+1)(j_{\beta}(j_{\beta}+1))^{\frac{1}{2}}\delta_{j_{\gamma},0}\times\nonumber\\
&~~~\times R_B(j_{\alpha},j_{\alpha},j_{\gamma})R_{H+}(j_{\beta},j_{\beta},j_{\gamma}) \label{eq:BF},\\
\sum_{m's,j_{\gamma}}B^{\alpha\beta\gamma}B^{\bar{\alpha}\bar{\beta}\bar{\gamma}} &= \frac{(j_{\alpha}+1)^2(j_{\beta}+1)^2}{2\pi^2}. \label{eq:BB3}
\end{align}

%------- end of "Angular momentum summations" subsection ----------

%------- start of "1-loop self energy diagrams" section --------

\section{1-loop self energy diagrams}\label{app:1Loop}

\subsection{Flat space self-energy diagrams}\label{sapp:1LoopFlat}
\subsubsection{Dimensional regularization}\label{ssapp:1LoopFlatDR}
We write the flat space 1PI self-energy diagrams (figure 1) which were computed in dimensional regularization, in terms of the integrals $I_{m,n}\equiv I_{m,n}(0)$ (\ref{eq:Imn}) ,

\paragraph{\underline{Vector $p^2$ expansion}: $-\frac{1}{2}\langle A_i(0,p)A_j(0,-p)\rangle^{1PI}_{p^2\delta_{ij}}$.}
\begin{align}
\mathbf{SE1a} &= \frac{-4d^2-22d-20}{(d+3)(d+5)}I_{0,2}+\frac{2d^2+14d+4}{(d+3)(d+5)}I_{0,3}-\frac{8(2+d)}{(d+3)(d+5)}I_{0,4},\\
\mathbf{SE1b} &= \frac{d^2+3d-6}{(d+3)(d+5)}I_{1,1},\\
\mathbf{SE1c} &= \frac{2N_s}{d+3}\left[I_{0,3}-\frac{4}{d+5}I_{0,4}\right],\\
\mathbf{SE1d} &= 2N_f\left[I_{1,3}-\frac{4}{d+3}I_{1,4}+I_{0,3}-\frac{4}{d+5}I_{0,4}\right].
\end{align}

\paragraph{\underline{Vector $\omega^2$ expansion}: $-\frac{1}{2}\langle A_i(\omega,0)A_j(-\omega,0)\rangle^{1PI}_{\omega^2\delta_{ij}}$}
\begin{align}
\mathbf{SE1a} &=\frac{d+2}{d+3}\left[2I_{0,3}-8I_{1,4}\right],\\
\mathbf{SE1b} &= -\frac{d+2}{d+3}I_{0,1},\\
\mathbf{SE1c} &= \frac{2N_s}{d+3}\left[I_{0,3}-4I_{1,4}\right],\\
\mathbf{SE1d} &= 2N_f\left[\frac{d+1}{d+3}I_{0,3}+3I_{1,3}-4\frac{d+1}{d+3}I_{1,4}-4I_{2,4}\right].
\end{align}

\paragraph{\underline{Scalar $p^2$ expansion}: $-\frac{1}{2}\langle \phi^a(0,p)\phi^b(0,-p)\rangle^{1PI}_{p^2}$.}
\begin{align}
\mathbf{SE2a} &= -\delta^{ab}\frac{4(d+2)}{d+3}I_{0,2},\\
\mathbf{SE2b} &= \delta^{ab}\frac{d-1}{d+3}I_{1,1},\\
\mathbf{SE2c} &= -\mathrm{tr}(\rho^{a\dag}\rho^b+\rho^{b\dag}\rho^a)\left[-I_{0,2}+\frac{2}{d+3}I_{0,3}\right].
\end{align}

\paragraph{\underline{Scalar $\omega^2$ expansion}: $-\frac{1}{2}\langle \phi^a(\omega,0)\phi^b(-\omega,0)\rangle^{1PI}_{\omega^2}$.}

\begin{align}
\mathbf{SE2a} &= 0,\\
\mathbf{SE2b} &= -\delta^{ab}I_{0,1},\\
\mathbf{SE2c} &= -\mathrm{tr}(\rho^{a\dag}\rho^b+\rho^{b\dag}\rho^a)\left[-I_{0,2}+2I_{1,3}\right].
\end{align}

\paragraph{\underline{Fermion $p$ expansion}: $\langle \psi_I(0,p)\psi^{\dag J}(0,p)\rangle^{1PI}_{p_i\sigma^i}$.}
\begin{align}
\mathbf{SE3a} &= \delta_I^{~J}\frac{2}{d+3}\left[-(d^2+3d+2)I_{0,2}+2(d+2)I_{0,3}\right],\\
\mathbf{SE3b} &= \delta_I^{~J}\left[-2I_{0,1}+\frac{4}{d+3}I_{0,2}\right],\\
\mathbf{SE3c} &= -2(\rho^a\rho^{a\dag})_I^{~J}\left[I_{0,2}-\frac{2}{d+3}I_{0,3}\right].
\end{align}

\paragraph{\underline{Fermion $\omega$ expansion}: $\langle \psi_I(\omega,0)\psi^{\dag J}(\omega,0)\rangle^{1PI}_{\omega}$.}
\begin{align}
\mathbf{SE3a} &= \delta_I^{~J}2(d+2)\left[-I_{0,2}+2I_{1,3}\right],\\
\mathbf{SE3b} &= \delta_I^{~J}\left[2I_{0,1}-4I_{1,2}\right]=0,\\
\mathbf{SE3c} &= -2(\rho^a\rho^{a\dag})_I^{~J}\left[I_{0,2}-2I_{1,3}\right].
\end{align}

\subsubsection{Cutoff regularization}\label{ssapp:1LoopFlatCO}
We used the following definitions of regulator dependent quantities in the main text,

\begin{align}
\ln\left(\frac{\cal{A}_{nkl}M}{a}\right) &\equiv \int_0^{\infty} dq\frac{qR_g^n(q)R_s^k(q)R_f^l(q)}{q^2+\frac{a^2}{M^2}},\label{eq:RegDepFunA}\\
\cal{B}_1^{s,g} &\equiv \frac{1}{2\pi}\int_0^{\infty} dq R'_f(q)R_{s,g}(q),\label{eq:RegDepFunB}\\
\cal{F}_2^{g,s,f} &\equiv \frac{1}{4\pi^2}\int_0^{\infty}dq qR_{g,s,f}(q)R''_{g,s,f}(q),\label{eq:RegDepFunF}\\
\cal{C}_2^{nkl} &\equiv \frac{1}{4\pi^2}\int_0^{\infty}dq qR_g^n(q)R_s^k(q)R_f^l(q)\label{eq:RegDepFunC}.
\end{align}

The cutoff scheme results for the flat space self energy diagrams in terms of the regulator dependent functions (\ref{eq:RegDepFunA})-(\ref{eq:RegDepFunC}) are,
\paragraph{\underline{Vector $p^2$ expansion}: $-\frac{1}{2}\langle A_i(0,p)A_j(0,-p)\rangle^{1PI}_{p^2\delta_{ij}}$.}
\begin{align}
\mathbf{SE1a} &= -\frac{1}{40\pi^2}-\frac{9}{40\pi^2}\ln\left(\frac{\cal{A}_{200}M}{a}\right)-\frac{1}{15}\cal{F}_2^g,\\
\mathbf{SE1b} &= \frac{1}{10\pi^2}\ln\left(\frac{\cal{A}_{100}M}{a}\right),\\
\mathbf{SE1c} &= \frac{N_s}{48\pi^2}\ln\left(\frac{\cal{A}_{020}M}{a}\right)+\frac{N_s}{240\pi^2}-\frac{N_s}{30}\cal{F}_2^s,\\
\mathbf{SE1d} &= \frac{N_f}{12\pi^2}\ln\left(\frac{\cal{A}_{002}M}{a}\right)+\frac{N_f}{30\pi^2}-\frac{4N_f}{15}\cal{F}_2^f.
\end{align}

\paragraph{\underline{Vector $\omega^2$ expansion}: $-\frac{1}{2}\langle A_i(\omega,0)A_j(-\omega,0)\rangle^{1PI}_{\omega^2\delta_{ij}}$}
\begin{align}
\mathbf{SE1a} &= \frac{1}{24\pi^2}\ln\left(\frac{\cal{A}_{200}M}{a}\right),\\
\mathbf{SE1b} &= -\frac{1}{6\pi^2}\ln\left(\frac{\cal{A}_{100}M}{a}\right),\\
\mathbf{SE1c} &= \frac{N_s}{48\pi^2}\ln\left(\frac{\cal{A}_{020}M}{a}\right),\\
\mathbf{SE1d} &= \frac{N_f}{12\pi^2}\ln\left(\frac{\cal{A}_{002}M}{a}\right).
\end{align}

\paragraph{\underline{Scalar $p^2$ expansion}: $-\frac{1}{2}\langle \phi^a(0,p)\phi^b(0,-p)\rangle^{1PI}_{p^2}$.}
\begin{align}
\mathbf{SE2a} &= -\delta^{ab}\frac{1}{3\pi^2}\ln\left(\frac{\cal{A}_{110}M}{a}\right),\\
\mathbf{SE2b} &= \delta^{ab}\frac{1}{12\pi^2}\ln\left(\frac{\cal{A}_{010}M}{a}\right),\\
\mathbf{SE2c} &= -\mathrm{tr}\left(\rho^{a\dag}\rho^b+\rho^{b\dag}\rho^a\right)\left[-\frac{1}{16\pi^2}\ln\left(\frac{\cal{A}_{002}M}{a}\right)-\frac{1}{48\pi^2}+\frac{\cal{F}_2^f}{6}\right].
\end{align}

\paragraph{\underline{Scalar $\omega^2$ expansion}: $-\frac{1}{2}\langle \phi^a(\omega,0)\phi^b(-\omega,0)\rangle^{1PI}_{\omega^2}$.}
\begin{align}
\mathbf{SE2a} &= 0,\\
\mathbf{SE2b} &= -\frac{1}{4\pi^2}\delta^{ab}\ln\left(\frac{\cal{A}_{010}M}{a}\right),\\
\mathbf{SE2c} &= +\frac{1}{16\pi^2}\mathrm{tr}\left(\rho^{a\dag}\rho^b+\rho^{b\dag}\rho^a\right)\ln\left(\frac{\cal{A}_{002}M}{a}\right).
\end{align}

\paragraph{\underline{Fermion $p$ expansion}: $\langle \psi_I(0,p)\psi^{\dag J}(0,p)\rangle^{1PI}_{p_i\sigma^i}$.}
\begin{align}
\mathbf{SE3a} &= \delta_I^{~J}\left[\frac{1}{12\pi^2}\ln\left(\frac{\cal{A}_{101}M}{a}\right)-\frac{1}{3\pi}\cal{B}_1^g\right],\\
\mathbf{SE3b} &= -\delta_I^{~J}\left(\frac{1}{3\pi^2}\ln\left(\frac{\cal{A}_{001}M}{a}\right)-\frac{1}{6\pi^2}\right),\\
\mathbf{SE3c} &= -(\rho^a\rho^{a\dag})_I^{~J}\left[\frac{1}{8\pi^2}\ln\left(\frac{\cal{A}_{011}M}{a}\right)+\frac{1}{6\pi}\cal{B}_1^s\right].
\end{align}

\paragraph{\underline{Fermion $\omega$ expansion}: $\langle \psi_I(\omega,0)\psi^{\dag J}(\omega,0)\rangle^{1PI}_{\omega}$.}
\begin{align}
\mathbf{SE3a}  &= -\frac{1}{4\pi^2}\delta_I^{~J}\ln\left(\frac{\cal{A}_{101}M}{a}\right),\\
\mathbf{SE3b}  &= 0,\\
\mathbf{SE3c}  &= -\frac{1}{8\pi^2}(\rho^a\rho^{a\dag})_I^{~J}\ln\left(\frac{\cal{A}_{011}M}{a}\right).
\end{align}

\subsection{Curved space self-energy diagrams}\label{sapp:1LoopCurved}
\subsubsection{Dimensional regularization}\label{ssapp:1LoopCurvedDR}

When applying dimensional regularization on $S^3$ we get additional vertices included in the action (\ref{eq:CubicAction}), (\ref{eq:QuarticAction}), coming from the fact that the gauge field has additional components. We needed the following additional cubic vertices,

\begin{align}
\cal{L'}_{3} &= ig\mathrm{tr}\left(-\frac{C^{\beta\gamma\alpha}}{j_{\alpha}(j_{\alpha}+2)}[a^{\beta},A^{\gamma}]\partial_t\partial_a\cal{A}_a^{\alpha}-\frac{C^{\beta\alpha\gamma}}{j_{\alpha}(j_{\alpha}+2)}[a^{\beta}\partial_a\cal{A}_a^{\gamma}]\partial_tA^{\alpha}+D^{\alpha\beta\gamma}[\partial_aA^{\beta},A^{\alpha}]\cal{A}_a^{\gamma}+\right.\nonumber\\
&~~~+\frac{(j_{\alpha}+1)^2-(j_{\beta}+1)^2}{j_{\gamma}(j_{\gamma}+2)}D^{\alpha\beta\gamma}A_{\alpha}A_{\beta}\partial_a\cal{A}_a^{\gamma}+2C^{\alpha\beta\gamma}\cal{A}_a^{\alpha}A^{\beta}\cal{A}_a^{\gamma}+\nonumber\\
&~~~+\frac{C^{\alpha\beta\gamma}}{j_{\alpha}(j_{\alpha}+2)}[\partial_b\cal{A}_b^{\alpha},\partial_aA^{\beta}]\cal{A}^{\gamma}_a-\frac{C^{\gamma\alpha\beta}}{j_{\beta}(j_{\beta}+2)}[A^{\alpha},\partial_a\partial_b\cal{A}_b^{\beta}]\cal{A}_a^{\gamma}-\nonumber\\
&~~~-\frac{(j_{\alpha}+1)^2C^{\beta\alpha\gamma}}{j_{\beta}(j_{\beta}+2)j_{\gamma}(j_{\gamma}+2)}A^{\alpha}\partial_a\cal{A}_a^{\beta}\partial_b\cal{A}_b^{\gamma}-j_{\alpha}(j_{\alpha}+2)\hat{B}^{\alpha\beta\gamma}[\phi^{\alpha}_{\bar{a}},\partial_a\cal{A}^{\beta}_a]\phi^{\gamma}_{\bar{a}}-\nonumber\\
&~~~\left.-B^{\alpha\beta\gamma}[\cal{A}^{\beta}_b,\phi^{\alpha}_{\bar{a}}]\partial_b\phi^{\gamma}_{\bar{a}}\right).
\end{align}

The additional quartic vertices which were necessary for the computation are,

\begin{align}
\cal{L}'_4 &= -\frac{1}{2}g^2\mathrm{tr}\left(B^{\alpha\gamma\bar{\lambda}}D^{\beta\delta\lambda}[\cal{A}_a^{\alpha},A^{\beta}][\cal{A}_a^{\gamma},A^{\delta}]+\hat{B}^{\beta\delta\bar{\lambda}}D^{\alpha\gamma\lambda}[A^{\alpha},\partial_a\cal{A}_a^{\beta}][A^{\gamma},\partial_b\cal{A}_b^{\delta}]+\right.\nonumber\\
&~~~+C^{\lambda\alpha\gamma}C^{\bar{\lambda}\delta\beta}[A^{\alpha},\partial_a\cal{A}_a^{\beta}][\partial_b\cal{A}_b^{\gamma},A^{\delta}]+\frac{C^{\lambda\alpha\gamma}C^{\bar{\lambda}\beta\delta}}{j_{\gamma}(j_{\gamma}+2)j_{\delta}(j_{\delta}+2)}[A^{\alpha},A^{\beta}][\partial_a\cal{A}_a^{\gamma},\partial_b\cal{A}_b^{\delta}]+\nonumber\\
&~~~\left.+B^{\alpha\beta\bar{\lambda}}B^{\gamma\delta\lambda}[\cal{A}^{\alpha}_b,\phi^{\beta}_{\bar{a}}][\cal{A}^{\gamma}_b,\phi^{\delta}_{\bar{a}}]+B^{\beta\delta\bar{\lambda}}\hat{B}^{\alpha\gamma\lambda}[\partial_a\cal{A}^{\alpha}_a,\phi^{\beta}_{\bar{a}}][\partial_b\cal{A}_b^{\gamma},\phi_{\bar{a}}^{\delta}]\right).
\end{align}

Above we used an additional spherical harmonic integral,

\begin{align}
\hat{B}^{\alpha\beta\gamma} &= \frac{1}{j_{\alpha}(j_{\alpha}+2)j_{\beta}(j_{\beta}+2)}\int_{S^3}\vec{\partial}S^{\alpha}\cdot\vec{\partial}S^{\beta}S^{\gamma}.
\end{align}

The solutions to the vector self-energy diagrams are,

\begin{align}
\mathbf{SE1c} &= -\frac{N_s}{\pi^2}([\frac{1}{\epsilon}+\frac{1}{2}\ln(\pi)-\frac{\gamma}{2}+1]+\gamma+\frac{1}{12})+\cal O(\epsilon),\\
\mathbf{SE1d} &= -\frac{4N_f}{\pi^2}\left\{\left[\frac{1}{\epsilon}+\frac{1}{2}\ln(\pi)-\frac{1}{2}\gamma+1\right]+\gamma-\frac{2}{3}\right\}+\cal O(\epsilon),\\
\mathbf{SE1k} &= \frac{N_s}{4\pi^2}+\cal O(\epsilon).
\end{align}

The other diagrams in figures 1 and 3 were computed in \cite{Aharony:2006rf} (see equation (3.59) in that paper).

\subsubsection{Cutoff regularization}\label{ssapp:1LoopCurvedCO}

These are the expressions for the self-energy diagrams on $S^3$ in the cutoff scheme diagram by diagram,

\begin{align}
SE1c &= \frac{N_s}{\pi^2}\left[4\pi^2M^2\cal{C}_2^{020} - \frac{1}{12} - \ln\left(\cal A_{020} M\right) - \gamma \right], \\
SE1d &= \frac{4N_f}{\pi^2}\left( 4\pi^2M^2\cal{C}_2^{002} - \frac{1}{12} + 2\pi^2\cal{F}_2^f -\frac{1}{8} - \ln(\cal A_{002}M) - \gamma \right),\\
SE1k &= -\frac{3N_s}{\pi^2}\left[4\pi^2M^2\cal{C}_2^{010} - \frac{1}{12}\right] , \\
SE2a &= 0, \\
SE2b &= \frac{\delta^{ab}}{2\pi^2}\left[\frac{7}{4} - 4\pi^2M^2\cal{C}_2^{010} + \frac{1}{12} - \ln\left(\cal A_{010} M\right) - \gamma  \right], \\
SE2c &= \frac{\mathrm{tr}(\rho^{a\dag}\rho^b + \rho^{b\dag}\rho^a)}{2\pi^2}\left[4\pi^2 M^2\cal{C}_2^{002} + \frac{5}{12} - \frac{1}{4}\left(\ln(\cal{A}_{002}M)+\gamma+2\ln(2)\right) \right], \\
SE2d &= \mathbf{Q}^{(ab)cc}\left[M^2\cal{C}_2^{010} - \frac{1}{48\pi^2}\right],\\
SE2f &= -\frac{\delta^{ab}}{\pi^2}\left(4\pi^2M^2\cal{C}_2^{100} - \frac{1}{12} - \ln(\cal A_{100}M) - \gamma \right).
\end{align}

The other diagrams in figures 1 and 3 were computed in \cite{Aharony:2006rf} (see equation (4.14) in that paper).
%-------- end of "1-loop self energy diagrams" section ---------

%-------- start of "2-loop vacuum diagrams" section ---------

\section{2-loop vacuum diagrams}\label{app:2LoopDiagrams}
\subsection{Planar diagrams}
By expanding the propagators (\ref{eq:BosPropOferConvention}), (\ref{eq:fProp}) in the diagrams (\ref{eq:L2})-(\ref{eq:L8}), and using the identities in appendix \ref{sapp:AMSums}, we extract the $f_n(x)$ terms in the effective action defined at (\ref{eq:2LoopEffAction})\footnote{In this section we use $\bar{a},\bar{b},\ldots=1,\ldots,N_s$ for the scalars internal indices.},

\begin{align}
f_{1,+}^{L1} &= \sum_{a=1}^{\infty}\sum_{b=1}^{\infty}\sum_{\frac{c}{2}=\frac{|a-b|+1}{2}}^{\frac{a+b-1}{2}}\frac{1}{2(a+1)(b+1)(c+1)(a+b+c+3)}\left[(a+b+c+3)^2R_{E+}^2(a,b,c)+\right. \nonumber \\
&\left.+(a+b-c+1)^2R_{E-}^2(a,b,c)\!+\!(-a+b+c+1)^2R_{E-}^2(b,c,a)\!+\!(a-b+c+1)^2R_{E-}^2(c,a,b)\right]\times\nonumber \\
&\times \left[ \frac{(a+1)x^{c+1}}{-a+b+c+1}-\frac{(a+c+2)} {a-b+c+1}R_g\left(\frac{a+1}{M}\right)R_g\left(\frac{c+1}{M}\right)\right]x^{b+1}\label{eq:fnL1},
\end{align}

\begin{align}\label{eq:fnL2}
f_{1,+}^{L2} &= -2N_s\sum_{a,b=1}^\infty \sum_{\frac{c}{2} = \frac{|a-b|+1}{2}}^{\frac{a+b-1}{2}}\frac{R_C^2(a,b,c)}{(a+1)(b+1)(c+1)(a+b+c+3)} \times \nonumber \\
&\left[ \frac{2(a+b+2)x^{c+1}}{a+b-c+1}R_g\left(\frac{b+1}{M}\right)R_s\left(\frac{a+1}{M}\right) - \frac{2(a+1)x^{b+c+2}}{-a+b+c+1}R_s\left(\frac{a+1}{M}\right) + \right. \nonumber \\
& \left. + \! \frac{(a+c+2)x^{b+1}}{a-b+c+1}\!R_s\!\left(\frac{a+1}{M}\!\right)\!R_s\left(\frac{c+1}{M}\!\right)\!-\!  \frac{(b+1)x^{a+c+2}}{a-b+c+1}R_g\!\left(\frac{b+1}{M}\!\right) \right],
\end{align}

\begin{align}\label{eq:fnL3}
f^{L3}_{1,+} &= -4N_f\! \sum_{a,b=1}^\infty \! \sum_{\frac{c}{2} = \frac{|a-b-1|+1}{2}}^{\frac{a+b-2}{2}}\bigg\{ \frac{R_{G2}^2(a,b,c)\!}{(c+1)(a+b+c+2)(a+b-c)}\bigg(\!(a+b+1)x^{c+1} -\nonumber\\
&- (c+1)x^{a+b+1}\bigg)R_f\left(\frac{a+\frac{1}{2}}{M}\right)R_f\left(\frac{b+\frac{1}{2}}{M}\right)R_g\left(\frac{c+1}{M}\right)\bigg\}, \\
f_{1,-}^{L3} &= 2N_f \sum_{a,b=1}^\infty \Bigg\{\!\Bigg(\sum_{\frac{c}{2} = \frac{|a-b-1|+1}{2}}^{\frac{a+b-2}{2}} \frac{R_{G2}^2(a,b,c)}{(c+1)(a+b+c+2)}\left(x^{b+\frac{1}{2}} + x^{a+c+\frac{3}{2}}+x^{a+\frac{1}{2}} + x^{b+c+\frac{3}{2}}\right)- \nonumber \\
&\!\!\! -\!\! \sum_{\frac{c}{2} = \frac{|a-b|+1}{2}}^{\frac{a+b-1}{2}}\!  \frac{R_{G0}^2(a,b,c) \!+\! R_{G1}^2(a,b,c)}{(c+1)(a-b+c+1)}\left(\! x^{b+\frac{1}{2}}\!- \!x^{a+c+\frac{3}{2}}\right)\Bigg)\times\nonumber\\
&\times\!R_f\left(\frac{a+\frac{1}{2}}{M}\right)R_f\left(\frac{b+\frac{1}{2}}{M}\right)R_g\left(\frac{c+1}{M}\right) \Bigg\},
\end{align}

\begin{align}
f_{1,+}^{L4} &= - 2\mathrm{tr}\left(\rho^{\bar{a}\dag}\rho^{\bar{a}}\right)\!\!\sum_{a,b=1}^\infty\sum_{\frac{c}{2}=\frac{|a-b|}{2}}^{\frac{a+b-2}{2}}\left\{ \frac{R_{H+}^2(a,b,c)}{c+1} \!\left[\! \frac{(a+b+1)x^{c+1}\! -\! (c+1)x^{a+b+1}}{(a+b+c+2)(a+b-c)}\!\right]\right.\times\nonumber\\
&\left.\times R_f\left(\frac{a+\frac{1}{2}}{M}\right)\!R_g\left(\frac{b+\frac{1}{2}}{M}\right)R_s\left(\frac{c+1}{M}\right)\right\},\label{eq:fnL4p}  \end{align}
\begin{align}
f_{1,-}^{L4} &= 2\mathrm{tr}\left(\rho^{\bar{a}\dag}\rho^{\bar{a}}\right)\sum_{a,b=1}^\infty\left\{\left(\sum_{\frac{c}{2}=\frac{|a-b|}{2}}^{\frac{a+b-2}{2}}\left[\frac{R_{H+}^2(a,b,c)}{(c+1)(a+b+c+2)}\left(x^{a+\frac{1}{2}}+x^{b+c+\frac{3}{2}}\right)\right]+\right.\right.\nonumber\\
&\!\left.+\!\sum_{\frac{c}{2}=\frac{|a-b|+1}{2}}^{\frac{a+b-1}{2}}\!\left[\!\frac{R_{H-}^2(a,b,c)}{(c+1)(-a+b+c+1)}\left(x^{b+c+\frac{3}{2}}-x^{a+\frac{1}{2}}\right)\!\right]\right)\times\nonumber\\
&\left.\times R_f\left(\frac{a+\frac{1}{2}}{M}\right)R_f\left(\frac{b+\frac{1}{2}}{M}\right)R_s\left(\frac{c+1}{M}\right)\right\},\label{eq:fnL4m}\\
f_{1,+}^{L5a} &= \frac{1}{3\pi^2}\sum_{a,b=1}^{\infty}\frac{a(a+2)}{a+1}\frac{b(b+2)}{b+1}\left(x^{a+1}+x^{b+1}+x^{a+b+2}\right) R_g\left(\frac{a+1}{M}\right)R_g\left(\frac{b+1}{M}\right) \label{eq:fnL5a},\\
f_{1,+}^{L5b} &= \sum_{a,b=1}^{\infty}\sum_{\frac{c}{2}=\frac{||a-b|-1|+1}{2}}^{\frac{a+b}{2}}\frac{1}{c(c+2)}\left(R_{D+}^2 + R_{D-}^2 \right)\left[ \frac{b-a}{a+1}x^{a+b+2} + \right. \nonumber \\
&\left.+ \left[\frac{b+1}{a+1} + \frac{a+1}{b+1}\right]x^{b+1}R_g\left(\frac{a+1}{M}\right)\right],\label{eq:fnL5b}\\
f_{1,+}^{L6a} &= -\frac{Q^{\bar{a}\bar{a}\bar{b}\bar{b}}}{8\pi^2}\sum_{a,b=1}^{\infty}ab\left(x^a + x^b + x^{a+b}\right)R_s\left(\frac{a}{M}\right) R_s\left(\frac{b}{M}\right)\label{eq:fnL6a},\\
f_{1,+}^{L6b} &= \frac{N_s}{4\pi^2}\sum_{a,b=0}^{\infty}\sum_{\frac{c}{2}= \frac{||a-b|-1|+1}{2}}^{\frac{a+b}{2}}\frac{c+1}{c(c+2)}\left[(b+1)(b-a)x^{a+b+2}+ \right.\nonumber \\
&\left.+ ((a+1)^2+(b+1)^2)x^{b+1}R_s\left(\frac{a+1}{M}\right)\right]\label{eq:fnL6b},\\
f_{1,+}^{L7} &= \frac{N_s}{2\pi^2}\sum_{a,b=1}^{\infty}\left(a-\frac{1}{a}\right)b\left[x^a+x^b+x^{a+b}\right] R_g\left(\frac{a}{M}\right)R_s\left(\frac{b}{M}\right)\label{eq:fnL7},
\end{align}
\begin{align}
f_{1,+}^{L8} &= 2N_f\sum_{a,b=1}^{\infty}\sum_{\frac{c}{2} = \frac{|a-b|+1}{2}}^{\frac{a+b-1}{2}}\frac{R_{H-}^2(a,b,c)}{c(c+2)} x^{a+b+1}, \label{eq:fnL8p}\\
f_{1,-}^{L8} &= 2N_f\sum_{a,b=1}^{\infty}\left\{\left[\sum_{\frac{c}{2} = \frac{||a-b|-1|+1}{2}}^{\frac{a+b}{2}}\frac{R_{H+}^2(a,b,c)}{c(c+2)}-\sum_{\frac{c}{2} = \frac{|a-b|+1}{2}}^{\frac{a+b-1}{2}}\frac{R_{H-}^2(a,b,c)}{c(c+2)} \right]x^{b+1/2}R_f\left(\frac{a + \frac{1}{2} }{M}\right)\right\}\label{eq:fnL8m}.
\end{align}

\subsection{Non-planar diagrams}
Define the functions,
\begin{align}
&\cal K_{nm}(y,z;w) \equiv \left[y^n + z^n\right]\left[w^m - (yz)^m\right] \\
&\cal L_{nm}(y,z,w) \equiv (yz)^n\left(y^m+z^m\right) + (zw)^n\left(z^m+w^m\right) + (wy)^n\left(w^m+y^m\right).
\end{align}

In terms of those the $f_{nm}^{Li}$'s are,

\begin{align}
f_{nm}^{L1}(x)&=-\sum_{a,b=1}^{\infty}\sum_{\frac{c}{2}={\frac{|a-b|+1}{2}}}^{\frac{a+b-1}{2}}\!\!\left\{\frac{\frac{1}{3}(a+b+c+3)^2R^2_{E+}+(a+b-c+1)^2R^2_{E-}}{8(a+1)(b+1)(c+1)}\left[\frac{\cal{L}_{nm}(x^{a+1},x^{b+1},x^{c+1})}{a+b+c+3}+\right.\right.\nonumber\\
&\left.\left.+\frac{\cal{K}_{nm}(x^{a+1},x^{b+1},x^{c+1})}{a+b-c+1}\!+\!\frac{\cal{K}_{nm}(x^{b+1},x^{c+1},x^{a+1})}{-a+b+c+1}\!+\!\frac{\cal{K}_{nm}(x^{c+1},x^{a+1},x^{b+1})}{a-b+c+1}\right]\right\},\label{eq:fnmL1}\\
f_{nm}^{L2}(x)&=-\sum_{a,b=1}^{\infty}\sum_{\frac{c}{2}={\frac{|a-b|+1}{2}}}^{\frac{a+b-1}{2}}\left\{\frac{N_sR^2_C}{2(a+1)(b+1)(c+1)}\left[\frac{\cal{L}_{nm}(x^{a+1},x^{b+1},x^{c+1})}{a+b+c+3}+\right.\right.\nonumber\\
&\left.\left.+\frac{\cal{K}_{nm}(x^{a+1},x^{b+1},x^{c+1})}{a+b-c+1}\!+\!\frac{\cal{K}_{nm}(x^{b+1},x^{c+1},x^{a+1})}{-a+b+c+1}\!+\!\frac{\cal{K}_{nm}(x^{c+1},x^{a+1},x^{b+1})}{a-b+c+1}\right]\right\},\label{eq:fnmL2}\\
f_{nm}^{L3}(x)&=-\frac{N_f}{2}\sum_{a,b=1}^{\infty}\!\bigg\{\!\sum_{\frac{c}{2}={\frac{||a-b|-1|+1}{2}}}^{\frac{a+b}{2}}\!\left[\!\frac{2R_{G2}^2}{c+1}\!\left(\frac{\cal{L}_{nm}(-x^{a+\frac{1}{2}},-x^{b+\frac{1}{2}},x^{c+1})}{a+b+c+2}\!+\!\frac{\cal{K}_{nm}(-x^{a+\frac{1}{2}},-x^{b+\frac{1}{2}},x^{c+1})}{a+b-c}\right)\!\right]-\nonumber\\
&-\!\!\sum_{\frac{c}{2}={\frac{|a-b|+1}{2}}}^{\frac{a+b-1}{2}}\!\left[\frac{R_{G0}^2+R_{G1}^2}{c+1}\left(\frac{\cal{K}_{nm}(-x^{b+\frac{1}{2}},x^{c+1},-x^{a+\frac{1}{2}})}{-a+b+c+1}\!+\!\frac{\cal{K}_{nm}(x^{c+1},-x^{a+\frac{1}{2}},-x^{b+\frac{1}{2}})}{a-b+c+1}\right)\right]\bigg\},\label{eq:fnm3}\\
f_{nm}^{L4}(x)&=\frac{\mathrm{tr(\rho^{\bar{a}\dag}\rho^{\bar{a}})}}{2}\left\{\!\sum_{\frac{c}{2}={\frac{|a-b|+1}{2}}}^{\frac{a+b-1}{2}}\!\left[\frac{R^2_{H-}}{c+1}\left(\frac{\cal{K}_{nm}(-x^{b+\frac{1}{2}},x^{c+1},-x^{a+\frac{1}{2}})}{-a+b+c+1}\!+\!\frac{\cal{K}_{nm}(x^{c+1},-x^{a+\frac{1}{2}},-x^{b+\frac{1}{2}})}{a-b+c+1}\right)\right]\!\!-\right.\nonumber\\
&-\left.\sum_{a,b=1}^{\infty}\!\sum_{\frac{c}{2}={\frac{|a-b|}{2}}}^{\frac{a+b-2}{2}}\!\left[\frac{R_{H+}^2}{c+1}\left(\frac{\cal{L}_{nm}(-x^{a+\frac{1}{2}},-x^{b+\frac{1}{2}},x^{c+1})}{a+b+c+2}\!+\!\frac{\cal{K}_{nm}(-x^{a+\frac{1}{2}},-x^{b+\frac{1}{2}},x^{c+1})}{a+b-c}\right)\right]\right\},\\
f_{nm}^{L5a}(x)&=\frac{1}{6\pi^2}\sum_{a,b=1}^{\infty}\frac{a(a+2)b(b+2)}{(a+1)(b+1)}x^{(a+1)n+(b+1)m}(1+x^{(a+1)m}+x^{(b+1)n}),\label{eq:fnmL5a}
\end{align}
\begin{align}
f_{nm}^{L5b}(x)&=\sum_{a,b=1}^{\infty}\sum_{\frac{c}{2}=\frac{||a-b|-1|+1}{2}}^{\frac{a+b}{2}}\frac{R^2_{D+}+R^2_{D-}}{2c(c+2)}\cdot\frac{x^{(a+1)n+(b+1)m}}{a+1}(a+b+2+(b-a)(x^{(a+1)m}+x^{(b+1)n})),\label{eq:fnmL5b}\\
f_{nm}^{L6a}(x)&=-\frac{\mathbf{Q}^{\bar{a}\bar{a}\bar{b}\bar{b}}}{16\pi^2}\sum_{a,b=0}^{\infty}(a+1)(b+1)x^{(a+1)n+(b+1)m}(1+x^{(a+1)m}+x^{(b+1)n}),\label{eq:fnmL6a}\\
f_{nm}^{L6b}(x)&=\frac{N_s}{4}\sum_{a,b=0}^{\infty}\sum_{\frac{c}{2}={\frac{||a-b|-1|+1}{2}}}^{\frac{a+b}{2}}\frac{R^2_{B}}{c(c+2)}\cdot\frac{x^{(a+1)n+(b+1)m}}{a+1}(a+b+2+(b-a)(x^{(a+1)m}+x^{(b+1)n})),\label{eq:fnmL6b}\\
f_{nm}^{L7}(x)&=\frac{N_s}{4\pi^2}\sum_{a=1,b=0}^{\infty}\frac{a(a+2)(b+1)}{a+1}x^{(a+1)n+(b+1)m}(1+x^{(a+1)m}+x^{(b+1)n}),\label{eq:fnmL7}\\
f_{nm}^{L8}(x)&=-N_f\sum_{a,b=1}^{\infty}\left\{\sum_{\frac{c}{2}=\frac{||a-b|-1|+1}{2}}^{\frac{a+b}{2}}\frac{R^2_{H+}}{c(c+2)}(-1)^{m+n}x^{(a+\frac{1}{2})n+(b+\frac{1}{2})m}-\right.\nonumber\\
&\left.-\sum_{\frac{c}{2}=\frac{|a-b|+1}{2}}^{\frac{a+b-1}{2}}\frac{R^2_{H-}}{c(c+2)}(-1)^mx^{(a+b+1)n}(x^{(a+\frac{1}{2})m}+x^{(b+\frac{1}{2})m})\right\}\label{eq:fnmL8}.
\end{align}

The summations over $n,m$ can be performed explicitly, and in particular one has,

\begin{align}
&\cal K(y,z;w) \equiv \sum_{m,n=1}^\infty \cal K_{mn}(y,z;w) = \left[\frac{y}{1-y} + \frac{z}{1-z}\right]\left[\frac{w}{1-w} - \frac{yz}{1-yz}\right] \\
&\cal L(y,z,w) \equiv \sum_{m,n=1}^\infty \cal L_{mn}(y,z,w) = \frac{yz}{1-yz}\left(\frac{y}{1-y} + \frac{z}{1-z}\right) + \frac{zw}{1-zw}\left(\frac{z}{1-z} + \frac{w}{1-w}\right) + \nonumber \\
&~~~ + \frac{wy}{1-wy}\left(\frac{w}{1-w} + \frac{y}{1-y}\right).
\end{align}

In terms of those the expressions for $\widetilde{F}^{np}_2$ defined in (\ref{eq:Fnp}) are,

\begin{align}
\widetilde{F}^{np}_{L1}(x)&=\sum_{a,b=1}^{\infty}\sum_{\frac{c}{2}={\frac{|a-b|+1}{2}}}^{\frac{a+b-1}{2}}\!\!\left\{\frac{\frac{1}{3}(a+b+c+3)^2R^2_{E+}+(a+b-c+1)^2R^2_{E-}}{4(a+1)(b+1)(c+1)}\left[\frac{\cal{L}(x^{a+1},x^{b+1},x^{c+1})}{a+b+c+3}+\right.\right.\nonumber\\
&\left.\left.+\frac{\cal{K}(x^{a+1},x^{b+1},x^{c+1})}{a+b-c+1}\!+\!\frac{\cal{K}(x^{b+1},x^{c+1},x^{a+1})}{-a+b+c+1}\!+\!\frac{\cal{K}(x^{c+1},x^{a+1},x^{b+1})}{a-b+c+1}\right]\right\},\label{eq:FnpL1}
\end{align}
\begin{align}
\widetilde{F}^{np}_{L2}(x)&=\sum_{a,b=1}^{\infty}\sum_{\frac{c}{2}={\frac{|a-b|+1}{2}}}^{\frac{a+b-1}{2}}\left\{\frac{N_sR^2_C}{(a+1)(b+1)(c+1)}\left[\frac{\cal{L}(x^{a+1},x^{b+1},x^{c+1})}{a+b+c+3}+\right.\right.\nonumber\\
&\left.\left.+\frac{\cal{K}(x^{a+1},x^{b+1},x^{c+1})}{a+b-c+1}\!+\!\frac{\cal{K}(x^{b+1},x^{c+1},x^{a+1})}{-a+b+c+1}\!+\!\frac{\cal{K}(x^{c+1},x^{a+1},x^{b+1})}{a-b+c+1}\right]\right\},\label{eq:FnpL2}\\
\widetilde{F}^{np}_{L3}(x)&=N_f\sum_{a,b=1}^{\infty}\!\bigg\{\!\sum_{\frac{c}{2}={\frac{||a-b|-1|+1}{2}}}^{\frac{a+b}{2}}\!\left[\!\frac{2R_{G2}^2}{c+1}\!\left(\frac{\cal{L}(-x^{a+\frac{1}{2}},-x^{b+\frac{1}{2}},x^{c+1})}{a+b+c+2}\!+\!\frac{\cal{K}(-x^{a+\frac{1}{2}},-x^{b+\frac{1}{2}},x^{c+1})}{a+b-c}\right)\!\right]-\nonumber\\
&-\!\!\sum_{\frac{c}{2}={\frac{|a-b|+1}{2}}}^{\frac{a+b-1}{2}}\!\left[\frac{R_{G0}^2+R_{G1}^2}{c+1}\left(\frac{\cal{K}(-x^{b+\frac{1}{2}},x^{c+1},-x^{a+\frac{1}{2}})}{-a+b+c+1}\!+\!\frac{\cal{K}(x^{c+1},-x^{a+\frac{1}{2}},-x^{b+\frac{1}{2}})}{a-b+c+1}\right)\right]\bigg\},\label{eq:Fnp3}\\
\widetilde{F}^{np}_{L4}(x)&=\mathrm{tr}(\rho^{\bar{a}\dag}\rho^{\bar{a}})\!\sum_{a,b=1}^{\infty}\!\left\{\sum_{\frac{c}{2}={\frac{|a-b|}{2}}}^{\frac{a+b-2}{2}}\!\left[\frac{R_{H+}^2}{c+1}\left(\frac{\cal{L}(-x^{a+\frac{1}{2}},-x^{b+\frac{1}{2}},x^{c+1})}{a+b+c+2}\!+\!\frac{\cal{K}(-x^{a+\frac{1}{2}},-x^{b+\frac{1}{2}},x^{c+1})}{a+b-c}\right)\right]\!\!-\right.\nonumber\\
&-\left.\sum_{\frac{c}{2}={\frac{|a-b|+1}{2}}}^{\frac{a+b-1}{2}}\!\left[\frac{R^2_{H-}}{c+1}\left(\frac{\cal{K}(-x^{b+\frac{1}{2}},x^{c+1},-x^{a+\frac{1}{2}})}{-a+b+c+1}\!+\!\frac{\cal{K}(x^{c+1},-x^{a+\frac{1}{2}},-x^{b+\frac{1}{2}})}{a-b+c+1}\right)\right]\right\},\label{eq:FnpL4}\\
\widetilde{F}^{np}_{L5a}(x)&=-\frac{1}{3\pi^2}\sum_{a,b=1}^{\infty}\frac{a(a+2)b(b+2)}{(a+1)(b+1)}\frac{x^{a+b+2}(1+x^{a+1}+x^{b+1}-3x^{a+b+2})}{(1-x^{a+1})(1-x^{b+1})(1-x^{a+b+2})},\label{eq:FnpL5a}\\
\widetilde{F}^{np}_{L5b}(x)&=\sum_{a,b=1}^{\infty}\sum_{\frac{c}{2}=\frac{||a-b|-1|+1}{2}}^{\frac{a+b}{2}}\frac{R^2_{D+}+R^2_{D-}}{c(c+2)}\frac{x^{a+b+2}}{x^{a+b+2}-1}\left[1+\frac{1+b}{1+a}\cdot\frac{1+x^{a+1}+x^{b+1}-3x^{a+b+2}}{(1-x^{a+1})(1-x^{b+1})}\right],\label{eq:FnpL5b}\\
\widetilde{F}^{np}_{L6a}(x)&=\frac{\mathbf{Q}^{\bar{a}\bar{a}\bar{b}\bar{b}}}{8\pi^2}\sum_{a,b=0}^{\infty}(a+1)(b+1)\frac{x^{a+b+2}(1+x^{a+1}+x^{b+1}-3x^{a+b+2})}{(1-x^{a+1})(1-x^{b+1})(1-x^{a+b+2})},\label{eq:FnpL6a}\\
\widetilde{F}^{np}_{L6b}(x)&=-\frac{N_s}{2}\sum_{a,b=0}^{\infty}\sum_{\frac{c}{2}={\frac{||a-b|-1|+1}{2}}}^{\frac{a+b}{2}}\frac{R^2_{B}}{c(c+2)}\frac{x^{a+b+2}}{1-x^{a+b+2}}\left[1+\frac{1+b}{1+a}\cdot\frac{1+x^{a+1}+x^{b+1}-3x^{a+b+2}}{(1-x^{a+1})(1-x^{b+1})}\right],\label{eq:FnpL6b}\\
\widetilde{F}^{np}_{L7}(x)&=-\frac{N_s}{2\pi^2}\sum_{a=1,b=0}^{\infty}\frac{a(a+2)(b+1)}{a+1}\frac{x^{a+b+2}(1+x^{a+1}+x^{b+1}-3x^{a+b+2})}{(1-x^{a+1})(1-x^{b+1})(1-x^{a+b+2})},\label{eq:FnpL7}
\end{align}
\begin{align}
\widetilde{F}^{np}_{L8}(x)&=2N_f\sum_{a,b=1}^{\infty}\left\{\sum_{\frac{c}{2}=\frac{||a-b|-1|+1}{2}}^{\frac{a+b}{2}}\frac{R^2_{H+}}{c(c+2)}\frac{x^{a+b+1}}{(1+x^{a+\frac{1}{2}})(1+x^{b+\frac{1}{2}})}+\right.\nonumber\\
&\left.+\sum_{\frac{c}{2}=\frac{|a-b|+1}{2}}^{\frac{a+b-1}{2}}\frac{R^2_{H-}}{c(c+2)}\frac{x^{a+b+1}}{1-x^{a+b+1}}\left(\frac{x^{a+\frac{1}{2}}}{1+x^{a+\frac{1}{2}}}+\frac{x^{b+\frac{1}{2}}}{1+x^{b+\frac{1}{2}}}\right)\right\}\label{eq:FnpL8}.
\end{align}

In the above, when they were not written, the arguments of the reduced matrix elements are $R_{\#}=R_{\#}(a,b,c)$.
\subsection{Regulator dependent part of 2-loop diagrams}\label{sapp:RegDep2Loop}
The regulator dependent parts of the diagrams are obtained from the $f_{1}^{Li}$'s in (\ref{eq:fnL1})-(\ref{eq:fnL8m}) by expanding the summands in large $a$ and picking up the terms which diverge when we take the cutoff to infinity\footnote{Actually the $\sum RR'$ terms are regulator independent since $\int R(q)R'(q)dq=-\frac{1}{2}$ but we included those anyway.}.
The results in terms of the functions of temperature defined in (\ref{eq:xFunctions}) are,

%--------------------------------------- f_1 -------------------------------------%
\begin{align}
f_{L1}^{reg} &=  -\frac{1}{3\pi^2}\sum_{a=1}^\infty\left\{\left[f(x)a+\frac{8g_1(x)-7f(x)}{5}\frac{1}{a}\right]R_g^2\left(\frac{a}{M}\right)+\right.\nonumber\\
&+\left.\frac{g_1(x)-4f(x)}{10M}\left[R_g\left(\frac{a}{M}\right)R_g'\left(\frac{a}{M}\right)+\frac{a}{M}R_g\left(\frac{a}{M}\right)R_g''\left(\frac{a}{M}\right)\right]\right\},\\
%--------------------------------------- f_2 -------------------------------------%
f_{L2}^{reg} &= - \frac{N_s}{3\pi^2}\sum_{a=1}^\infty\left\{\frac{f(x)}{2}aR_s^2\left(\frac{a}{M}\right)+g_1(x)\frac{1}{a}R_s\left(\frac{a}{M}\right)R_g\left(\frac{a}{M}\right)+\right.\nonumber\\
&\left.+\frac{g_1(x)-4f(x)}{20M}\left[R_s\left(\frac{a}{M}\right)R_s'\left(\frac{a}{M}\right)+\frac{a}{M}R_s\left(\frac{a}{M}\right)R_s''\left(\frac{a}{M}\right)\right]\right\},\\
%--------------------------------------- f_3 -------------------------------------%
f_{L3+}^{reg}&=-\frac{N_f}{3\pi^2}\sum_{a=1}^\infty\left\{f(x)(2a-1)R_f^2\left(\frac{a-\frac{1}{2}}{M}\right)+\frac{2g_1(x)-3f(x)}{5M}\left[R_f\left(\frac{a-\frac{1}{2}}{M}\right)R_f'\left(\frac{a-\frac{1}{2}}{M}\right)+\right.\right.\nonumber\\
&\left.\left.+\frac{a}{M}R_f\left(\frac{a-\frac{1}{2}}{M}\right)R_f''\left(\frac{a-\frac{1}{2}}{M}\right)\right]\right\},\\
f_{L3-}^{reg} &= -\frac{N_f}{\pi^2}\frac{2x^{\frac{3}{2}}(1+x)}{(1-x)^4}\sum_{a=1}^\infty\left\{\frac{1}{a}R_g\left(\frac{a}{M}\right)R_f\left(\frac{a-\frac{1}{2}}{M}\right)-\frac{1}{2M}R_g\left(\frac{a}{M}\right)R_f'\left(\frac{a-\frac{1}{2}}{M}\right)\right\},\\
%--------------------------------------- f_4 -------------------------------------%
f_{L4+}^{reg} &= -\frac{\text{tr}\left(\rho^{\bar{a}\dag}\rho^{\bar{a}}\right)}{4\pi^2}\sum_{a=1}^{\infty}\left\{k(x)(2a-1)R_f^2\left(\frac{a-\frac{1}{2}}{M}\right)+\frac{g_1(x)}{3M}\left[R_f\left(\frac{a-\frac{1}{2}}{M}\right)R'_f\left(\frac{a-\frac{1}{2}}{M}\right)+\right.\right.\nonumber\\
&\left.\left.+\frac{a}{M}R_f\left(\frac{a-\frac{1}{2}}{M}\right) R''_f\left(\frac{a-\frac{1}{2}}{M}\right)\right]\right\},
\end{align}
\begin{align}
f_{L4-}^{reg} &= \frac{\text{tr}\left(\rho^{\bar{a}\dag}\rho^{\bar{a}}\right)}{2\pi^2M}\frac{x^{\frac{3}{2}}(1+x)}{(1-x)^4}\sum_{a=1}^{\infty}R_s\left(\frac{a}{M}\right)R'_f\left(\frac{a-\frac{1}{2}}{M}\right),\\
%--------------------------------------- f_5a ------------------------------------%
f_{L5a}^{reg} &= \frac{2}{3\pi^2}f(x)\sum_{a=1}^{\infty}\left\{\left[a-\frac{1}{a}\right]R_g\left(\frac{a}{M}\right)\right\},\\
%--------------------------------------- f_5b ------------------------------------%
f_{L5b}^{reg} &= \sum_{a=1}^\infty\left\{\left[f(x)a+\frac{8g_1(x)-7f(x)}{5}\frac{1}{a}\right]R_g\left(\frac{a}{M}\right)\right\},\\
%--------------------------------------- f_6a ------------------------------------%
f_{L6a}^{reg} &= -\frac{1}{4\pi^2}k(x)Q^{\bar{a}\bar{a}\bar{b}\bar{b}}\sum_{b=1}^\infty b R_s\left(\frac{b}{M}\right),\\
%--------------------------------------- f_6b ------------------------------------%
f_{L6b}^{reg} &=  \frac{N_s}{4\pi^2}\sum_{a=1}^{\infty}\left[k(x)aR_s\left(\frac{a}{M}\right)+2\frac{2g_1(x)+3k(x)}{3}\frac{1}{a} R_s\left(\frac{a}{M}\right)\right],\\
%--------------------------------------- f_7a ------------------------------------%
f_{L7a}^{reg} &= \frac{N_s}{2\pi^2}\sum_{a=1}^{\infty}\left\{k(x) \left(a-\frac{1}{a}\right)R_g\left(\frac{a}{M}\right) + f(x)a R_s\left(\frac{a}{M}\right)\right\},\\
%--------------------------------------- f_8 -------------------------------------%
f_{L8+}^{reg} &= 0 \\
f_{L8-}^{reg}
&= \frac{N_f}{\pi^2}\frac{2x^{\frac{3}{2}}(1+x)}{(1-x)^4}\sum_{a=1}^{\infty}\frac{1}{a}R_f\left(\frac{a - \frac{1}{2} }{M}\right).
\end{align}

Or in integral form using the Euler-Maclaurin formula and the regulator dependent integrals (\ref{eq:RegDepFunA})-(\ref{eq:RegDepFunC}),

\begin{align}
f_{L1}^{reg} &= -\frac{4}{3} f(x) \left(M^2\cal C_2^{200} - \frac{1}{48\pi^2}\right) + \frac{g_1(x)-4f(x)}{60\pi^2} - 2\frac{g_1(x)-4f(x)}{15}\cal F_2^g - \nonumber \\
& - \frac{8g_1(x)-7f(x)}{15\pi^2}\left[\ln\left(\cal A_{200} M\right) + \gamma_E\right], \\
f_{L2}^{reg} &= -\frac{N_s}{\pi^2}\left[\frac{1}{6}f(x)\left(4\pi^2M^2\cal C_2^{020} - \frac{1}{12}\right) + \frac{1}{3}g_1(x)\left(\ln\left(\cal A_{110}M\right) + \gamma_E\right) + \right.
\nonumber \\
&~~~\left. +  \frac{g_1(x)-4f(x)}{60}\left\{-\frac{1}{2} + 4\pi^2\cal F_2^s\right\}\right],\\
f_{L3,+}^{reg} &= -\frac{N_f}{15\pi^2}\left\{ 5f(x)\left[8\pi^2M^2\cal C_2^{002} + \frac{1}{12} \right] - \frac{2g_1(x) - 3f(x)}{2} + \left[2g_1(x) - 3f(x)\right]4\pi^2\cal F_2^f \right\},  \\
f_{L3,-}^{reg} &= \frac{N_f}{\pi^2}\frac{2x^{\frac{3}{2}}(1+x)}{(1-x)^4}\left\{\pi\cal B_1^g - \left[\ln\left(\cal A_{101}M\right) + \gamma_E\right] \right\},  \\
f_{L4,+}^{reg} &= -\frac{\text{tr}(\rho^{\bar{a}\dag}\rho^{\bar{a}})}{12\pi^2}\left\{k(x)\left[24\pi^2M^2\cal C_2^{002} + \frac{1}{4} \right] - \frac{g_1(x)}{2} + g_1(x)4\pi^2\cal F_2^f\right\},  \\
f_{L4,-}^{reg} &= \frac{\text{tr}(\rho^{\bar{a}\dag}\rho^{\bar{a}})}{\pi}\frac{x^{\frac{3}{2}}(1+x)}{(1-x)^4}\cal B_1^s,\end{align}
\begin{align}
f_{L5a}^{reg} &= \frac{8}{3}f(x) \left(M^2 \cal C_2^{100} - \frac{1}{48\pi^2}\right) - \frac{2}{3\pi^2}f(x)\left[\ln\left(\cal A_{100} M\right) + \gamma_E\right] , \\
f_{L5b}^{reg} &= \frac{4}{3}f(x) \left(M^2 \cal C_2^{100} - \frac{1}{48\pi^2}\right) + \frac{8g_1(x) + 3f(x)}{15\pi^2} \left[\ln\left(\cal A_{100} M\right) +\gamma_E\right] ,\\ %
f_{L6a}^{reg} &= -k(x)Q^{\bar{a}\bar{a}\bar{b}\bar{b}}\left(M^2\cal C_2^{010} - \frac{1}{48\pi^2}\right) , \\
f_{L6b}^{reg} &= N_s\left\{ k(x)\left(M^2\cal C_2^{010} - \frac{1}{48\pi^2}\right) + \frac{1}{2\pi^2}\left(k(x) + \frac{2}{3}g_1(x)\right) \left[\ln\left(\cal A_{010} M\right) + \gamma_E\right] \right\} ,\\
f_{L7a}^{reg} &= N_s\left[ 2f(x)\left(M^2\cal C_2^{010}-\frac{1}{48\pi^2}\right) + 2k(x)\left(M^2\cal C_2^{100}-\frac{1}{48\pi^2} \right) - \right. \nonumber \\
&~~~\left. -  \frac{1}{2\pi^2}k(x)\left(\ln\left(\cal A_{100}M\right) + \gamma_E\right) \right], \\
f_{L8,+}^{reg} &= 0 ,\\
f_{L8,-}^{reg} &=  \frac{N_f}{\pi^2}\frac{2x^{\frac{3}{2}}(1+x)}{(1-x)^4}\left[ \ln\left(\cal A_{001} M\right) + \gamma_E\right].
\end{align}

%-------- end of "2-loop vacuum diagrams" section ---------
\pagebreak

%\bibliography{ThesisBib}
%\bibliographystyle{plain}

\end{document}